# TIGER: A Tuning-Insensitive Approach for Optimally Estimating Gaussian Graphical Models

Han Liu* and Lie Wang†

July 5th, 2012

**Abstract**

We propose a new procedure for estimating high dimensional Gaussian graphical models. Our approach is asymptotically tuning-free and non-asymptotically tuning-insensitive: it requires very few efforts to choose the tuning parameter in finite sample settings. Computationally, our procedure is significantly faster than existing methods due to its tuning-insensitive property. Theoretically, the obtained estimator is simultaneously minimax optimal for precision matrix estimation under different norms. Empirically, we illustrate the advantages of our method using thorough simulated and real examples. The R package `bigmatrix` implementing the proposed methods is available on the Comprehensive R Archive Network: http://cran.r-project.org/.

## 1 Introduction

We consider the problem of learning high dimensional Gaussian graphical models: let $\boldsymbol{x}_1, \ldots, \boldsymbol{x}_n$ be $n$ data points from a $d$-dimensional random vector $\boldsymbol{X} = (X_1, ..., X_d)^T$ with $\boldsymbol{X} \sim N_d(\boldsymbol{0}, \boldsymbol{\Sigma})$. We want to estimate an undirected graph denoted by $G = (V, E)$, where $V$ contains nodes corresponding to the $d$ variables in $\boldsymbol{X}$, and the edge set $E$ describes the conditional independence relationships between $X_1, ..., X_d$. Let $\boldsymbol{X}_{\backslash \{j,k\}} := \{X_\ell : \ell \neq i, j\}$. We say the joint distribution of $\boldsymbol{X}$ is Markov to $G$ if $X_j$ is independent of $X_k$ given $\boldsymbol{X}_{\backslash \{j,k\}}$ for all $(j, k) \notin E$. For Gaussian distributions, the graph $G$ is known to be encoded by the precision matrix $\boldsymbol{\Theta} := \boldsymbol{\Sigma}^{-1}$. More specifically, no edge connects $X_j$ and $X_k$ if and only if $\boldsymbol{\Theta}_{jk} = 0$. The graph estimation problem is then reduced to the estimation of the precision matrix $\boldsymbol{\Theta}$. Such a problem is also called *covariance selection* (Dempster, 1972).

---

*Department of Operations Research and Financial Engineering, Princeton University, Princeton, NJ 08544, USA; e-mail: `hanliu@princeton.edu` Research supported by NSF Grant III–1116730.

†Department of Mathematics, Massachusetts Institute of Technology, Cambridge, MA 02139, USA; e-mail: `liewang@math.mit.edu` Research supported by NSF Grant DMS-1005539.

In low dimensions where $d < n$, Drton and Perlman (2007, 2008) develop a multiple testing procedure for identifying the sparsity pattern of the precision matrix. In high dimensions where $d \gg n$, Meinshausen and Bühlmann (2006) propose a neighborhood pursuit approach for estimating Gaussian graphical models by solving a collection of sparse regression problems using the Lasso (Tibshirani, 1996; Chen et al., 1998). This approach can be viewed as a pseudo-likelihood approximation of the full likelihood. A related approach is to directly estimate $\boldsymbol{\Theta}$ by penalizing the likelihood using the $L_1$-penalty (Banerjee et al., 2008; Yuan and Lin, 2007; Friedman et al., 2008). To further reduce the estimation bias, Lam and Fan (2009); Jalali et al. (2012); Shen et al. (2012) propose either greedy algorithms or non-convex penalties for sparse precision matrix estimation. Under certain conditions, Ravikumar et al. (2011); Rothman et al. (2008) study the theoretical properties of the penalized likelihood methods. Yuan (2010) and Cai et al. (2011a) also propose the graphical Dantzig selector and CLIME respectively, which can be solved by linear programming and are more amenable to theoretical analysis than the penalized likelihood approach. More recently, Liu and Luo (2012) and Sun and Zhang (2012) propose the SCIO and scaled-Lasso methods, which estimate the sparse precision matrix in a column-by-column fashion and have good theoretical properties.

Besides Gaussian graphical models, Liu et al. (2012) propose a semiparametric procedure named *nonparanormal* SKEPTIC which extends the Gaussian family to the more flexible semiparametric Gaussian copula family. Instead of assuming $\boldsymbol{X}$ follows a Gaussian distribution, they assume there exists a set of monotone functions $f_1, \ldots, f_d$, such that the transformed data $f(\boldsymbol{X}) := (f_1(X_1), \ldots, f_d(X_d))^T$ is Gaussian. More details can be found in Liu et al. (2012) and Lafferty et al. (2012). Zhao et al. (2012) developed a scalable software package to implement the nonparanormal algorithms. Other nonparametric graph estimation methods include forest graphical models (Liu et al., 2011) or conditional graphical models (Liu et al., 2010a).

Most of these methods require choosing some tuning parameters that control the bias-variance tradeoff. Theoretical justifications of these methods are usually built on some oracle choices of tuning parameters that cannot be implemented in practice. It remains an challenging problem on choosing the regularization parameter in a data-dependent way. Popular techniques include the $C_p$-statistic (Mallows, 1973), AIC (Akaike information criterion, Akaike (1973)), BIC (Bayesian information criterion, Schwarz (1978)), extended BIC (Chen and Chen, 2008, 2012; Foygel and Drton, 2010), RIC (Risk inflation criterion, Foster and George (1994)), cross validation (Efron, 1982), and covariance penalization (Efron, 2004). Most of these methods require data splitting and have been only justified for low dimensional settings. Significant progress has been made recently on developing likelihood-free regularization selection techniques, including permutation methods (Wu et al., 2007;

Boos et al., 2009; Lysen, 2009) and subsampling methods (Lange et al., 2004; Ben-david et al., 2006; Meinshausen and Bühlmann, 2010; Bach, 2008). Meinshausen and Bühlmann (2010) and Bach (2008) and Liu et al. (2010b) also propose to select the tuning parameters using subsampling. However, these subsampling based methods are computationally expensive and are still lack of theoretical guarantees.

In this paper we propose a new procedure for estimating high dimensional Gaussian graphical models. Our method, named TIGER (Tuning-Insensitive Graph Estimation and Regression), owes a "tuning-insensitive property": it automatically adapts to the unknown sparsity pattern and is asymptotically tuning-free. In finite sample settings, we only need to pay very few efforts on tuning the regularization parameter. The main idea is to estimate the precision matrix $\Theta$ in a column-by-column fashion. For each column, the computation is reduced to a sparse regression problem. This idea has been adopted by many methods, including the neighborhood pursuit (Meinshausen and Bühlmann, 2006), graphical Dantzig selector (Yuan, 2010), CLIME (Cai et al., 2011a), SCIO (Liu and Luo, 2012), and the scaled-Lasso method (Sun and Zhang, 2012). These methods differ from each other mainly by how they solve the sparse regression subproblem: the graphical Dantzig selector and CLIME use the Dantzig selector, the SCIO and neighborhood pursuit use the Lasso, while Sun and Zhang (2012) use the scaled-Lasso (Sun and Zhang, 2012). Unlike these existing methods, the TIGER solves this sparse regression problem using the SQRT-Lasso (Belloni et al., 2012). By using the SQRT-Lasso regression, the TIGER owes the tuning-insensitive property and improves upon existing methods both theoretically and empirically.

The main advantage of the TIGER over existing methods is its asymptotic tuning-free property, which allows us to use the entire dataset to efficiently learn and select the model. In contrast, it is well known that the cross-validation and subsampling methods are computationally expensive. Moreover, they may potentially waste valuable data which could otherwise be exploited to learn a better model (Bishop et al., 2003). With such a tuning-insensitive property, the TIGER also allows us to conduct more objective scientific data analysis.

Another advantage of the TIGER is its computational simplicity and scalability for large datasets. For problems with large dimensionality $d$, the TIGER divides the whole problem into many subproblems, in each subproblem it estimates one column of the precision matrix by solving a simple SQRT-Lasso problem. The final matrix estimator is formed by combining the vector solutions into a matrix. This procedure can be solved in a parallel fashion and achieves a linear scale up with the number of CPU cores. An additional performance improvement comes from the tuning-insensitive property of the TIGER. Most existing methods exploit cross-validation to choose tuning parameters, which requires computing the solutions over a full regularization path. In contrast, the TIGER solves the SQRT-Lasso subproblem only for several fixed tuning parameters. For sparse problems, this is significantly faster than existing methods.

In the current paper, we prove theoretical guarantees of the TIGER esitmator $\widehat{\Theta}$ of the true precision matrix $\Theta$. In particular, let $\|\Theta\|_{\max} := \max_{jk} |\Theta_{jk}|$ and $\|\Theta\|_1 := \max_j \sum_k |\Theta_{jk}|$. Under the assumption that the condition number of $\Theta$ is bounded by a constant, we establish the elementwise sup-norm rate of convergence:

$$\|\widehat{\Theta} - \Theta\|_{\max} = O_P\left(\|\Theta\|_1 \sqrt{\frac{\log d}{n}}\right). \quad (1)$$

If we further assume $\|\Theta\|_1 \leq M_d$ where $M_d$ may scale with $d$, the obtained rate in (1) is minimax optimal over the model class consisting of precision matrices with bounded condition numbers. This result allows us to effectively conduct graph estimation without the need of irrepresentable conditions.

Let $I(\cdot)$ be the indicator function and $s := \sum_{j \neq k} I(\Theta_{jk} \neq 0)$ be the number of nonzero off-diagonal elements of $\Theta$. The result in (1) implies that the Frobenious norm error between $\widehat{\Theta}$ and $\Theta$ satisfies:

$$\|\widehat{\Theta} - \Theta\|_{\mathsf{F}} := \sqrt{\sum_{i,j} |\widehat{\Theta}_{jk} - \Theta_{jk}|^2} = O_P\left(\|\Theta\|_1 \sqrt{\frac{(d+s)\log d}{n}}\right). \quad (2)$$

Similarly, if we assume $\|\Theta\|_1 \leq M_d$ where $M_d$ may scale with $d$, the rate in (2) is minimax optimal for the Frobenious norm error in the same model class consisting of precision matrices with bounded condition numbers.

Let $\|\Theta\|_2$ be the largest eigenvalue of $\Theta$ (i.e., $\|\Theta\|_2$ is the spectral norm of $\Theta$) and $k := \max_{i=1,\ldots,d} \sum_j I(\Theta_{ij} \neq 0)$. We also prove that

$$\|\widehat{\Theta} - \Theta\|_2 \leq \|\widehat{\Theta} - \Theta\|_1 = O_P\left(k\|\Theta\|_2 \sqrt{\frac{\log d}{n}}\right). \quad (3)$$

Under the same condition that $\|\Theta\|_1 \leq M_d$ where $M_d$ may scale with $d$, this spectral norm rate in (3) is also minimax optimal over the same model class as before.

Besides these theoretical results, we also establish a relationship between the SQRT-Lasso and the scaled-Lasso proposed by Sun and Zhang (2012). More specifically, the objective function of the scaled-Lasso can be viewed as a variational upper bound of the SQRT-Lasso. This relationship allows us to develop a very efficient algorithm for the TIGER .

Computationally, the TIGER is significantly faster than existing methods since very few tunings are needed. In particularly, we propose an iterative algorithm with initial values searched by the Alternating Direction Method of Multipliers (ADMM). For each reduced sparse regression subproblem, the computational complexity is of the same order as solving one single Lasso with a sparse solution. Under the parallel computing framework, our algorithm is even faster and more scalable than the `glasso` and `huge` packages (Friedman et al., 2008; Zhao et al., 2012). Empirically, we present thorough numerical simulations to compare the graph recovery and parameter estimation performance of our method with other approaches. A real data experiment on a gene expression dataset is also provided to back up our theory. The R package `bigmatrix` implementing the proposed methods is available on the Comprehensive R Archive Network: http://cran.r-project.org/.

The rest of the paper is organized as follows. In Section 2, we introduce basic notations and backgrounds on Gaussian graphical models. In Section 3, we present the TIGER estimator and its computational algorithm. In Section 4, we present the theoretical properties including the rates of convergence for parameter estimation and graph recovery. We also provide further discussions on the connections and differences of our results with other related methods. In Section 5, we demonstrate its numerical performance through synthetic and real datasets. The proofs of the main results are given in the appendix.

## 2 Background

Let $\boldsymbol{x}_1, \ldots, \boldsymbol{x}_n$ be $n$ data points from a $d$-dimensional Gaussian random vector $\boldsymbol{X} := (X_1, \ldots, X_d)^T \sim N_d(\boldsymbol{0}, \boldsymbol{\Sigma})$. We denote $\boldsymbol{x}_i := (x_{i1}, \ldots, x_{id})^T$. As has been discussed in the previous section, we define precision matrix to be $\boldsymbol{\Theta} := \boldsymbol{\Sigma}^{-1}$. In this section, we start with some notations followed by an introduction of solving Gaussian graphical models in a column-by-column approach.

### 2.1 Notations

Let $\boldsymbol{v} := (v_1, \ldots, v_d)^T \in \mathbb{R}^d$ and $I(\cdot)$ be the indicator function, for $0 < q < \infty$, we define

$$\|\boldsymbol{v}\|_q := \Big(\sum_{j=1}^{d} |v_j|^q\Big)^{1/q}. \qquad (4)$$

We also define $\|\boldsymbol{v}\|_0 := \sum_{j=1}^{d} I(v_j \neq 0)$ and $\|\boldsymbol{v}\|_\infty := \max_j |v_j|$.

Let $\mathbf{A} \in \mathbb{R}^{d \times d}$ be a symmetric matrix and $I, J \subset \{1, \ldots, d\}$ be two sets, we denote $\mathbf{A}_{I,J}$ to be the submatrix of $\mathbf{A}$ with rows and columns indexed by $I$ and $J$. Let $\mathbf{A}_{*j}$ be the $j^{\text{th}}$ column of $\mathbf{A}$ and $\mathbf{A}_{*\setminus j}$ be the submatrix of $\mathbf{A}$ with the $j^{\text{th}}$ column $\mathbf{A}_{*j}$ removed.

We define the following matrix norms:

$$\|\mathbf{A}\|_q := \max_{\|v\|_q=1} \|\mathbf{A}\boldsymbol{v}\|_q \ , \|\mathbf{A}\|_{\max} := \max_{jk} |\mathbf{A}_{jk}|, \ \text{and} \ \|\mathbf{A}\|_{\mathsf{F}} = \Big(\sum_{j,k} |\mathbf{A}_{jk}|^2\Big)^{1/2}. \qquad (5)$$

It is easy to see that when $q = \infty$, $\|\mathbf{A}\|_\infty = \|\mathbf{A}\|_1$. We also denote $\Lambda_{\max}(\mathbf{A})$ and $\Lambda_{\min}(\mathbf{A})$ to be the largest and smallest eigenvalues of $\mathbf{A}$.

## 2.2 Gaussian Graphical Model and Column-by-Column Regression

Let $\boldsymbol{X} \sim N_d(\boldsymbol{0}, \boldsymbol{\Sigma})$, the conditional distribution of $X_j$ given $\boldsymbol{X}_{\setminus j}$ satisfies

$$X_j \,|\, \boldsymbol{X}_{\setminus j} \sim N_{d-1}\left(\boldsymbol{\Sigma}_{\setminus j,j}(\boldsymbol{\Sigma}_{\setminus j, \setminus j})^{-1}\boldsymbol{X}_{\setminus j},\, \boldsymbol{\Sigma}_{jj} - \boldsymbol{\Sigma}_{\setminus j,j}(\boldsymbol{\Sigma}_{\setminus j,\setminus j})^{-1}\boldsymbol{\Sigma}_{\setminus j,j}\right). \tag{6}$$

Let $\boldsymbol{\alpha}_j := (\boldsymbol{\Sigma}_{\setminus j,\setminus j})^{-1}\boldsymbol{\Sigma}_{\setminus j,j} \in \mathbb{R}^{d-1}$ and $\sigma_j^2 := \boldsymbol{\Sigma}_{jj} - \boldsymbol{\Sigma}_{\setminus j,j}(\boldsymbol{\Sigma}_{\setminus j,\setminus j})^{-1}\boldsymbol{\Sigma}_{\setminus j,j}$. We have

$$X_j = \boldsymbol{\alpha}_j^T \boldsymbol{X}_{\setminus j} + \epsilon_j, \tag{7}$$

where $\epsilon_j \sim N\left(0,\, \sigma_j^2\right)$ is independent of $\boldsymbol{X}_{\setminus j}$. By the block matrix inversion formula, we have

$$\boldsymbol{\Theta}_{jj} = (\mathrm{Var}(\epsilon_j))^{-1} = \sigma_j^{-2}, \tag{8}$$

$$\boldsymbol{\Theta}_{\setminus j,j} = -(\mathrm{Var}(\epsilon_j))^{-1}\boldsymbol{\alpha}_j = -\sigma_j^{-2}\boldsymbol{\alpha}_j. \tag{9}$$

Therefore, we can recover $\boldsymbol{\Theta}$ in a column by column manner by regressing $X_j$ on $\boldsymbol{X}_{\setminus j}$ for $j = 1, 2, \cdots, d$. For example, let

$$\mathbf{X} := \begin{pmatrix} x_{11} & \cdots & x_{1d} \\ \vdots & \cdots & \vdots \\ x_{n1} & \cdots & x_{nd} \end{pmatrix} \in \mathbb{R}^{n \times d} \tag{10}$$

be the data matrix. We denote by $\boldsymbol{\alpha}_j := (\alpha_{j1}, \ldots, \alpha_{j(d-1)})^T \in \mathbb{R}^{d-1}$. Meinshausen and Bühlmann (2006) propose to iteratively estimate each $\boldsymbol{\alpha}_j$ by solving the Lasso regression:

$$\widehat{\boldsymbol{\alpha}}_j = \operatorname*{argmin}_{\boldsymbol{\alpha}_j \in \mathbb{R}^{d-1}} \frac{1}{2n}\|\mathbf{X}_{*j} - \mathbf{X}_{*\setminus j}\boldsymbol{\alpha}_j\|_2^2 + \lambda_j \|\boldsymbol{\alpha}_j\|_1, \tag{11}$$

where $\lambda_j$ is a tuning parameter. Once $\widehat{\boldsymbol{\alpha}}_j$ is given, we get the neighborhood edges by reading out the nonzero coefficients of $\boldsymbol{\alpha}_j$. The final graph estimate $\widehat{G}$ is obtained by either the "AND" or "OR" rule on combining the neighborhoods for all the $d$ nodes. However, the neighborhood pursuit method of Meinshausen and Bühlmann (2006) only estimates the graph $G$ and does not estimate the inverse covariance matrix $\boldsymbol{\Theta}$.

To explicitly estimate $\boldsymbol{\Theta}$, Yuan (2010) proposes to estimate $\boldsymbol{\alpha}_j$ by solving the Dantzig selector:

$$\widehat{\boldsymbol{\alpha}}_j = \operatorname*{argmin}_{\boldsymbol{\alpha}_j \in \mathbb{R}^{d-1}}\|\boldsymbol{\alpha}_j\|_1 \text{ subject to } \|\widehat{\boldsymbol{\Sigma}}_{\setminus j,j} - \widehat{\boldsymbol{\Sigma}}_{\setminus j,\setminus j}\boldsymbol{\alpha}_j\|_\infty \le \gamma_j, \tag{12}$$

where $\widehat{\boldsymbol{\Sigma}} := \frac{1}{n}\mathbf{X}\mathbf{X}^T$ is the sample covariance matrix and $\gamma_j$ is a tuning parameter. Once $\widehat{\boldsymbol{\alpha}}_j$ is given, we can estimate $\sigma_j^2$ by

$$\widehat{\sigma}_j^2 = \left[1 - 2\widehat{\boldsymbol{\alpha}}_j^T\widehat{\boldsymbol{\Sigma}}_{\setminus j,j} + \widehat{\boldsymbol{\alpha}}_j^T\widehat{\boldsymbol{\Sigma}}_{\setminus j,\setminus j}\widehat{\boldsymbol{\alpha}}_j\right]^{-1}. \tag{13}$$

We then get the estimate $\widehat{\boldsymbol{\Theta}}$ of $\boldsymbol{\Theta}$ by plugging $\widehat{\boldsymbol{\alpha}}_j$ and $\widehat{\sigma}_j^2$ into (8) and (9). Yuan (2010) analyzes the $L_1$-norm error $\|\widehat{\boldsymbol{\Theta}} - \boldsymbol{\Theta}\|_1$ and shows its minimax optimality over certain model space. However, no graph estimation result is provided for this approach.

In another work, Sun and Zhang (2012) propose to estimate $\boldsymbol{\alpha}_j$ and $\sigma_j$ by solving a scaled-Lasso problem:

$$\widehat{\boldsymbol{b}}_j, \widehat{\sigma}_j = \operatorname*{argmin}_{\boldsymbol{b}=(b_1,\ldots,b_d)^T, \sigma} \left\{ \frac{\boldsymbol{b}^T \widehat{\boldsymbol{\Sigma}} \boldsymbol{b}}{2\sigma} + \frac{\sigma}{2} + \lambda \sum_{k=1}^d \widehat{\boldsymbol{\Sigma}}_{kk} |b_k| \text{ subject to } b_j = -1 \right\}. \tag{14}$$

Once $\widehat{\boldsymbol{b}}_j$ is obtained, $\boldsymbol{\alpha}_j = \widehat{\boldsymbol{b}}_{\backslash j}$. Sun and Zhang (2012) analyze the spectral-norm rate of convergence of the obtained precision matrix estimator. They did not investigate the elementwise sup-norm and graph recovery performance. In the next section, we will show that the scaled-Lasso estimator is highly related to our proposed procedure and will discuss the relationship in more details.

To estimate both precision matrix $\boldsymbol{\Theta}$ and graph $G$, Cai et al. (2011a) proposes the CLIME estimator, which directly estimates the $j^{\text{th}}$ column of $\boldsymbol{\Theta}$ by solving

$$\widehat{\boldsymbol{\Theta}}_{*j} = \operatorname*{argmin}_{\boldsymbol{\Theta}_{*j}} \|\boldsymbol{\Theta}_{*j}\|_1 \text{ subject to } \|\widehat{\boldsymbol{\Sigma}} \boldsymbol{\Theta}_{*j} - \boldsymbol{e}_j\|_\infty \leq \delta_j, \text{ for } j = 1, \ldots, d, \tag{15}$$

where $\boldsymbol{e}_j$ is the $j^{\text{th}}$ canonical vector and $\delta_j$ is a tuning parameter. Cai et al. (2011a) show that this convex optimization can be formulated into a linear program and has the potential to scale to large problems. Once $\widehat{\boldsymbol{\Theta}}$ is obtained, we use another tuning parameter $\tau$ to threshold $\boldsymbol{\Theta}$ to estimate the graph $G$. In a follow-up work of the CLIME, Liu and Luo (2012) propose the SCIO estimator, which solves the $j^{\text{th}}$ column of $\boldsymbol{\Theta}$ by

$$\widehat{\boldsymbol{\Theta}}_{*j} = \operatorname*{argmin}_{\boldsymbol{\Theta}_{*j}} \left\{ \frac{1}{2} \boldsymbol{\Theta}_{*j}^T \widehat{\boldsymbol{\Sigma}} \boldsymbol{\Theta}_{*j} - \boldsymbol{e}_j^T \boldsymbol{\Theta}_{*j} + \lambda_j \|\boldsymbol{\Theta}_{*j}\|_1 \right\}. \tag{16}$$

The justifications of most of these graph estimation methods are built on some theoretical choices of tuning parameters that cannot be implemented in practice. For example, in the neighborhood pursuit method and the graphical Dantzig selector, the tuning parameter $\lambda_j$ and $\gamma_j$ depend on $\sigma_j^2$, which is unknown. Practically, we usually set $\lambda = \lambda_1 = \cdots = \lambda_d$ and $\gamma = \gamma_1 = \cdots = \gamma_d$ to reduce the number of tuning parameters. However, as we will illustrate in later sections, such a choice makes the estimating procedure non-adaptive to inhomogeneous graphs. The tuning parameters of the CLIME and SCIO depend on $\|\boldsymbol{\Theta}\|_1$, which is unknown. In general, these methods employ cross validation to conduct data-dependent tuning parameter selection and Liu and Luo (2012) provide theoretical analysis of the cross-validation estimator. However, as we discussed before, cross-validation is computationally expensive and a waste of valuable training data. In the next section, we will describe a tuning-insensitive procedure that simultaneously estimates the precision matrix $\boldsymbol{\Theta}$ and graph $G$ with the optimal rates of convergence.

# 3 Method

In this section we introduce the use of the SQRT-Lasso from Belloni et al. (2012) for simultaneously estimating the graph $G$ and precision matrix $\boldsymbol{\Theta} := \boldsymbol{\Sigma}^{-1}$.

The SQRT-Lasso is a penalized optimization algorithm for solving high dimensional linear regression problems. For a linear regression problem $\boldsymbol{y} = \mathbf{X}\boldsymbol{\beta} + \boldsymbol{\epsilon}$, where $\boldsymbol{y} \in \mathbb{R}^n$ is the response, $\mathbf{X} \in \mathbb{R}^{n \times d}$ is the design matrix, $\boldsymbol{\beta} \in \mathbb{R}^d$ is the vector of unknown coefficients, and $\boldsymbol{\epsilon} \in \mathbb{R}^n$ is the noise vector. The SQRT-Lasso estimates $\boldsymbol{\beta}$ by solving

$$\widehat{\boldsymbol{\beta}} = \arg\min_{\boldsymbol{\beta} \in \mathbb{R}^d} \left\{ \frac{1}{\sqrt{n}} \|\boldsymbol{y} - \mathbf{X}\boldsymbol{\beta}\|_2 + \lambda\|\boldsymbol{\beta}\|_1 \right\}, \tag{17}$$

where $\lambda$ is the tuning parameter. It is shown in Belloni et al. (2012) that the choice of $\lambda$ for the SQRT-Lasso method is asymptotically universal and does not depend on any unknown parameter. In contrast, most of other methods, including the Lasso and Dantzig selector, rely heavily on a known standard deviation of the noise. Moreover, the SQRT-Lasso method achieves near oracle performance for the estimation of $\boldsymbol{\beta}$.

## 3.1 TIGER for Graph and Precision Matrix Estimation

In the discussion of this section, we always condition on the observed data $\boldsymbol{x}_1, \ldots, \boldsymbol{x}_n$. Let $\widehat{\boldsymbol{\Gamma}} := \text{diag}(\widehat{\boldsymbol{\Sigma}})$ be a $d$-dimensional diagonal matrix with the diagonal elements be the same as those in $\widehat{\boldsymbol{\Sigma}}$. We define

$$\boldsymbol{Z} := (Z_1, \ldots, Z_d)^T = \boldsymbol{X}\widehat{\boldsymbol{\Gamma}}^{-1/2}. \tag{18}$$

By (7), we have

$$Z_j \widehat{\boldsymbol{\Gamma}}_{jj}^{1/2} = \boldsymbol{\alpha}_j^T \widehat{\boldsymbol{\Gamma}}_{\setminus j, \setminus j}^{1/2} \boldsymbol{Z}_{\setminus j} + \epsilon_j, \tag{19}$$

We define

$$\boldsymbol{\beta}_j := \widehat{\boldsymbol{\Gamma}}_{\setminus j, \setminus j}^{1/2} \widehat{\boldsymbol{\Gamma}}_{jj}^{-1/2} \boldsymbol{\alpha}_j \quad \text{and} \quad \tau_j^2 = \sigma_j^2 \widehat{\boldsymbol{\Gamma}}_{jj}^{-1}. \tag{20}$$

Therefore, we have

$$Z_j = \boldsymbol{\beta}_j^T \boldsymbol{Z}_{\setminus j} + \widehat{\boldsymbol{\Gamma}}_{jj}^{-1/2} \epsilon_j. \tag{21}$$

We define $\widehat{\mathbf{R}}$ to be the sample correlation matrix: $\widehat{\mathbf{R}} := \big(\text{diag}(\widehat{\boldsymbol{\Sigma}})\big)^{-1/2} \widehat{\boldsymbol{\Sigma}} \big(\text{diag}(\widehat{\boldsymbol{\Sigma}})\big)^{-1/2}$.

Motivated by the model in (21), we propose the following precision matrix estimator.

For $j = 1, \ldots, d$, we estimate the $j^{\text{th}}$ column of $\boldsymbol{\Theta}$ by solving:

$$\widehat{\boldsymbol{\beta}}_j := \underset{\boldsymbol{\beta}_j \in \mathbb{R}^{d-1}}{\text{argmin}} \left\{ \sqrt{1 - 2\boldsymbol{\beta}_j^T \widehat{\mathbf{R}}_{\setminus j, j} + \boldsymbol{\beta}_j^T \widehat{\mathbf{R}}_{\setminus j, \setminus j} \boldsymbol{\beta}_j} + \lambda \|\boldsymbol{\beta}_j\|_1 \right\}, \tag{22}$$

$$\widehat{\tau}_j := \sqrt{1 - 2\widehat{\boldsymbol{\beta}}_j^T \widehat{\mathbf{R}}_{\setminus j, j} + \widehat{\boldsymbol{\beta}}_j^T \widehat{\mathbf{R}}_{\setminus j, \setminus j} \widehat{\boldsymbol{\beta}}_j}, \tag{23}$$

$$\widehat{\boldsymbol{\Theta}}_{jj} = \widehat{\tau}_j^{-2} \widehat{\boldsymbol{\Gamma}}_{jj}^{-1} \quad \text{and} \quad \widehat{\boldsymbol{\Theta}}_{\setminus j, j} = -\widehat{\tau}_j^{-2} \widehat{\boldsymbol{\Gamma}}_{jj}^{-1/2} \widehat{\boldsymbol{\Gamma}}_{\setminus j, \setminus j}^{-1/2} \widehat{\boldsymbol{\beta}}_j. \tag{24}$$

For the estimator in (22), $\lambda$ is a tuning parameter. In the next section, we show that by choosing $\lambda = \pi\sqrt{\dfrac{\log d}{2n}}$, the obtained estimator achieves the optimal rates of convergence in the asymptotic setting. Therefore, our procedure is asymptotically tuning-parameter free. For finite samples, we set

$$\lambda := \zeta\pi\sqrt{\frac{\log d}{2n}}, \qquad (25)$$

and $\zeta$ can be chosen from a range $[\sqrt{2}/\pi, 1]$. Since the choice of $\zeta$ does not depend on any unknown parameters or quantities, we call the procedure *tuning-insensitive*. Empirically, we found that in most cases we encountered, we can simply set $\zeta = \sqrt{2}/\pi$ and the resulting estimator works well in finite sample settings. More details can be found in the simulation section.

If a symmetric precision matrix estimate is preferred, we conduct the following correction: $\widetilde{\Theta}_{jk} \leftarrow \min\{\widehat{\Theta}_{jk}, \widehat{\Theta}_{kj}\}$ for all $k \neq j$. Another symmetrization method is

$$\widetilde{\Theta} \leftarrow \frac{\widehat{\Theta} + \widehat{\Theta}^T}{2}. \qquad (26)$$

As has been shown by Cai et al. (2011a), if $\widehat{\Theta}$ is a good estimator, then $\widetilde{\Theta}$ will also be a good estimator: they achieve the same rates of convergence in the asymptotic settings.

Let $\mathbf{Z} \in \mathbb{R}^{n \times d}$ be the normalized data matrix, i.e., $\mathbf{Z}_{*j} = \mathbf{X}_{*j}\boldsymbol{\Sigma}_{jj}^{-1/2}$ for $j = 1, \ldots, d$. An equivalent form of (22) is

$$\widehat{\boldsymbol{\beta}}_j = \underset{\boldsymbol{\beta}_j \in \mathbb{R}^{d-1}}{\operatorname{argmin}} \left\{ \frac{1}{\sqrt{n}} \|\mathbf{Z}_{*j} - \mathbf{Z}_{*\backslash j}\boldsymbol{\beta}_j\|_2 + \lambda\|\boldsymbol{\beta}_j\|_1 \right\}, \qquad (27)$$

$$\widehat{\tau}_j = \frac{1}{\sqrt{n}}\|\mathbf{Z}_{*j} - \mathbf{Z}_{*\backslash j}\widehat{\boldsymbol{\beta}}_j\|_2. \qquad (28)$$

Once we have $\widehat{\Theta}$, the estimated graph $\widehat{G} := (V, \widehat{E})$ where $(j, k) \in \widehat{E}$ if and only if $\widehat{\Theta}_{jk}\widehat{\Theta}_{kj} \neq 0$.

## 3.2 Relationship with the Scaled-Lasso Estimator

In this subsection, we show that the scaled-Lasso from Sun and Zhang (2012) can be viewed as a variational upper bound of the objective function of the SQRT-Lasso and they are solving the same problem[1].

More specifically, we consider the following optimization:

$$\widehat{\boldsymbol{\beta}}_j := \underset{\boldsymbol{\beta}_j \in \mathbb{R}^{d-1}, \tau_j \geq 0}{\operatorname{argmin}} \left\{ \frac{1 - 2\boldsymbol{\beta}_j^T\widehat{\mathbf{R}}_{\backslash j, j} + \boldsymbol{\beta}_j^T\widehat{\mathbf{R}}_{\backslash j, \backslash j}\boldsymbol{\beta}_j}{2\tau_j} + \frac{\tau_j}{2} + \lambda\|\boldsymbol{\beta}_j\|_1 \right\}, \qquad (29)$$

---
[1]This relationship is proposed and kindly provided by Professor Cun-hui Zhang.

**Proposition 1.** *The optimization in* (22) *and* (29) *provide the same solution* $\widehat{\boldsymbol{\beta}}_j$.

*Proof.* For any $a, b > 0$, we have $a^2 + b^2 \geq 2ab$ and the equality is attained if and only $a = b$. Therefore, we have

$$\frac{1 - 2\boldsymbol{\beta}_j^T \widehat{\mathbf{R}}_{\setminus j,j} + \boldsymbol{\beta}_j^T \widehat{\mathbf{R}}_{\setminus j,\setminus j} \boldsymbol{\beta}_j}{2\tau_j} + \frac{\tau_j}{2} \geq \sqrt{1 - 2\boldsymbol{\beta}_j^T \widehat{\mathbf{R}}_{\setminus j,j} + \boldsymbol{\beta}_j^T \widehat{\mathbf{R}}_{\setminus j,\setminus j} \boldsymbol{\beta}_j}. \quad (30)$$

This shows that the objective function in (29) is an upper bound of the objective function in (22). The equality is attained if and only if

$$\tau_j = \sqrt{1 - 2\boldsymbol{\beta}_j^T \widehat{\mathbf{R}}_{\setminus j,j} + \boldsymbol{\beta}_j^T \widehat{\mathbf{R}}_{\setminus j,\setminus j} \boldsymbol{\beta}_j}. \quad (31)$$

We finish the proof. $\square$

This relationship between the TIGER and scaled-Lasso provides an efficient algorithm as described in the next subsection.

### 3.3 Computational Algorithm

Equation (22) is jointly convex with respect to $\boldsymbol{\beta}_j$ and $\tau_j$ and can be solved by a coordinate-descent procedure. In the $t^{\text{th}}$ iteration, for a given $\tau_j^{(t)}$, we first solve the subproblem

$$\boldsymbol{\beta}_j^{(t+1)} := \underset{\boldsymbol{\beta}_j \in \mathbb{R}^{d-1}}{\operatorname{argmin}} \left\{ \frac{1 - 2\boldsymbol{\beta}_j^T \widehat{\mathbf{R}}_{\setminus j,j} + \boldsymbol{\beta}_j^T \widehat{\mathbf{R}}_{\setminus j,\setminus j} \boldsymbol{\beta}_j}{2\tau_j^{(t)}} + \lambda \|\boldsymbol{\beta}_j\|_1 \right\}, \quad (32)$$

This is a Lasso problem and can be efficiently solved by the coordinate descent algorithm developed by Friedman et al. (2007). Once $\boldsymbol{\beta}_j^{(t+1)}$ is obtained, we can calculate $\tau_j^{(t+1)}$ as

$$\tau_j^{(t+1)} = \sqrt{1 - 2\big(\boldsymbol{\beta}_j^{(t+1)}\big)^T \widehat{\mathbf{R}}_{\setminus j,j} + \big(\boldsymbol{\beta}_j^{(t+1)}\big)^T \widehat{\mathbf{R}}_{\setminus j,\setminus j} \big(\boldsymbol{\beta}_j^{(t+1)}\big)}. \quad (33)$$

We iterate these two steps until the algorithm converges.

On thing to note is that the above algorithm converges fast if a good initial value of $\tau_j$ is provided. For example, if $\tau_j$ is close to $\widehat{\tau}_j$, the algorithm converges in only 3 to 5 iterations. In fact, with a good initial value of $\tau_j$, the computation is roughly the same as running one single tuning parameter of the Lasso with a sparse solution. However, with a bad initial value of $\tau_j$, the computational complexity can be as heavy as calculating the full regularization path of a Lasso problem, which is less efficient.

To obtain a good initial estimate of $\tau_j$, we propose an augmented Lagrange method as follows: We first reparameterize (27) and (28) as

$$\widehat{\boldsymbol{\beta}}_j, \widehat{\boldsymbol{\gamma}} = \underset{\boldsymbol{\beta}_j \in \mathbb{R}^{d-1}, \boldsymbol{\gamma} \in \mathbb{R}^n}{\operatorname{argmin}} \left\{ \frac{1}{\sqrt{n}} \|\boldsymbol{\gamma}\|_2 + \lambda \|\boldsymbol{\beta}_j\|_1 \text{ subject to } \boldsymbol{\gamma} = \mathbf{Z}_{*j} - \mathbf{Z}_{*\setminus j} \boldsymbol{\beta}_j \right\}, \quad (34)$$

$$\widehat{\tau}_j = \frac{1}{\sqrt{n}} \|\widehat{\boldsymbol{\gamma}}\|_2. \quad (35)$$

We consider the following augmented Lagrangian function

$$\mathcal{L}(\boldsymbol{\beta}_j, \boldsymbol{\gamma}, \boldsymbol{\alpha}) := \frac{1}{\sqrt{n}}\|\boldsymbol{\gamma}\|_2 + \lambda\|\boldsymbol{\beta}_j\|_1 + \rho\langle\boldsymbol{\alpha}, \boldsymbol{\gamma} - \mathbf{Z}_{*j} + \mathbf{Z}_{*\setminus j}\boldsymbol{\beta}_j\rangle + \frac{\rho}{2}\|\boldsymbol{\gamma} - \mathbf{Z}_{*j} + \mathbf{Z}_{*\setminus j}\boldsymbol{\beta}_j\|_2^2 \quad (36)$$

where $\rho > 0$ is the penalty parameter for the violation of the linear constraints. For simplicity, throughout this paper we assume that $\rho > 0$ is fixed (In implementations, we simply set $\rho = 1$). $\boldsymbol{\alpha} \in \mathbb{R}^n$ is the Lagrange multiplier vector but rescaled by $\rho$ for computational and notational convenience. This reparametrization decouples the computational dependency in the optimization problem. Therefore a complicated problem can be split into multiple simpler sub-problems, each of which can be solved easily.

The augmented Lagrangian method works in an iterative fashion. Suppose we have the solution $\boldsymbol{\beta}_j^{(t)}$, $\boldsymbol{\gamma}^{(t)}$, $\boldsymbol{\alpha}^{(t)}$ at the $t$-th iteration, the algorithm proceeds as follows:

**Step 1.** Update $\boldsymbol{\beta}_j$ by

$$\boldsymbol{\beta}_j^{(t+1)} = \underset{\boldsymbol{\beta}_j \in \mathbb{R}^{d-1}}{\operatorname{argmin}} \left\{ \lambda\|\boldsymbol{\beta}_j\|_1 + \frac{\rho}{2}\|\boldsymbol{\alpha}^{(t)} + \boldsymbol{\gamma}^{(t)} - \mathbf{Z}_{*j} + \mathbf{Z}_{*\setminus j}\boldsymbol{\beta}_j\|_2^2 \right\}. \quad (37)$$

Let $\boldsymbol{u}^{(t)} := \mathbf{Z}_{*j} - \boldsymbol{\alpha}^{(t)} - \boldsymbol{\gamma}^{(t)}$. and $\lambda_\rho := \lambda/\rho$. The problem in (37) is equivalent to

$$\boldsymbol{\beta}_j^{(t+1)} = \underset{\boldsymbol{\beta}_j \in \mathbb{R}^{d-1}}{\operatorname{argmin}} \left\{ \lambda_\rho\|\boldsymbol{\beta}_j\|_1 + \frac{1}{2}\|\boldsymbol{u}^{(t)} - \mathbf{Z}_{*\setminus j}\boldsymbol{\beta}_j\|_2^2 \right\}. \quad (38)$$

This is a Lasso subproblem which can be efficiently solved by the coordinate descent algorithm (Friedman et al., 2007).

**Step 2.** Given $\boldsymbol{\beta}_j^{(t+1)}$, we then update $\boldsymbol{\gamma}$ by

$$\boldsymbol{\gamma}^{(t+1)} = \underset{\boldsymbol{\gamma} \in \mathbb{R}^n}{\operatorname{argmin}} \left\{ \frac{1}{\sqrt{n}}\|\boldsymbol{\gamma}\|_2 + \frac{\rho}{2}\|\boldsymbol{\gamma} - \mathbf{Z}_{*j} + \mathbf{Z}_{*\setminus j}\boldsymbol{\beta}_j^{(t+1)} + \boldsymbol{\alpha}^{(t)}\|_2^2 \right\}. \quad (39)$$

The problem (39) has the closed-form solution by soft-thresholding,

$$\boldsymbol{\gamma}^{(t+1)} = \left(\mathbf{Z}_{*j} - \mathbf{Z}_{*\setminus j}\boldsymbol{\beta}_j^{(t+1)} - \boldsymbol{\alpha}^{(t)}\right) \cdot \left(1 - \frac{1}{\rho\sqrt{n}\|\mathbf{Z}_{*j} - \mathbf{Z}_{*\setminus j}\boldsymbol{\beta}_j^{(t+1)} - \boldsymbol{\alpha}^{(t)}\|_2}\right)_+, \quad (40)$$

where $(x)_+ := \max\{0, x\}$ is the hinge function.

**Step 3.** Given $\boldsymbol{\gamma}^{(t+1)}$ and $\boldsymbol{\beta}_j^{(t+1)}$, we update the Lagrange multiplier $\boldsymbol{\alpha}$ by

$$\boldsymbol{\alpha}^{(t+1)} = \boldsymbol{\alpha}^{(t)} + \boldsymbol{\gamma}^{(t+1)} - \mathbf{Z}_{*j} + \mathbf{Z}_{*\setminus j}\boldsymbol{\beta}_j^{(t+1)}. \quad (41)$$

Since we have rescaled the Lagrange multiplier $\boldsymbol{\alpha}$ by $\rho$, the updating equation for $\alpha$ is independent of $\rho$.

The algorithm stops when the following convergence criterion is satisfied

$$\max\left\{\frac{\|\boldsymbol{\beta}_j^{(t+1)} - \boldsymbol{\beta}_j^{(t)}\|_2}{\|\boldsymbol{\beta}_j^{(t)}\|_2}, \frac{\|\boldsymbol{\beta}_j^{(t+1)} - \boldsymbol{\gamma}^{(t)}\|_2}{\|\boldsymbol{\beta}_j^{(t)}\|_2}\right\} \leq \epsilon, \quad (42)$$

where $\epsilon > 0$ is a precision tolerance parameter. Theoretically, we can set $\epsilon$ to be a very small value, e.g. $\epsilon = 10^{-6}$. This directly solves $\widehat{\beta}_j$. However, empirically, we found it is more efficient to set $\epsilon = 10^{-3}$ and only obtain a crude initial estimate $\widehat{\tau}_j^{\text{crude}}$. We then use $\widehat{\tau}_j^{\text{crude}}$ as the initial value and alternatively solve (32) and (33). Such a hybrid procedure delivers the best empirical performance.

### 3.4 Fine-tune the Regularization Parameter

To secure the best finite sample performance, we could also find-tune the $\zeta$ in (25) on a small interval $[\sqrt{2}/\pi, 1]$. Practically, due to the tuning-insensitive property of our procedure, we find it suffices to only cross-validate 3 values and pick the best one: $\zeta \in \{\sqrt{2}/\pi, 0.6, 1\}$. In general, all these three values guarantee that the solutions are relatively sparse, the algorithm runs very efficiently.

## 4 Theoretical Properties

In this section we investigate the theoretical properties of the proposed method. We begin with some notations and assumptions. We define a matrix class $\mathcal{M}(\xi_{\max}, k)$:

$$\mathcal{M}(\xi_{\max}, k) := \left\{ \boldsymbol{\Theta} = \boldsymbol{\Theta}^T \in \mathbb{R}^{d \times d} : \boldsymbol{\Theta} \succ \mathbf{0}, \ \frac{\Lambda_{\max}(\boldsymbol{\Theta})}{\Lambda_{\min}(\boldsymbol{\Theta})} \leq \xi_{\max}, \ \max_i \sum_j I(\boldsymbol{\Theta}_{ij} \neq 0) \leq k \right\}, \quad (43)$$

where $\xi_{\max}$ is a constant and $k$ may scale with the sample size $n$. We first list down three required assumptions:

(A1) $\boldsymbol{\Theta} \in \mathcal{M}(\xi_{\max}, k)$,

(A2) $k^2 \log d = o(n)$,

(A3) $\limsup_{n \to \infty} \max_{1 \leq j \leq d} \boldsymbol{\Sigma}_{jj}^2 \dfrac{\log d}{n} < \dfrac{1}{4}$.

All these assumptions are mild. Assumption (A1) only requires the inverse covariance matrix $\boldsymbol{\Theta} := \boldsymbol{\Sigma}^{-1}$ to have a bounded condition number. Assumption (A2) is equivalent to

$$\lim_{n \to \infty} k \sqrt{\frac{\log d}{n}} = 0. \quad (44)$$

In later analysis, we will show that this condition is necessary to secure the consistency of the precision matrix estimation under different matrix norms. Assumption (A3) constrains that the marginal variance of $X_j$ should not diverge too fast.

## 4.1 Precision Matrix Estimation Consistency

We study the estimation error of precision matrix $\widehat{\Theta} - \Theta$ under different norms, including spectral norm, matrix $L_1$-norm, elementwise sup-norm, and Frobenius norm. The rate under the elementwise sup-norm is important for graph recovery. The rate under the spectral norm is important since it leads to the consistency of the estimation of eigenvalues and eigenvectors, which may further be utilized to analyze the theoretical properties of downstream statistical inference. We analyze the rate of spectral norm by bounding the $L_1$-norm rate. The rate of Frobenius is also a useful measure on the accuracy of the estimation of $\Theta$ and can be viewed as the sum of squared errors for estimating individual rows. Our main results indicate that the TIGER procedure simultaneously achieves the optimal rates of convergence under all these different matrix norms. We present these results in two main theorems and compare our results with related work in the literature. The proofs of these theorems can be found in the appendix.

Theorem 2 provides the rates of convergence under the matrix $L_1$ and spectral norms.

**Theorem 2** ($L_1$ and spectral norm rates)**.** *We choose the regularization parameter $\lambda$ as in (25) with $\zeta = 1$. Under Assumptions (A1), (A2), and (A3), we have*

$$\left\|\widehat{\Theta} - \Theta\right\|_1 = O_P\left(k\|\Theta\|_2\sqrt{\frac{\log d}{n}}\right), \tag{45}$$

$$\left\|\widehat{\Theta} - \Theta\right\|_2 = O_P\left(k\|\Theta\|_2\sqrt{\frac{\log d}{n}}\right). \tag{46}$$

*Proof.* The proof of (45) can be found in Appendix D.1. The proof of (46) follows from the inequality that $\left\|\widehat{\Theta} - \Theta\right\|_2 \leq \left\|\widehat{\Theta} - \Theta\right\|_1$. □

If we further assume $\|\Theta\|_1 \leq M_d$ where $M_d$ may scale with $d$, i.e., we define the following new matrix class $\mathcal{M}(M_d, \xi_{\max}, k)$:

$$\mathcal{M}(M_d, \xi_{\max}, k) := \left\{\Theta \in \mathcal{M}(\xi_{\max}, k) : \|\Theta\|_1 \leq M_d\right\}. \tag{47}$$

The result of Theorem 2 implies that

$$\sup_{\Theta \in \mathcal{M}(M_d, \xi_{\max}, k)} \left\|\widehat{\Theta} - \Theta\right\|_2 = O_P\left(kM_d\sqrt{\frac{\log d}{n}}\right). \tag{48}$$

Based on the results of Cai et al. (2011b), Liu and Luo (2012), and Yuan (2010), this rate of convergence is minimax optimal on model class $\mathcal{M}(M_d, \xi_{\max}, k)$.

The next theorem provides the rates of convergence under the elementwise sup-norm and Frobenius norm. The elementwise sup-norm result is useful for the graph recovery from the precision matrix $\Theta$.

**Theorem 3** (Elementwise sup-norm and Frobenius-norm rates). *We choose the regularization parameter $\lambda$ as in (25) with $\zeta = 1$. Let $s$ be the total number of nonzero off-diagonal elements of $\Theta$. Under Assumptions (A1), (A2), and (A3), we have*

$$\|\widehat{\Theta} - \Theta\|_{\max} = O_P\Big(\|\Theta\|_1 \sqrt{\frac{\log d}{n}}\Big), \tag{49}$$

$$\|\widehat{\Theta} - \Theta\|_{\mathsf{F}} = O_P\Big(\|\Theta\|_1 \sqrt{\frac{(d+s)\log d}{n}}\Big). \tag{50}$$

*Proof.* The proof of (49) can be found in Appendix D.2. To prove (50), since $s$ is the total number of nonzero off-diagonal elements of $\Theta$, we have

$$\|\widehat{\Theta} - \Theta\|_{\mathsf{F}} \leq \sqrt{s+d} \cdot \|\widehat{\Theta} - \Theta\|_{\max} \leq C \cdot \|\Theta\|_1 \cdot \sqrt{\frac{(d+s)\log d}{n}}, \tag{51}$$

where the second inequality follows from (49). □

Again, based on the results in Cai et al. (2011b) and Liu and Luo (2012), we know that the TIGER achieves the minimax optimal rates of convergence under both elementwise sup-norm and Frobenius norm on model class $\mathcal{M}(M_d, \xi_{\max}, k)$.

In summary, the TIGER simultaneously achieves the optimal rates of convergence for precision matrix estimation under spectral norm, Frobenius norm, matrix $L_1$-norm, and elementwise sup-norm.

## 4.2 Graph Recovery Consistency

Next, we study the graph recovery property of the TIGER. Let $\widehat{\Theta}$ be the estimated precision matrix. Recall that we define the estimated graph $\widehat{G} := (V, \widehat{E})$ where $(j,k) \in \widehat{E}$ if and only if $\widehat{\Theta}_{jk}\widehat{\Theta}_{kj} \neq 0$. Similarly, the true graph $G := (V, E)$ where $(j,k) \in E$ if and only if $\Theta_{jk} \neq 0$. We have the following theorem on graph recovery consistency.

**Theorem 4** (Graph recovery consistency). *We choose the regularization parameter $\lambda$ as in (25) with $\zeta = 1$. We assume Assumptions (A1), (A2), and (A3) hold and for a sufficiently large constant $K$, such that the smallest nonzero element of $\Theta$ satisfies*

$$\Theta_{\mathrm{crit}} := \min_{(j,k) \in E} |\Theta_{jk}| \geq K \|\Theta\|_1 \sqrt{\frac{\log d}{n}}. \tag{52}$$

*We have $\liminf_{n \to \infty} \mathbb{P}\Big(E \subset \widehat{E}\Big) = 1$.*

Theoretically, it is possible that the TIGER delivers some precision matrix estimates with very small nonzero values. To achieve exact recovery, we can hard threshold the estimated precision matrix $\widehat{\Theta}$ to get a sparse precision matrix estimate $\widehat{\Theta}^A$ as has been discussed in Cai et al. (2011a). Let

$$\widehat{\Theta}^A_{jk} := \begin{cases} \widehat{\Theta}_{jk} & \text{if} \quad |\widehat{\Theta}_{jk}| > A\sqrt{\dfrac{\log d}{n}} \\ 0 & \text{if} \quad |\widehat{\Theta}_{jk}| \leq A\sqrt{\dfrac{\log d}{n}} \end{cases}.$$

It can be seen that under the same conditions of Theorem 4, there exists a constant $A$ ($A$ may depend on $K$) such that the above hard threshold estimator achieves exact recovery. More discussions can be found in Cai et al. (2011a). Unlike the CLIME and graphical Dantzig selector where the linear program solver may deliver very small nonzero values (but not exact zero). Our algorithm defined in Section 3 is based on soft-thresholding operator and is capable of delivering exact zero. Empirically, we found the TIGER procedure works very effectively in graph estimation even without this hard-thresholding step. So this hard-thresholding step is more of theoretical intents and is not necessary in applications.

It is worth pointing out that if we are only interested in estimating the graph, assumption (A2) can be relaxed to $k \log d = o(n)$, see for example Meinshausen and Bühlmann (2006) and Jalali et al. (2012). Such an extension is relatively straightforward and will be reported elsewhere.

### 4.3 Discussion

In this subsection, we briefly discuss the theoretical properties of the TIGER estimator and compare them with other existing results.

The SCIO method proposed in Liu and Luo (2012) also provides the rates of convergence for precision matrix estimation under various norms. One can see that the TIGER estimator and SCIO estimator achieve the same rates of convergence in terms of spectral norm, matrix $L_1$-norm, Frobenius norm, and elementwise sup-norm. However, here we consider a much larger matrix class since we only bound the condition number of the precision matrix while the SCIO bounds the largest and smallest eigenvalues from above and below, respectively. Moreover, the SCIO requires the irrepresentable condition, which is stronger than our condition and not required by the TIGER. In fact, it is still an open problem on whether the SCIO estimator can achieve the same rates as the TIGER on the model class $\mathcal{M}(M_d, \xi_{\max}, k)$.

Note that the graphical Dantzig selector proposed in Yuan (2010) also considers the model class where the largest and smallest eigenvalues are bounded from above and below. Therefore the results of the Graphical Dantzig selector are again on a smaller model class than the TIGER estimator. Moreover, the graph recovery performance of the graphical Dantzig selector is still an open problem.

When compared with the CLIME in Cai et al. (2011a), we see that the rate of convergence of the TIGER is faster since the spectral norm rate of the CLIME is $O_P\left(kM_d^2\sqrt{\frac{\log d}{n}}\right)$, and the spectral norm rate of the TIGER is $O_P\left(kM_d\sqrt{\frac{\log d}{n}}\right)$.

When compared with the glasso method, see for example Rothman et al. (2008) and Ravikumar et al. (2011), the TIGER estimator achieves the minimax optimal rates of convergence under spectral norm, Frobenius norm, matrix $L_1$-norm and elementwise sup-norm. In contrast, the only theoretical result of the glasso is in terms of Frobenius norm. it is still an open problem on whether the glasso can achieve the same spectral norm rate of convergence as the TIGER method.

The scaled-Lasso estimator proposed in Sun and Zhang (2012) provides the rates of convergence under spectral norm and matrix $L_1$ norm of the precision matrix estimation. However, the scale-Lasso estimator considers a different model class where the smallest eigenvalue of the correlation matrix is bounded away from zero by a constant. It is not clear on the optimality of the obtained rates of convergence over that model class. Besides, Sun and Zhang (2012) does not provide the elementwise sup-norm analysis and hence does not have the graph recovery result. Though the TIGER has a close relationship with the scaled-Lasso, the theoretical analysis of the current paper is dramatically different from that in Sun and Zhang (2012).

Another related work is Meinshausen and Bühlmann (2006), where they only focus on the graph recovery and do not have precision matrix estimation result.

## 5 Experimental Results

We compare the numerical performance of the TIGER and other methods (glasso and CLIME) in parameter estimation and graph recovery using simulated and real datasets. The TIGER and CLIME algorithms are implemented in the R package `bigmatrix`, and it is publicly available through CRAN. The glasso is implemented in the R package `huge` (ver. 1.2.3).

### 5.1 Numerical Simulations

In our numerical simulations, we consider 6 settings to compare these methods: (i) $n = 200, d = 100$; (ii) $n = 200, d = 200$; (iii) $n = 200, d = 400$; (iv) $n = 400, d = 100$; (v) $n = 400, d = 200$; (vi) $n = 400, d = 400$. We adopt the following models for generating undirected graphs and precision matrices. A typical run of the generated graphs and the heatmaps of the precision matrices are illustrated in Figure 1.

∗ **Scale-free graph**. The degree distribution of the scale-free graph follows a power law.

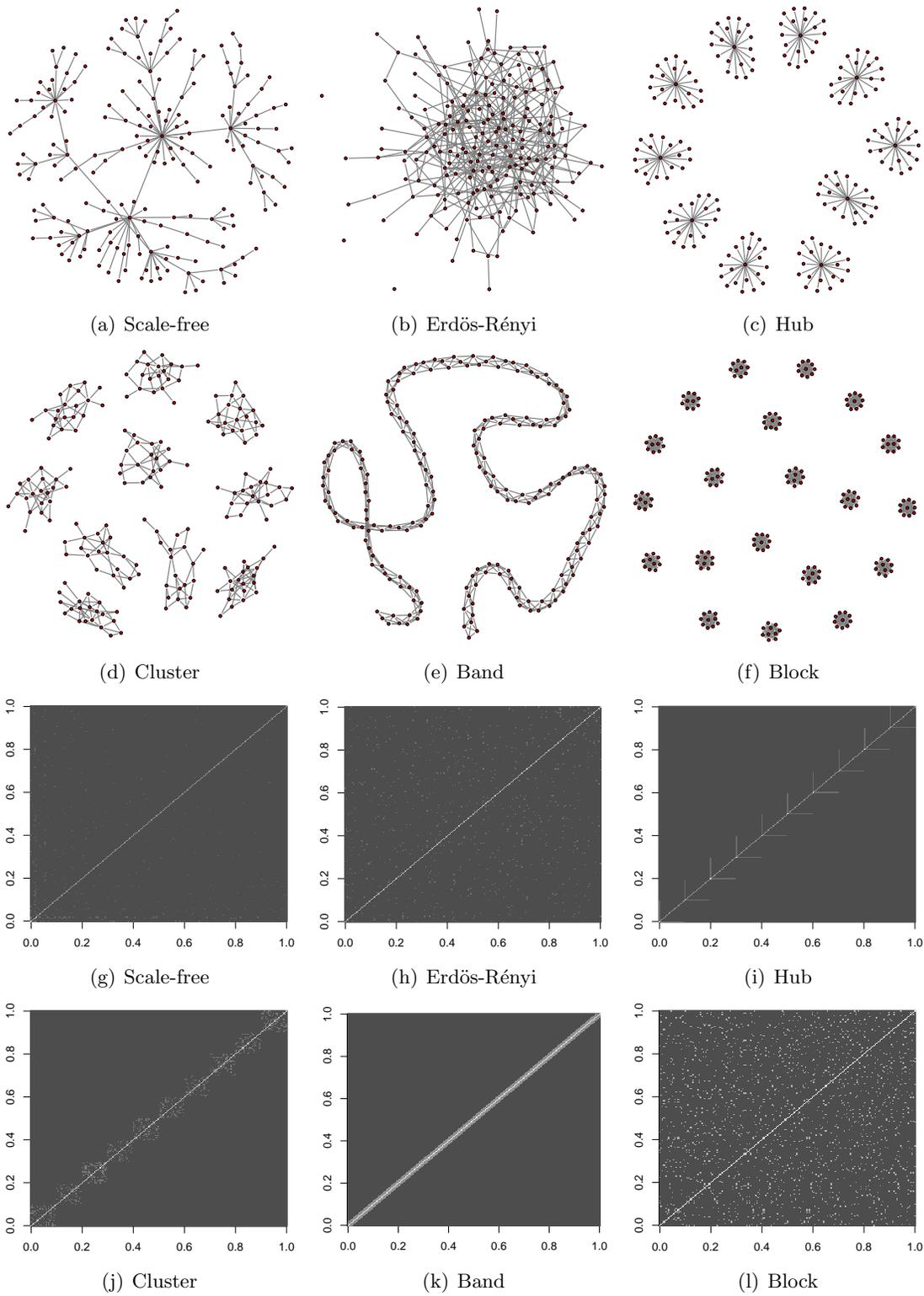

Figure 1: An illustration of the 6 graph patterns and the heatmaps of their corresponding precision matrices adopted in the simulations. To ease visualization, we only present example graphs with $d = 200$ nodes.

The graph is generated by the preferential attachment mechanism. The graph begins with an initial small chain graph of 2 nodes. New nodes are added to the graph one at a time. Each new node is connected to one existing node with a probability that is proportional to the number of degrees that the existing node already has. Formally, the probability $p_i$ that the new node is connected to an existing node $i$ is, $p_i = \frac{k_i}{\sum_j k_j}$, where $k_i$ is the degree of node $i$. The resulting graph has $d$ edges ($d = 200$ or $d = 400$). Once the graph is obtained, we generate an adjacency matrix $\mathbf{A}$ by setting the nonzero off-diagonal elements to be 0.3 and the diagonal elements to be 0. We calculate its smallest eigenvalue $\Lambda_{\min}(\mathbf{A})$. The precision matrix is constructed as

$$\boldsymbol{\Theta} = \mathbf{D}\big[\mathbf{A} + \big(|\Lambda_{\min}(\mathbf{A})| + 0.2\big) \cdot \mathbf{I}_d\big]\mathbf{D}, \tag{53}$$

where $\mathbf{D} \in \mathbb{R}^{d \times d}$ is a diagonal matrix with $\mathbf{D}_{jj} = 1$ for $j = 1, \ldots, d/2$ and $\mathbf{D}_{jj} = 3$ for $j = d/2 + 1, \ldots, d$. The covariance matrix $\boldsymbol{\Sigma} := \boldsymbol{\Theta}^{-1}$ is then computed to generate the multivariate normal data: $\boldsymbol{x}_1, \ldots, \boldsymbol{x}_n \sim N_d(\mathbf{0}, \boldsymbol{\Sigma})$.

∗ **Erdös-Rényi random graph**. We add an edge between each pair of nodes with probability 0.02 independently. The resulting graph has approximately 400 edges when $d = 200$ and 1,596 edges when $d = 400$. Once the graph is obtained, we construct the adjacency matrix $\mathbf{A}$ and generate the precision matrix $\boldsymbol{\Theta}$ using (53) but setting $\mathbf{D}_{jj} = 1$ for $j = 1, \ldots, d/2$ and $\mathbf{D}_{jj} = 1.5$ for $j = d/2 + 1, \ldots, d$. We then invert $\boldsymbol{\Theta}$ to get the covariance matrices $\boldsymbol{\Sigma}$ and generate the multivariate normal data: $\boldsymbol{x}_1, \ldots, \boldsymbol{x}_n \sim N_d(\mathbf{0}, \boldsymbol{\Sigma})$.

∗ **Hub graph**. The $d$ nodes are evenly partitioned into $d/20$ disjoint groups with each group contains 20 nodes. Within each group, one node is selected as the hub and we add edges between the hub and the other 19 nodes in that group. The resulting graph has 190 edges when $d = 200$ and 380 edges when $d = 400$. Once the graph is obtained, we generate the precision and covariance matrices in the same way as the Erdös-Rényi random graph model.

∗ **Cluster graph**. Similar to the hub model, the $d$ nodes are evenly partitioned into $d/20$ disjoint groups with each group contains 20 nodes. The subgraph of each group is an Erdös-Rényi random graph with the probability parameter 0.2. The resulting graph has approximately 380 edges when $d = 200$ and 760 edges when $d = 400$. Once the graph is obtained, we generate the precision and covariance matrices in the same way as the Erdös-Rényi random graph model.

∗ **Band graph**. Each node is assigned a coordinate $j$ with $j = 1, ..., d$. Two nodes are connected by an edge whenever the corresponding coordinates are at distance less than or equal to 3. The resulting graph has approximately 594 edges when $d = 200$ and 1,194 edges when $d = 400$. Once the graph is obtained, we generate the precision and covariance

matrices in the same way as the Erdös-Rényi random graph model.

∗ **Block graph**. The precision matrix $\boldsymbol{\Theta}$ is a block diagonal matrix with block size $d/20$. Each block has off-diagonal entries equal to 0.5 and diagonal entries equal to 1. Such a matrix is guaranteed to be positive definite. The resulting matrix is then randomly permuted by rows and columns. The resulting graph has approximately 900 edges when $d = 200$ and $3,800$ edges when $d = 400$. The covariance matrix $\boldsymbol{\Sigma} := \mathbf{D}^{-1}\boldsymbol{\Theta}^{-1}\mathbf{D}^{-1}$ is then computed to generate multivariate normal data, where $\mathbf{D}$ is a diagonal matrix with $\mathbf{D}_{jj} = 1$ for $j = 1, \ldots, d/2$ and $\mathbf{D}_{jj} = 1.5$ for $j = d/2 + 1, \ldots, d$.

## 5.2 Graph Recovery Performance

We first compare the TIGER with the CLIME and glasso on their graph recovery performance. Let $G = (V, E)$ be a $d$-dimensional graph. We denote by $|E|$ the number of edges in the graph $G$. We use the false positive and false negative rates to evaluate the graph recovery performance. Let $\widehat{G}^\lambda = (V, \widehat{E}^\lambda)$ be an estimated graph using a regularization parameter $\lambda$ in any of these procedures. The number of false positives when using the regularization parameter $\lambda$ is defined as $\mathrm{FP}(\lambda) :=$ the number of edges in $\widehat{E}^\lambda$ but not in $E$. The number of false negatives with $\lambda$ is defined as $\mathrm{FN}(\lambda) :=$ the number of edges in $E$ but not in $\widehat{E}^\lambda$. We further define the false negative rate (FNR) and false positive rate (FPR) as

$$\mathrm{FNR}(\lambda) := \frac{\mathrm{FN}(\lambda)}{|E|} \quad \text{and} \quad \mathrm{FPR}(\lambda) := \mathrm{FP}(\lambda)/\left[\binom{d}{2} - |E|\right]). \tag{54}$$

To illustrate the overall performance of the TIGER, CLIME and glasso methods over the full paths, the receiver operating characteristic (ROC) curves are drawn using $\big(\mathrm{FNR}(\lambda), 1 - \mathrm{FPR}(\lambda)\big)$. For the TIGER method, we found that using or without using the second hard-thresholding step provides the same graph estimates. So, the presented result does not use the second hard-thresholding step. The ROC curves for these models are presented in Figures 2, 3, 4.

From the ROC curves on the scale-free model in Figure 2, we see that the graph recovery performance of the TIGER is significantly better than those of the CLIME and glasso in higher dimensional settings (when $d = 200$ or $d = 400$ and $d \geq n$). This result suggests that the TIGER is more adaptive to inhomogeneous noise models. From Figure 2, we also see that the TIGER has similar graph recovery performance as the CLIME on the band model. Both the TIGER and CLIME significantly outperform the glasso. In particular, in the high dimensional setting when $d = 400$ and $n = 200$, the TIGER outperforms both CLIME and glasso. This result suggests that the TIGER is more reliable when facing higher dimensional problems on this model.

For the other models, from the ROC curves in Figures 3 and 4, we see that the three methods perform similarly on these settings, and for Erdös-Rényi random graph models, the TIGER outperforms the CLIME in the settings when $n \leq d$. This means that the TIGER is effective for a wide range of models. In summary, the above numerical results suggest that the TIGER is a very competitive graph estimation method in high dimensions.

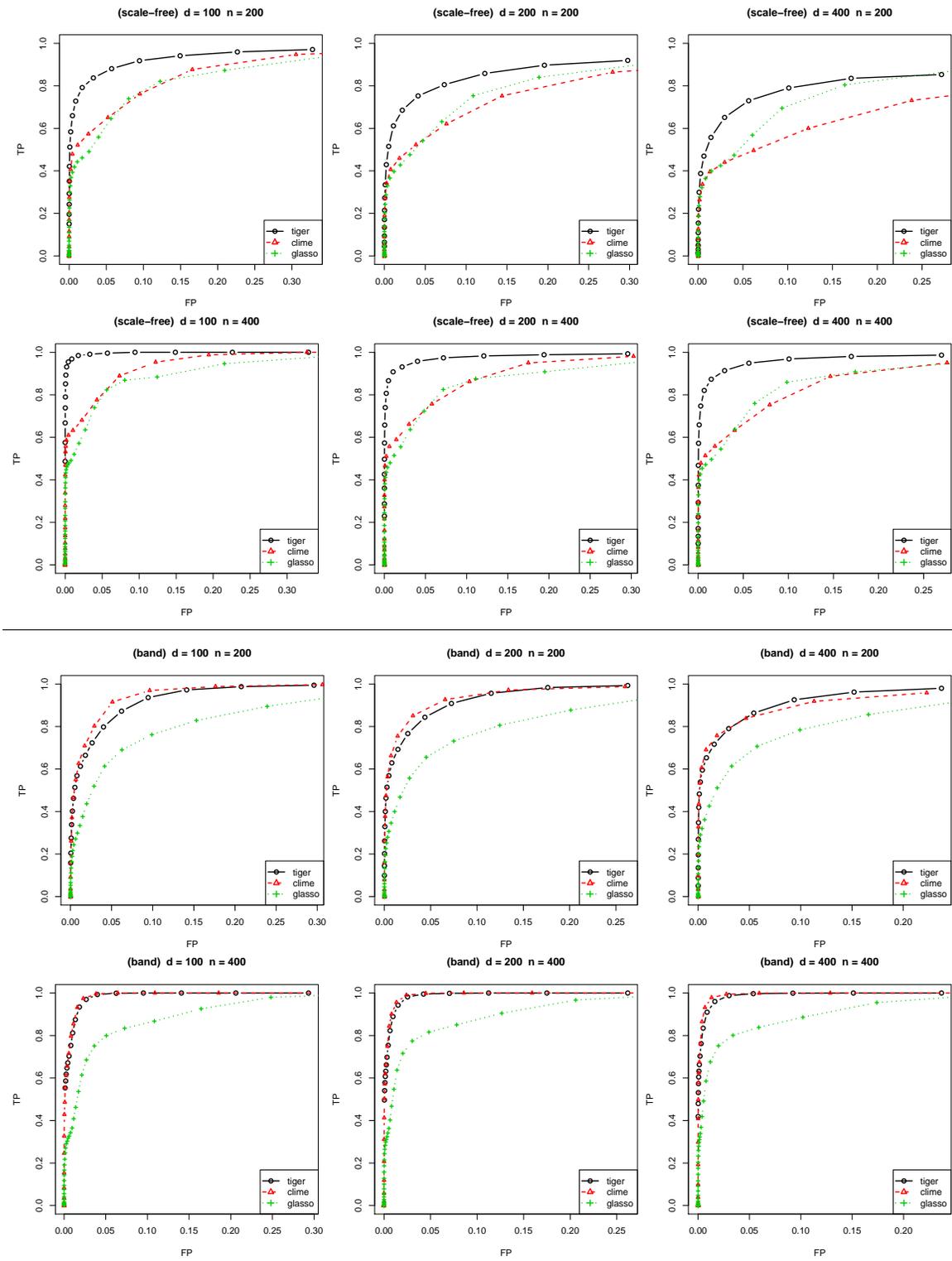

Figure 2: ROC curves for the Scale-free and Band models (Best visualized in color).

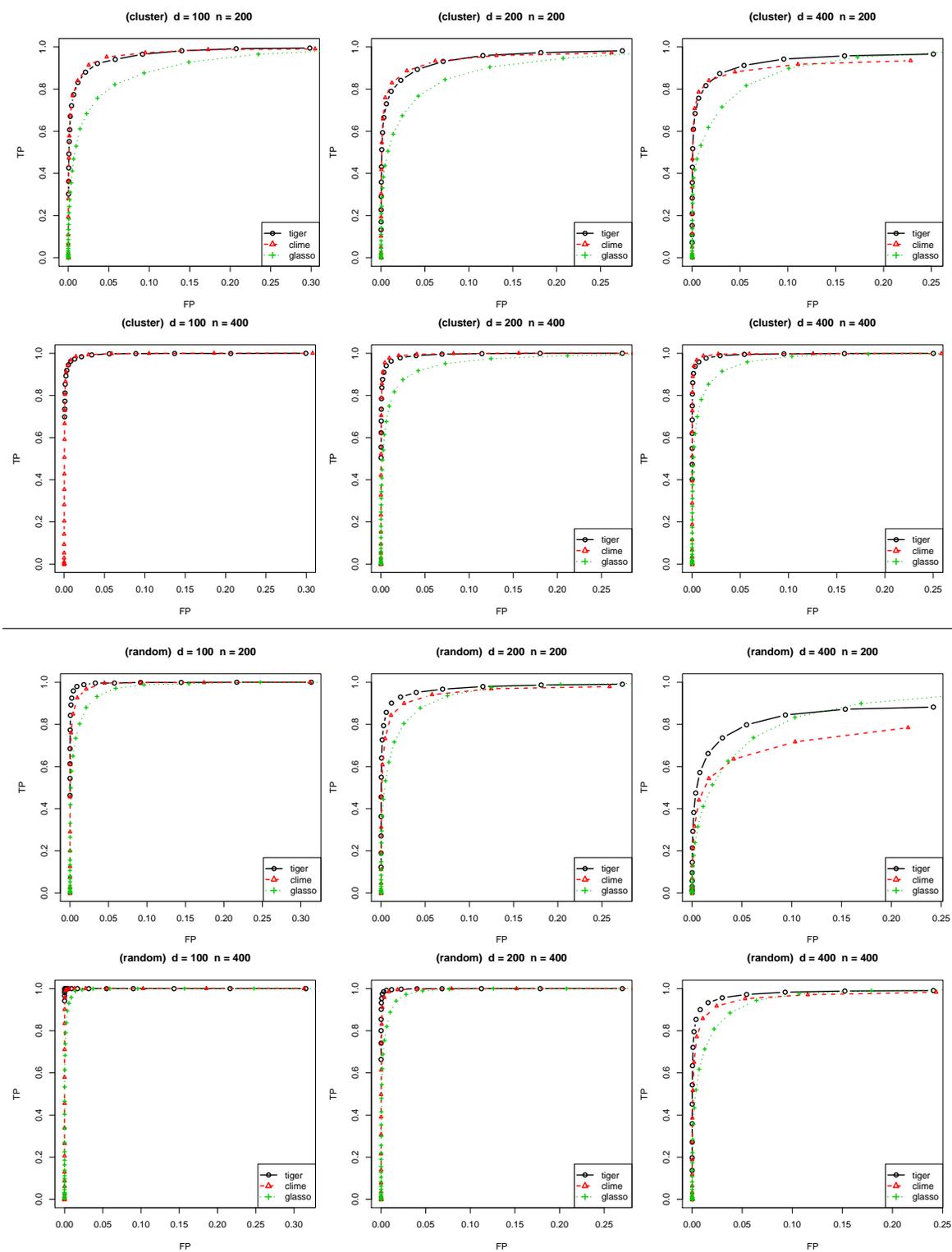

Figure 3: ROC curves for the cluster and Erdös-Rényi random graph models (Best visualized in color).

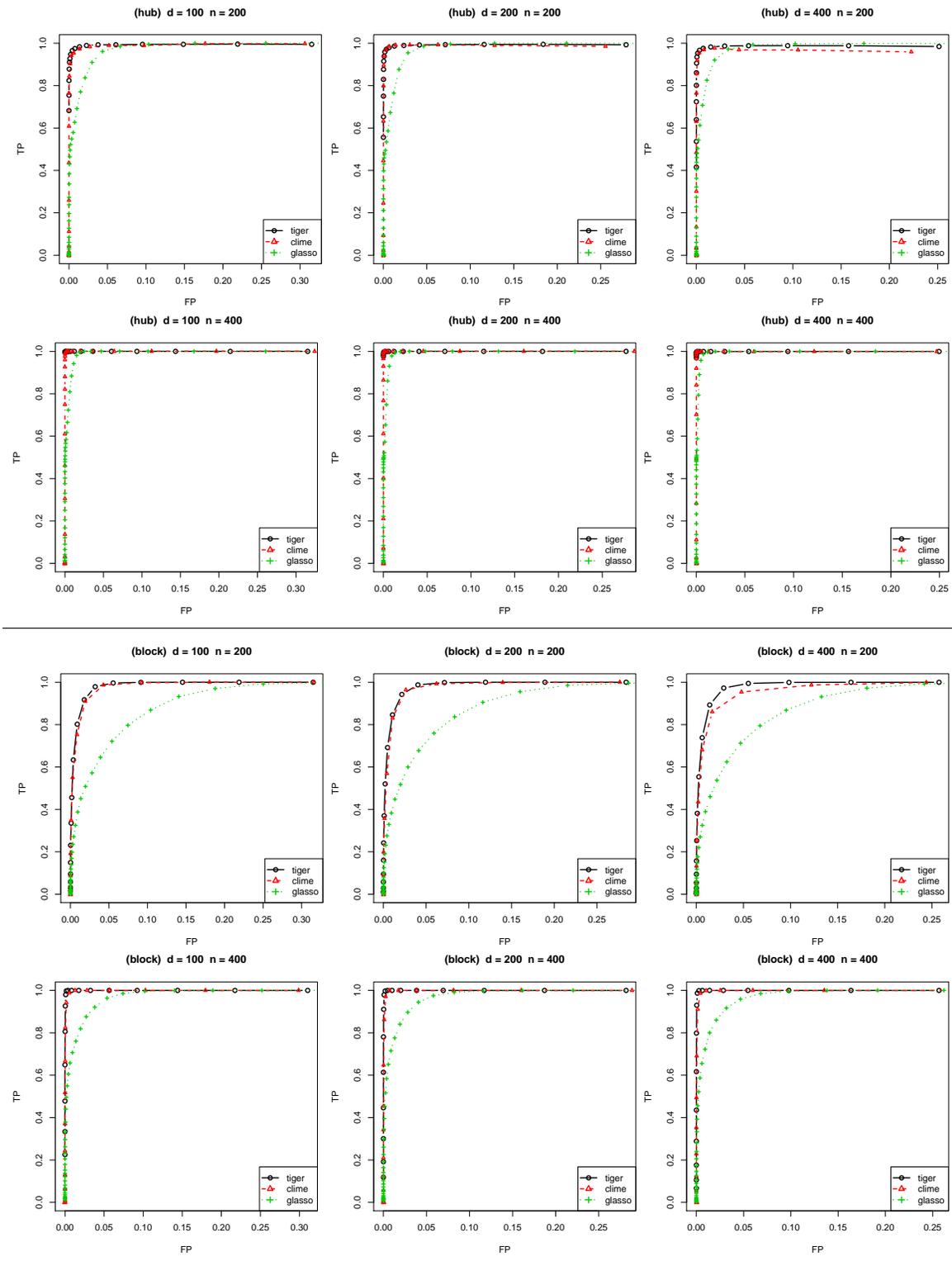

Figure 4: ROC curves for the hub and block models (Best visualized in color).

## 5.3 Tuning-Insensitive Regularization Path

We are interested in studying the tuning-insensitive property of the TIGER. For conciseness, we only discuss the TIGER and CLIME in this section and compare their regularization paths. Recall that

$$\lambda = \zeta \cdot \frac{\pi}{\sqrt{2}} \sqrt{\frac{\log d}{n}}, \tag{55}$$

we see that $\zeta$ and $\lambda$ have a one-to-one mapping. In Figure 5 (a) and (b), we plot the curves of Frobenius-norm errors vs. the tuning parameter $\zeta$ for the TIGER and CLIME. We define $\text{FNR}(\zeta)$ and $\text{FPR}(\zeta)$ in the same way as in (54). We also define the graph recovery accuracy as

$$\text{Accuracy}(\zeta) := 1 - \text{FPR}(\zeta) - \text{FNR}(\zeta). \tag{56}$$

In Figure 5 (c) and (d), we plot the curves of the graph recovery accuracy vs. the tuning parameter $\zeta$ for the TIGER and CLIME. The vertical axis of these plots are calibrated so that the results are directly comparable. These plots illustrate the tuning-insensitive property of the TIGER regularization path. For the TIGER, we found it is empirically safe to only consider the regularization path over the range $\zeta \in [\sqrt{2}/\pi, 1]$. From both subplots (a) and (c), the regularization paths are flat and do not change dramatically with the change of $\zeta$ (In another word, the procedure is insensitive to the tuning parameter). In contrast, for the CLIME, we need to search over a much larger range of $\zeta$ to find the optimal value and the paths are more irregular. In the subplots (b) and (d), we visualize the regularization paths of the CLIME over $\zeta \in [0.125, 2]$, these are only a sub-fraction of the whole regularization paths of the CLIME and the paths are irregular (or more sensitive to the choice of $\zeta$). Therefore, it is much easier to choose a reasonable tuning parameter for the TIGER than for the CLIME. In most cases, the choice of $\zeta = 1$ for TIGER provides a sparser graph estimate than the true graph. This is due to the fact that the asymptotic analysis in the previous section involves many union bounds, which may be too conservative in finite sample settings. As will be illustrated in the next subsection, we found that in most settings, simply setting $\zeta = \sqrt{2}/\pi$ yields a reasonably good graph and precision matrix estimates. Such a choice is used as a default in the `bigmatrix` package on CRAN.

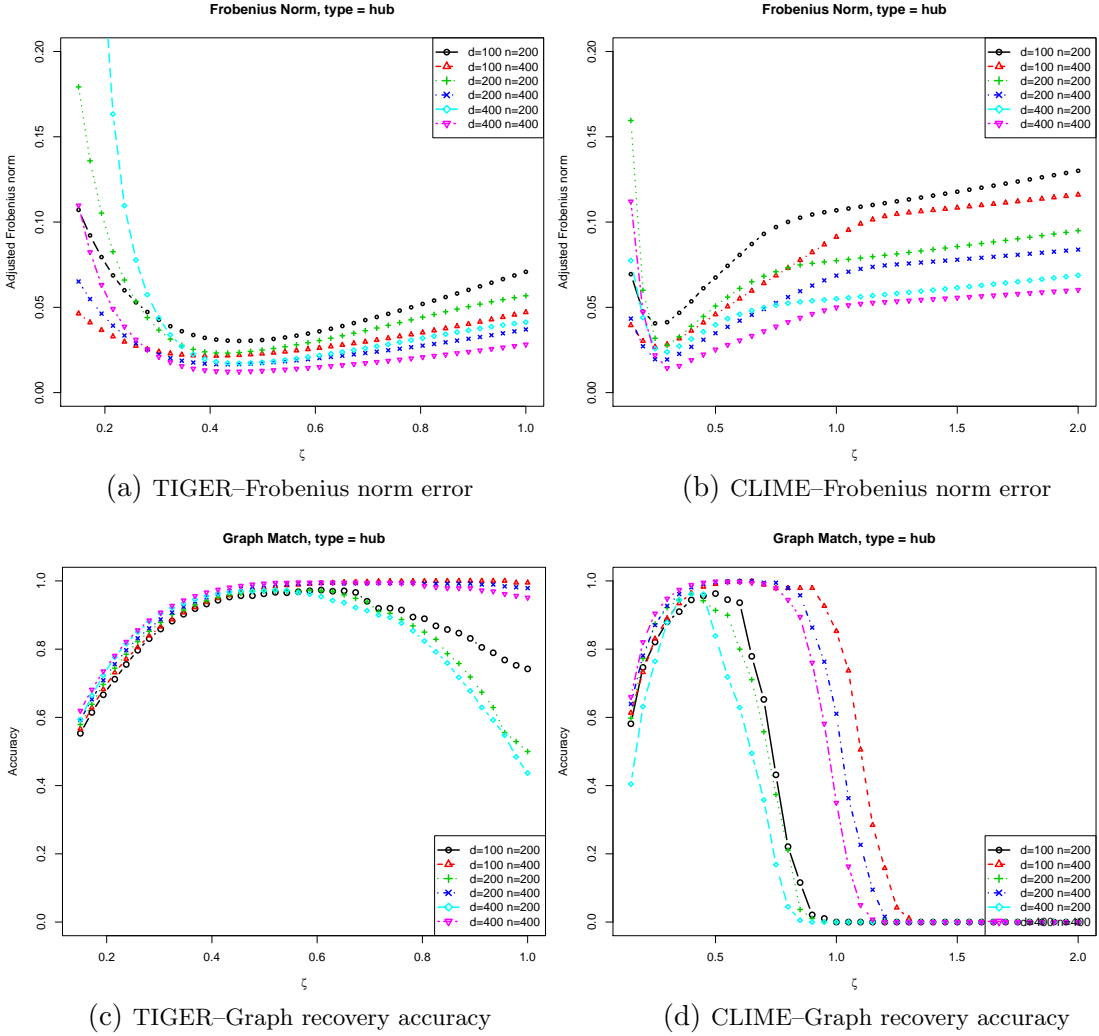

Figure 5: Comparison of the regularization paths of the TIGER and CLIME on the hub graph model. The vertical axis of these plots (Forebius divide by the dimensionality $d$) are calibrated so that the results are directly comparable. These plots illustrate the tuning-insensitive property of the TIGER regularization path. For the TIGER, we only need to consider the range $\zeta \in [\sqrt{2}/\pi, 1]$. From both subplots (a) and (c), the regularization paths are flat and do not change dramatically with $\zeta$. In contrast, for the CLIME, we need to search over a much larger range of $\zeta$ to find the optimal value and the paths are more irregular (the subplots (b) and (d) only show a part of this range).

## 5.4 Quantitative Evaluation on Parameter Estimation Performance

We then compare the TIGER with CLIME and glasso on their parameter estimation performance. For each model described before, we generate a training sample of $n$. The tuning parameters of the glasso and CLIME are automatically chosen in a data-dependent way. More specifically, for a given sample size $n$, we generate the same number of independent data points from the same distribution as a validation set. For each tuning parameter, the glasso or CLIME estimated precision matrix $\widehat{\Theta}$ is calculated on training data. The optimal tuning parameter is chosen by minimizing the held-out negative log-likelihood loss

$$\mathcal{L}_n(\widehat{\Theta}) := \text{tr}(\widehat{\Sigma}\widehat{\Theta}) - \log|\widehat{\Theta}| \tag{57}$$

on the validation set. For the TIGER, the tuning parameter $\zeta$ in (25) is simply chosen to be $\zeta = \sqrt{2}/\pi$ so that the procedure is completely tuning-free. For dimensionality $d = 100, 200, 400$, we consider the spectral norm error $\|\widehat{\Theta} - \Theta\|_2$ and Frobenius norm error $\|\widehat{\Theta} - \Theta\|_\mathsf{F}$ of all the 6 models described in the previous subsections.

The results are reported in Tables 1 and 2. In these tables we present the mean and standard deviation (in the parenthesis) of the spectral and Frobenius norm errors based on 50 random trials. We see that in almost all cases, the TIGER and CLIME outperform the glasso. In most cases, the TIGER outperforms the CLIME. Our results, though obtained in different experimental settings, are consistent with the results in Sun and Zhang (2012) for the scaled-Lasso based matrix inversion method.

One possible reason for the superior empirical performance of the TIGER over CLIME and glasso is that the data-dependent tuning selection procedure described in the previous section does not work well for the CLIME and glasso. To gain more insight, we also report the oracle estimation results in Tables 3 and 4. In these tables we present the mean and standard deviation (in the parenthesis) of the spectral and Frobenius norm errors based on 50 random trials. For all three methods, we draw the full regularization paths and select the best tuning parameter by minimizing the corresponding spectral or Frobenius norm errors to the true precision matrix. From these tables, we see that again the TIGER and CLIME outperform glasso and the TIGER is slightly better than CLIME.

## 5.5 Gene Network

This dataset includes 118 gene expression arrays from Arabidopsis thaliana originally appeared in Wille et al. (2004). Our analysis focuses on gene expression from 39 genes involved in two isoprenoid metabolic pathways: 16 from the mevalonate (MVA) pathway are located in the cytoplasm, 18 from the plastidial (MEP) pathway are located in the chloroplast, and 5 are located in the mitochondria. While the two pathways generally operate indepen-

Table 1: Quantitative comparisons of the TIGER, Glasso, and CLIME on the scale-free, hub, and band models using data-dependent model selection method.

| Model | $n$ | $d$ | Spectrum Norm | | | Frobenius Norm | | |
|---|---|---|---|---|---|---|---|---|
| | | | TIGER | CLIME | GLasso | TIGER | CLIME | GLasso |
| scale-free | 200 | 100 | 3.71370 (0.4651) | 4.41206 (0.4149) | 6.00048 (0.1838) | 11.5245 (0.7596) | 14.3260 (1.0260) | 21.1852 (0.4794) |
| | | 200 | 4.11834 (0.6495) | 4.19539 (0.3070) | 6.54951 (0.3823) | 16.3318 (0.8017) | 17.0776 (0.5752) | 35.1912 (2.4153) |
| | | 400 | 4.43263 (0.6412) | 4.58973 (0.5466) | 7.21601 (0.2184) | 23.4459 (0.7896) | 24.1157 (1.0119) | 50.9695 (0.3715) |
| | 400 | 100 | 2.77888 (0.3167) | 3.53363 (0.5060) | 5.30674 (0.1676) | 8.3591 (0.5206) | 11.4838 (0.7872) | 17.4945 (0.3696) |
| | | 200 | 2.68762 (0.2668) | 3.29772 (0.2480) | 5.23455 (0.2602) | 11.7521 (0.4706) | 14.0718 (0.5345) | 23.3348 (0.3524) |
| | | 400 | 3.31452 (0.5758) | 3.57110 (0.3068) | 6.32043 (0.2675) | 16.9996 (0.6411) | 19.1326 (0.7619) | 41.8133 (0.3717) |
| hub | 200 | 100 | 2.67040 (0.4446) | 4.31716 (0.4979) | 6.51814 (0.3070) | 5.4347 (0.3356) | 7.3545 (0.5204) | 10.6739 (0.4329) |
| | | 200 | 3.02307 (0.4151) | 5.35713 (0.3096) | 7.41337 (0.1374) | 8.2277 (0.3049) | 12.7079 (0.4182) | 18.1376 (0.1653) |
| | | 400 | 3.34315 (0.3017) | 6.11506 (0.2093) | 7.69167 (0.1061) | 12.0676 (0.2801) | 19.8038 (0.3058) | 26.4038 (0.2002) |
| | 400 | 100 | 1.82245 (0.2619) | 2.46543 (0.3995) | 5.28610 (0.1834) | 3.7161 (0.1811) | 4.5920 (0.2968) | 8.5380 (0.2247) |
| | | 200 | 2.07601 (0.2105) | 3.11260 (0.2491) | 6.23933 (0.1208) | 5.6517 (0.2366) | 7.5627 (0.3261) | 15.0739 (0.1747) |
| | | 400 | 2.23719 (0.1675) | 4.02006 (0.4557) | 6.43531 (0.0913) | 8.1420 (0.1207) | 13.0960 (1.1531) | 22.0039 (0.1230) |
| band | 200 | 100 | 5.72715 (0.1546) | 4.37815 (0.3626) | 6.36789 (0.0850) | 16.7205 (0.2793) | 12.8498 (0.8868) | 17.4244 (0.1506) |
| | | 200 | 6.04373 (0.1381) | 5.81030 (0.1378) | 7.32236 (0.0517) | 24.6770 (0.2373) | 23.3002 (0.3368) | 28.8514 (0.1293) |
| | | 400 | 6.28046 (0.0760) | 6.76049 (0.0738) | 7.87172 (0.1168) | 36.0083 (0.1940) | 38.6665 (0.1779) | 44.7691 (0.7423) |
| | 400 | 100 | 4.34163 (0.1847) | 2.92088 (0.2552) | 5.53820 (0.0871) | 12.5012 (0.2989) | 8.2567 (0.5046) | 14.7940 (0.1483) |
| | | 200 | 4.69878 (0.1302) | 3.62779 (0.1494) | 5.71676 (0.0740) | 19.0534 (0.2633) | 14.3844 (0.2863) | 21.9130 (0.1285) |
| | | 400 | 5.01406 (0.0709) | 4.24658 (0.4150) | 6.82527 (0.0392) | 28.6523 (0.2313) | 24.1215 (2.0819) | 37.6074 (0.0953) |

Table 2: Quantitative comparisons of the TIGER, Glasso, and CLIME on the block, Erdös-Rényi random, and cluster models using data-dependent model selection method.

| Model | $n$ | $d$ | Spectrum Norm | | | Frobenius Norm | | |
|---|---|---|---|---|---|---|---|---|
| | | | TIGER | CLIME | GLasso | TIGER | CLIME | GLasso |
| block | 200 | 100 | 3.88080 | 2.92791 | 4.62777 | 12.7803 | 9.8588 | 14.1474 |
| | | | (0.2123) | (0.1907) | (0.1221) | (0.2231) | (0.2728) | (0.3767) |
| | | 200 | 4.21258 | 3.50847 | 5.05219 | 19.1984 | 16.4653 | 21.8517 |
| | | | (0.2073) | (0.2014) | (0.0408) | (0.1841) | (0.2043) | (0.0455) |
| | | 400 | 4.54196 | 4.72388 | 5.44260 | 28.8940 | 29.7674 | 34.0288 |
| | | | (0.1456) | (0.1344) | (0.0387) | (0.1471) | (0.1169) | (0.2657) |
| | 400 | 100 | 2.61796 | 1.91879 | 3.63244 | 8.4651 | 6.1005 | 10.7808 |
| | | | (0.1746) | (0.2812) | (0.0969) | (0.2232) | (0.6898) | (0.0964) |
| | | 200 | 2.86024 | 2.04201 | 4.26378 | 12.8697 | 9.7477 | 17.7396 |
| | | | (0.1613) | (0.1219) | (0.0476) | (0.2201) | (0.2173) | (0.0674) |
| | | 400 | 3.12185 | 3.03083 | 4.70495 | 19.4939 | 19.6861 | 28.0988 |
| | | | (0.1423) | (0.1365) | (0.0333) | (0.2277) | (0.1791) | (0.0522) |
| random | 200 | 100 | 1.40361 | 1.63446 | 2.66316 | 4.9173 | 5.5390 | 7.4979 |
| | | | (0.2093) | (0.2246) | (0.2626) | (0.2158) | (0.2072) | (0.5315) |
| | | 200 | 1.92515 | 2.13847 | 2.97502 | 9.3623 | 10.4678 | 13.2284 |
| | | | (0.1352) | (0.1984) | (0.2074) | (0.1845) | (0.5063) | (0.8328) |
| | | 400 | 3.03486 | 3.57549 | 4.16129 | 17.6548 | 20.1956 | 23.3014 |
| | | | (0.0710) | (0.0417) | (0.0313) | (0.1165) | (0.1937) | (0.0777) |
| | 400 | 100 | 0.96871 | 1.08246 | 1.94886 | 3.3962 | 3.7713 | 5.4841 |
| | | | (0.1265) | (0.1420) | (0.1879) | (0.1332) | (0.1521) | (0.2622) |
| | | 200 | 1.38675 | 1.45379 | 2.20816 | 6.6106 | 7.2891 | 9.6509 |
| | | | (0.1133) | (0.1517) | (0.0634) | (0.1273) | (0.2580) | (0.1012) |
| | | 400 | 2.21101 | 2.49335 | 3.02710 | 13.3298 | 14.9634 | 17.0508 |
| | | | (0.0792) | (0.2862) | (0.0468) | (0.2506) | (0.7566) | (0.1373) |
| cluster | 200 | 100 | 3.84966 | 3.45717 | 5.35790 | 8.9219 | 8.3992 | 11.7727 |
| | | | (0.3371) | (0.3465) | (0.1325) | (0.2710) | (0.3329) | (0.1427) |
| | | 200 | 3.66157 | 3.90672 | 5.11136 | 11.6676 | 12.1507 | 16.0086 |
| | | | (0.2469) | (0.3981) | (0.2043) | (0.2153) | (0.7474) | (0.5784) |
| | | 400 | 2.99469 | 3.34376 | 4.11403 | 15.1022 | 16.5420 | 20.2334 |
| | | | (0.1527) | (0.1257) | (0.0848) | (0.1233) | (0.1500) | (0.1057) |
| | 400 | 100 | 2.74935 | 2.32058 | 4.21725 | 6.4102 | 5.7983 | 8.9115 |
| | | | (0.2604) | (0.2737) | (0.4733) | (0.2067) | (0.2324) | (0.8694) |
| | | 200 | 2.97759 | 2.80625 | 4.18294 | 8.7524 | 8.4317 | 11.9685 |
| | | | (0.1958) | (0.2788) | (0.0969) | (0.1863) | (0.3450) | (0.1067) |
| | | 400 | 2.20812 | 2.29522 | 3.58522 | 11.0885 | 11.4516 | 16.8512 |
| | | | (0.0627) | (0.0675) | (0.0630) | (0.1962) | (0.1888) | (0.1091) |

Table 3: Quantitative comparisons of the TIGER, Glasso, and CLIME on the scale-free, hub, and band models using oracle model selection method.

| Model | $n$ | $d$ | Spectrum Norm | | | Frobenius Norm | | |
|---|---|---|---|---|---|---|---|---|
| | | | TIGER | CLIME | GLasso | TIGER | CLIME | GLasso |
| scale-free | 200 | 100 | 3.49331 (0.2888) | 4.20880 (0.3896) | 4.44102 (0.2517) | 11.4419 (0.7051) | 14.0362 (0.8067) | 14.6163 (0.4395) |
| | | 200 | 3.55648 (0.4300) | 3.97078 (0.3312) | 4.06802 (0.3903) | 15.5481 (0.5613) | 16.8466 (0.4808) | 20.6089 (0.2790) |
| | | 400 | 3.93523 (0.5407) | 4.50636 (0.4290) | 4.36077 (0.2883) | 21.2401 (0.5478) | 23.6391 (0.2663) | 32.9509 (0.1983) |
| | 400 | 100 | 2.65555 (0.2640) | 2.92806 (0.2593) | 3.72758 (0.2702) | 8.3378 (0.5277) | 10.1875 (0.4220) | 11.5142 (0.4212) |
| | | 200 | 2.62826 (0.2368) | 3.20446 (0.1861) | 3.46773 (0.3240) | 11.7350 (0.4554) | 13.9074 (0.3791) | 16.2752 (0.2982) |
| | | 400 | 3.10365 (0.4002) | 3.42918 (0.2338) | 3.74972 (0.3319) | 16.7337 (0.5986) | 18.9504 (0.4950) | 25.1616 (0.2071) |
| hub | 200 | 100 | 2.42283 (0.3287) | 2.66693 (0.3362) | 4.13856 (0.2862) | 5.4124 (0.3213) | 6.6481 (0.3566) | 9.6412 (0.2294) |
| | | 200 | 2.80843 (0.2737) | 3.37883 (0.2853) | 4.18505 (0.2116) | 8.2170 (0.3008) | 10.9757 (0.3744) | 15.5200 (0.1934) |
| | | 400 | 3.18571 (0.2904) | 4.51681 (0.3600) | 4.06350 (0.0675) | 12.0676 (0.2801) | 15.5895 (0.2402) | 24.0977 (0.1782) |
| | 400 | 100 | 1.63670 (0.1863) | 1.70506 (0.2001) | 3.31738 (0.2704) | 3.6970 (0.1763) | 4.2848 (0.2483) | 7.6101 (0.2076) |
| | | 200 | 1.93008 (0.2119) | 2.23970 (0.2799) | 3.52276 (0.1672) | 5.6341 (0.2342) | 6.4934 (0.2451) | 12.3937 (0.1873) |
| | | 400 | 2.10478 (0.1478) | 2.49148 (0.2029) | 3.45920 (0.1311) | 8.1420 (0.1207) | 10.4606 (0.1850) | 19.0323 (0.1332) |
| band | 200 | 100 | 3.29816 (0.2127) | 3.27817 (0.2519) | 4.69050 (0.1342) | 12.2957 (0.4519) | 11.7620 (0.4908) | 14.0160 (0.2475) |
| | | 200 | 3.89225 (0.2489) | 4.53169 (0.2191) | 4.78844 (0.1114) | 19.7902 (0.3364) | 20.0180 (0.3778) | 22.8946 (0.2232) |
| | | 400 | 4.72424 (0.1313) | 5.57854 (0.1061) | 4.77270 (0.0857) | 31.1798 (0.2101) | 34.4919 (0.3571) | 37.4700 (0.1044) |
| | 400 | 100 | 2.09320 (0.1604) | 2.18392 (0.2292) | 3.87918 (0.1348) | 7.8187 (0.2798) | 7.4255 (0.2845) | 10.8361 (0.2176) |
| | | 200 | 2.44876 (0.1452) | 2.65924 (0.1878) | 4.00698 (0.1037) | 12.8840 (0.1946) | 12.3307 (0.2244) | 17.2196 (0.1598) |
| | | 400 | 2.82127 (0.1166) | 2.94520 (0.0979) | 4.08170 (0.0655) | 20.4888 (0.2645) | 21.4755 (0.2665) | 28.8954 (0.1768) |

Table 4: Quantitative comparisons of the TIGER, Glasso, and CLIME on the block, Erdös-Rényi random, and cluster models using oracle model selection method.

| Model | $n$ | $d$ | Spectrum Norm | | | Frobenius Norm | | |
|---|---|---|---|---|---|---|---|---|
| | | | TIGER | CLIME | GLasso | TIGER | CLIME | GLasso |
| block | 200 | 100 | 1.78662 | 1.96932 | 2.30284 | 7.0491 | 7.2065 | 8.7645 |
| | | | (0.1583) | (0.1773) | (0.1754) | (0.2670) | (0.3231) | (0.1882) |
| | | 200 | 2.26303 | 2.58484 | 2.72901 | 11.8684 | 13.6219 | 16.0190 |
| | | | (0.1872) | (0.1816) | (0.0660) | (0.2641) | (0.2550) | (0.1207) |
| | | 400 | 2.79826 | 3.67637 | 3.60729 | 20.5628 | 26.3292 | 25.4660 |
| | | | (0.2608) | (0.2448) | (0.0702) | (0.2978) | (0.2781) | (0.0911) |
| | 400 | 100 | 1.18086 | 1.22283 | 1.89098 | 4.5090 | 4.2706 | 6.2619 |
| | | | (0.1352) | (0.1702) | (0.1439) | (0.1631) | (0.1898) | (0.1538) |
| | | 200 | 1.33627 | 1.48764 | 1.94656 | 7.4712 | 7.9438 | 11.0249 |
| | | | (0.1202) | (0.1009) | (0.1143) | (0.1473) | (0.1856) | (0.1133) |
| | | 400 | 1.58491 | 2.04870 | 2.58467 | 11.8926 | 15.1659 | 19.7672 |
| | | | (0.1664) | (0.1761) | (0.0316) | (0.1673) | (0.1953) | (0.1090) |
| random | 200 | 100 | 1.30656 | 1.54958 | 1.93012 | 4.8291 | 5.4697 | 6.4292 |
| | | | (0.1552) | (0.1801) | (0.1224) | (0.2090) | (0.1878) | (0.1705) |
| | | 200 | 1.74892 | 2.05243 | 2.23740 | 9.1014 | 10.2764 | 11.3447 |
| | | | (0.1026) | (0.0972) | (0.0769) | (0.1862) | (0.2062) | (0.1505) |
| | | 400 | 2.55517 | 3.49539 | 3.01732 | 17.0088 | 20.1956 | 20.0869 |
| | | | (0.0663) | (0.1640) | (0.0501) | (0.1298) | (0.1937) | (0.0937) |
| | 400 | 100 | 0.87331 | 1.03685 | 1.35098 | 3.3014 | 3.7713 | 4.8714 |
| | | | (0.0838) | (0.0957) | (0.0712) | (0.1317) | (0.1521) | (0.1361) |
| | | 200 | 1.19717 | 1.37938 | 1.79435 | 6.2812 | 7.1464 | 8.7373 |
| | | | (0.0783) | (0.0963) | (0.0578) | (0.1174) | (0.1425) | (0.0880) |
| | | 400 | 1.67683 | 2.11043 | 2.45761 | 12.3481 | 14.0788 | 15.6806 |
| | | | (0.1227) | (0.0694) | (0.0596) | (0.2473) | (0.3149) | (0.1436) |
| cluster | 200 | 100 | 2.50462 | 2.61965 | 3.12584 | 7.9860 | 8.1190 | 9.9439 |
| | | | (0.2153) | (0.2687) | (0.2686) | (0.2865) | (0.3492) | (0.1874) |
| | | 200 | 2.72033 | 2.89910 | 2.91801 | 11.1472 | 11.5100 | 13.2443 |
| | | | (0.2079) | (0.2528) | (0.1019) | (0.2200) | (0.2111) | (0.1515) |
| | | 400 | 2.49864 | 3.30405 | 2.94846 | 14.8704 | 16.5420 | 18.3397 |
| | | | (0.1125) | (0.1874) | (0.1043) | (0.2025) | (0.1500) | (0.1082) |
| | 400 | 100 | 1.63625 | 1.74909 | 2.47721 | 5.3929 | 5.4465 | 7.1791 |
| | | | (0.1633) | (0.1612) | (0.2247) | (0.2439) | (0.2127) | (0.1576) |
| | | 200 | 1.84653 | 2.04163 | 2.29099 | 7.7920 | 7.7961 | 10.4819 |
| | | | (0.1555) | (0.1747) | (0.1570) | (0.1718) | (0.1957) | (0.1135) |
| | | 400 | 1.66908 | 1.75361 | 2.46020 | 10.3907 | 10.8631 | 14.6437 |
| | | | (0.1359) | (0.0699) | (0.0658) | (0.1888) | (0.1774) | (0.1132) |

dently, crosstalk is known to happen (Wille et al., 2004). Our goal is to recover the gene regulatory network, with special interest in crosstalk.

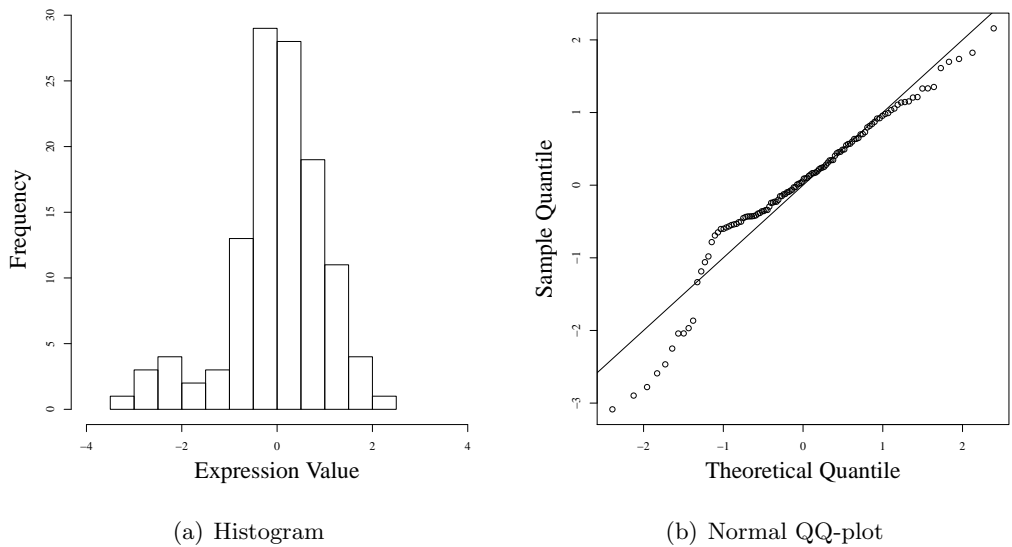

(a) Histogram

(b) Normal QQ-plot

Figure 6: The histogram and normal QQ plots of the marginal expression levels of the gene MECPS. We see the data are not exactly Gaussian distributed.

We first examine whether the data actually satisfies the Gaussian distribution assumption. In Figure 6 we plot the histogram and normal QQ plot of the expression levels of a gene named MECPS. From the histogram, we see the distribution is left-skewed compared to the Gaussian distribution. From the normal QQ plot, we see the empirical distribution has a heavier tail compared to Gaussian. To suitably apply the TIGER method on this dataset, we need to first transform the data so that its distribution is closer to Gaussian. Therefore, we Gaussianize the marginal expression values of each gene by converting them to the corresponding normal-scores. This is automatically done by the `huge.npn` function in the R package `huge` (Zhao et al., 2012).

We apply the TIGER on the transformed data using the default tuning parameter $\zeta = \sqrt{2}/\pi$. The estimated network is shown in Figure 7. We see the estimated network is very sparse with only 44 edges. We draw the within-pathway connections using solid lines and the between-pathway connections using dashed lines. Our result is consistent with previous investigations, which suggest that the connections from genes AACT1 and HMGR2 to gene MECPS indicate a primary sources of the crosstalk between the MEP and MVA pathways and these edges are presented in the estimated network. MECPS is clearly a hub gene for this pathway.

For the MEP pathway, the genes DXPS2, DXR, MCT, CMK, HDR, and MECPS are connected as in the true metabolic pathway. Similarly, for the MVA pathway, the genes

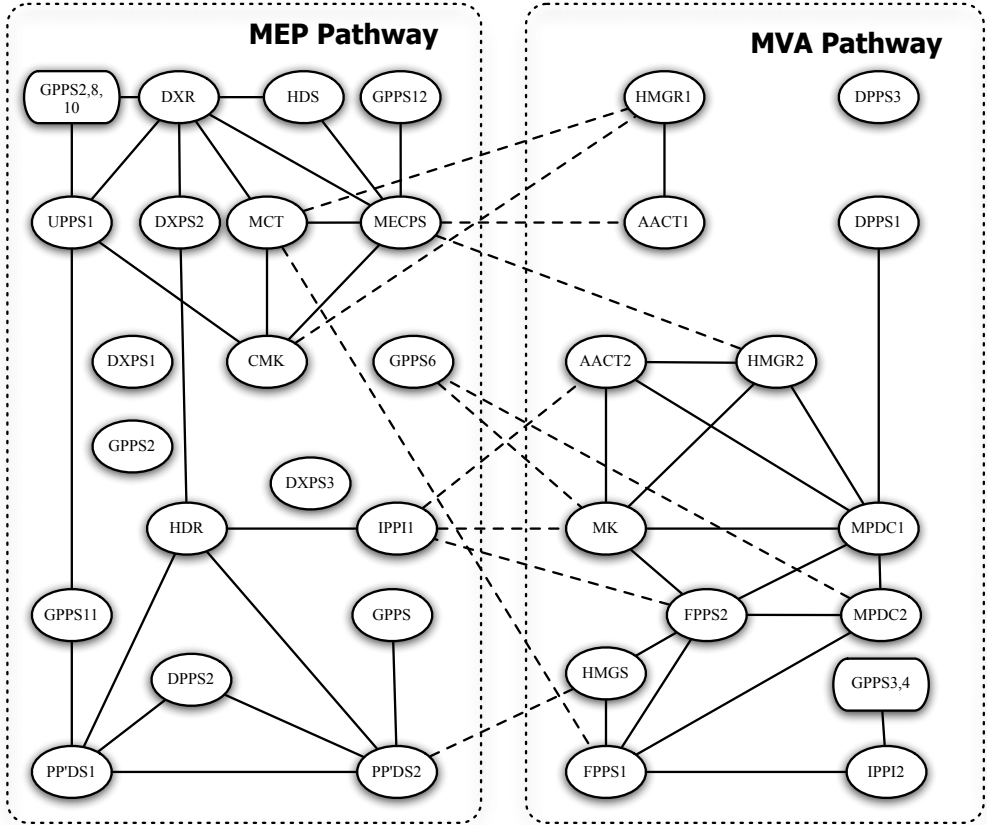

Figure 7: The estimated gene networks of the Arabadopsis dataset. The within-pathway links are denoted by solid lines and between-pathway links are denoted by dashed lines.

AACT2, HMGR2, MK, MPDC1, MPDC2, FPPS1 and FPP2 are closely connected. Our analysis suggests 11 cross-pathway links, which is consistent to previous investigation in Wille et al. (2004). This result suggests that there might exist rich inter-pathway crosstalks.

## 6 Conclusions

We introduce a tuning-insensitive approach named TIGER for estimating high dimensional Gaussian graphical models. Our method is asymptotically tuning-free and simultaneously achieves the minimax optimal rates of convergence in precision matrix estimation under different matrix norms (matrix $L_1$, spetral, Frobenius, and elementwise sup-norm). Computationally, our procedure is significantly faster than existing methods due to its tuning-insensitive property. The advantages of our estimators are also illustrated using both sim-ulated and real data examples. The TIGER approach is a very competitive alternative for estimating high dimensional Gaussian graphical models.

There are several possible directions to expand the current methods. First, it is interesting to extend the TIGER to the nonparanomral setting for estimating high dimensional Gaussian copula graphical models (Liu et al., 2012). Second, it is also interesting to extend the TIGER to more complex settings where latent variables or missing data might exist.

## Acknowledgement

We thank Professor Cun-hui Zhang for constructive comments and help on pointing out the connection between the SQRT-Lasso and scaled-Lasso. We also thank Professor Tony Cai, Professor Weidong Liu, and Professor Harrison Zhou for their help on providing the results on minimax lower bounds. We also thank Xingguo Li and Tuo Zhao for their help on implementing the softwares and running the experiments. Han Liu's research is supported by NSF Grant III–1116730. Lie Wang's research is supported by NSF Grant DMS-1005539.

## A  Appendix: Proofs

We first prove several important lemmas, followed by the proof of the main theorems. For notational simplicity, we use a generic constant $C$ whose value may change from line to line.

### A.1  Preliminaries

For any set $S \subset \{1, 2, \cdots, d\}$ and $|S| \leq k$, let $\boldsymbol{\Delta}_{\bar{c}}^d(S)$ denote a subset of $\mathbb{R}^d$ defined as

$$\boldsymbol{\Delta}_{\bar{c}}^d(S) := \{\boldsymbol{\beta} \in \mathbb{R}^d : \|\boldsymbol{\beta}_{S^c}\|_1 \leq \bar{c}\|\boldsymbol{\beta}_S\|_1, \boldsymbol{\beta} \neq \boldsymbol{0}\}. \tag{58}$$

Also, we define

$$\boldsymbol{\Delta}_{\bar{c}}^d(k) := \bigcup_{S \subset \{1,2,\cdots,d\}, |S| \leq k} \boldsymbol{\Delta}_{\bar{c}}^d(S). \tag{59}$$

Recall that the matrix class $\mathcal{M}(\xi_{\max}, k)$ is defined as

$$\mathcal{M}(\xi_{\max}, k) := \left\{\boldsymbol{\Theta} = \boldsymbol{\Theta}^T \in \mathbb{R}^{d \times d} : \boldsymbol{\Theta} \succ \boldsymbol{0}, \frac{\Lambda_{\max}(\boldsymbol{\Theta})}{\Lambda_{\min}(\boldsymbol{\Theta})} \leq \xi_{\max}, \max_j \sum_k I(\boldsymbol{\Theta}_{jk} \neq 0) \leq k\right\},$$

let $\boldsymbol{\Theta} := \boldsymbol{\Sigma}^{-1} \in \mathcal{M}(\xi_{\max}, k)$, we have

$$\frac{\Lambda_{\max}(\boldsymbol{\Theta})}{\Lambda_{\min}(\boldsymbol{\Theta})} = \frac{\Lambda_{\max}(\boldsymbol{\Sigma}^{-1})}{\Lambda_{\min}(\boldsymbol{\Sigma}^{-1})} = \frac{\Lambda_{\max}(\boldsymbol{\Sigma})}{\Lambda_{\min}(\boldsymbol{\Sigma})} \leq \xi_{\max}. \tag{60}$$

We define the population correlation matrix $\mathbf{R}$ as

$$\mathbf{R} := [\mathrm{diag}(\mathbf{\Sigma})]^{-1/2} \mathbf{\Sigma} [\mathrm{diag}(\mathbf{\Sigma})]^{-1/2}. \tag{61}$$

We also define $\tau_j := \sigma_j \widehat{\mathbf{\Gamma}}_{jj}^{-1/2}$. We recall that there are three assumptions:

(A1) $\mathbf{\Sigma}^{-1} \in \mathcal{M}(\xi_{\max}, k)$,

(A2) $k^2 \log d = o(n)$,

(A3) $\limsup_{n \to \infty} \max_{1 \leq j \leq d} \mathbf{\Sigma}_{jj}^2 \dfrac{\log d}{n} < \dfrac{1}{4}$.

We define

$$Q_j(\boldsymbol{\beta}_j) := \left\| \mathbf{Z}_{*j} - \mathbf{Z}_{*\setminus j} \boldsymbol{\beta}_j \right\|_2. \tag{62}$$

Therefore, $\widehat{\boldsymbol{\beta}}_j$ from (27) can be written as

$$\widehat{\boldsymbol{\beta}}_j = \operatorname*{argmin}_{\boldsymbol{\beta} \in \mathbb{R}^{d-1}} \left\{ \frac{1}{\sqrt{n}} Q_j(\boldsymbol{\beta}_j) + \lambda \|\boldsymbol{\beta}\|_1 \right\}, \quad \text{for } j = 1, \ldots, d. \tag{63}$$

In the whole proof, for notational simplicity, we always denote the tuning parameter $\lambda$ to be

$$\lambda = c \sqrt{\frac{2a \log d}{n}}, \tag{64}$$

where $c > 1$ and $a > 2$. It is easy to see that $\lambda = \dfrac{\pi}{\sqrt{2}} \sqrt{\dfrac{\log d}{n}}$ is a special case of this setup.

## A.2 Technical Lemmas

Throughout this paper we often use one of the following tail bounds for central $\chi^2$ random variables. We denote $\chi_d^2$ to be a Chi-square variable with $d$ degrees of freedom. Lemma 5 presents some well known results of $\chi_d^2$ and the proofs can be found in the original papers.

**Lemma 5** (Johnstone. (2000) and Laurent and Massart (1998))**.** *Let $X \sim \chi_d^2$. We have*

$$\max\left\{ \mathbb{P}\left( X - d \geq 2\sqrt{dt} + 2t \right), \mathbb{P}\left( X - d \leq -2\sqrt{dt} \right) \right\} \leq \exp(-t) \quad \text{for all } t \geq 0, \tag{65}$$

$$\mathbb{P}(|X - d| > dt) \leq \exp\left(-\frac{3}{16} dt^2\right), \quad \text{for all } t \in \left[0, \frac{1}{2}\right), \tag{66}$$

$$\mathbb{P}(X \leq (1-t)d) \leq \exp\left(-\frac{1}{4} dt^2\right), \quad \text{for all } t \in \left[0, \frac{1}{2}\right). \tag{67}$$

The next lemma bounds the tail of sample correlation for bivariate Gaussian random variables.

**Lemma 6.** Let $\boldsymbol{X} := (X_1, X_2)^T$ follows a bivariate normal distribution:

$$\begin{pmatrix} X_1 \\ X_2 \end{pmatrix} \sim N_2\left(\begin{pmatrix} 0 \\ 0 \end{pmatrix}, \begin{pmatrix} \boldsymbol{\Sigma}_{11} & \boldsymbol{\Sigma}_{12} \\ \boldsymbol{\Sigma}_{21} & \boldsymbol{\Sigma}_{22} \end{pmatrix}\right). \tag{68}$$

Let $\boldsymbol{x}_1, \ldots, \boldsymbol{x}_n \in \mathbb{R}^2$ be $n$ independent data points from $\boldsymbol{X}$ and $\widehat{\boldsymbol{\Sigma}} := \dfrac{1}{n}\sum_{i=1}^n \boldsymbol{x}_i \boldsymbol{x}_i^T$ be the sample covariance matrix. We define the sample and population correlations $\widehat{\rho}$ and $\rho$ as

$$\widehat{\rho} := (\widehat{\boldsymbol{\Sigma}}_{11})^{-1/2} \widehat{\boldsymbol{\Sigma}}_{12} (\widehat{\boldsymbol{\Sigma}}_{22})^{-1/2} \quad \text{and} \quad \rho := (\boldsymbol{\Sigma}_{11})^{-1/2} \boldsymbol{\Sigma}_{12} (\boldsymbol{\Sigma}_{22})^{-1/2}. \tag{69}$$

For any $t \in [0, \frac{1+|\rho|}{2})$, we have

$$\mathbb{P}\left(|\widehat{\rho} - \rho| > t\right) \leq 4 \exp\left[-\frac{3nt^2}{64(1+|\rho|)^2}\right]. \tag{70}$$

*Proof.* Suppose that $r$ is a small positive number such that $|\rho|r \leq t(1-r)$, we have

$$\mathbb{P}\left(|\widehat{\rho} - \rho| > t\right) \tag{71}$$

$$= \mathbb{P}\left(\left|(\widehat{\boldsymbol{\Sigma}}_{11})^{-1/2} \widehat{\boldsymbol{\Sigma}}_{12} (\widehat{\boldsymbol{\Sigma}}_{22})^{-1/2} - \rho\right| > t\right) \tag{72}$$

$$= \mathbb{P}\left(\left|\widehat{\boldsymbol{\Sigma}}_{12} - \rho(\widehat{\boldsymbol{\Sigma}}_{11}\widehat{\boldsymbol{\Sigma}}_{22})^{1/2}\right| > t(\widehat{\boldsymbol{\Sigma}}_{11}\widehat{\boldsymbol{\Sigma}}_{22})^{1/2}\right). \tag{73}$$

Let $r := \dfrac{t}{1+2|\rho|+t}$, we define the event

$$\mathcal{A} := \left\{\left|\widehat{\boldsymbol{\Sigma}}_{11} - \boldsymbol{\Sigma}_{11}\right| \leq r\boldsymbol{\Sigma}_{11} \text{ and } \left|\widehat{\boldsymbol{\Sigma}}_{22} - \boldsymbol{\Sigma}_{22}\right| \leq r\boldsymbol{\Sigma}_{22}\right\}. \tag{74}$$

On $\mathcal{A}$, we have

$$\left|(\widehat{\boldsymbol{\Sigma}}_{11}\widehat{\boldsymbol{\Sigma}}_{22})^{1/2} - (\boldsymbol{\Sigma}_{11}\boldsymbol{\Sigma}_{22})^{1/2}\right| \leq r(\boldsymbol{\Sigma}_{11}\boldsymbol{\Sigma}_{22})^{1/2}. \tag{75}$$

Therefore, using (75), we have

$$\mathbb{P}\left(|\widehat{\rho} - \rho| > t\right) \tag{76}$$

$$\leq \mathbb{P}\left(\left|\widehat{\boldsymbol{\Sigma}}_{12} - \rho(\widehat{\boldsymbol{\Sigma}}_{11}\widehat{\boldsymbol{\Sigma}}_{22})^{1/2}\right| > t(\widehat{\boldsymbol{\Sigma}}_{11}\widehat{\boldsymbol{\Sigma}}_{22})^{1/2} \mid \mathcal{A}\right) + \mathbb{P}(\mathcal{A}^c) \tag{77}$$

$$\leq \mathbb{P}\left(\left|\widehat{\boldsymbol{\Sigma}}_{12} - \rho(\boldsymbol{\Sigma}_{11}\boldsymbol{\Sigma}_{22})^{1/2}\right| > [t(1-r) - |\rho|r](\boldsymbol{\Sigma}_{11}\boldsymbol{\Sigma}_{22})^{1/2}\right) + \mathbb{P}(\mathcal{A}^c) \tag{78}$$

$$= \mathbb{P}\left(\left|\widehat{\boldsymbol{\Sigma}}_{12} - \boldsymbol{\Sigma}_{12}\right| > [t(1-r) - |\rho|r](\boldsymbol{\Sigma}_{11}\boldsymbol{\Sigma}_{22})^{1/2}\right) + \mathbb{P}(\mathcal{A}^c) \tag{79}$$

$$\leq \mathbb{P}\left(\left|\widehat{\boldsymbol{\Sigma}}_{12} - \boldsymbol{\Sigma}_{12}\right| > [t(1-r) - |\rho|r](\boldsymbol{\Sigma}_{11}\boldsymbol{\Sigma}_{22})^{1/2}\right) \tag{80}$$

$$+ \mathbb{P}\left(\left|\widehat{\boldsymbol{\Sigma}}_{11} - \boldsymbol{\Sigma}_{11}\right| > r\boldsymbol{\Sigma}_{11}\right) + \mathbb{P}\left(\left|\widehat{\boldsymbol{\Sigma}}_{22} - \boldsymbol{\Sigma}_{22}\right| > r\boldsymbol{\Sigma}_{22}\right).$$

Therefore, it suffices to analyze the terms $\mathbb{P}\left(|\widehat{\mathbf{\Sigma}}_{12} - \mathbf{\Sigma}_{12}| > \epsilon\right)$ and $\mathbb{P}\left(|\widehat{\mathbf{\Sigma}}_{11} - \mathbf{\Sigma}_{11}| > \epsilon\right)$. For this, we denote $\widetilde{\boldsymbol{x}}_i := (\widetilde{x}_{i1}, \widetilde{x}_{i2})$, where

$$\widetilde{x}_{i1} := \frac{x_{i1}}{(\mathbf{\Sigma}_{11})^{1/2}} \sim N(0,1) \text{ and } \widetilde{x}_{i2} := \frac{x_{i2}}{(\mathbf{\Sigma}_{22})^{1/2}} \sim N(0,1). \tag{81}$$

We have

$$\sum_{i=1}^{n} \left[\frac{(\widetilde{x}_{i1} + \widetilde{x}_{i2})^2}{2(1+\rho)}\right] \sim \chi_n^2 \text{ and } \sum_{i=1}^{n} \left[\frac{(\widetilde{x}_{i1} - \widetilde{x}_{i2})^2}{2(1-\rho)}\right] \sim \chi_n^2. \tag{82}$$

Furthermore, since $\sum_{i=1}^{n} \widetilde{x}_{i1}^2 \sim \chi_n^2$, we have

$$\mathbb{P}\left(|\widehat{\mathbf{\Sigma}}_{11} - \mathbf{\Sigma}_{11}| > \epsilon\right) = \mathbb{P}\left(\left|\frac{1}{n}\sum_{i=1}^{n} \widetilde{x}_{i1}^2 - 1\right| > \frac{\epsilon}{\mathbf{\Sigma}_{11}}\right) \tag{83}$$

$$= \mathbb{P}\left(\left|\sum_{i=1}^{n} \widetilde{x}_{i1}^2 - n\right| > \frac{n\epsilon}{\mathbf{\Sigma}_{11}}\right) \tag{84}$$

$$\leq \exp\left(-\frac{3n\epsilon^2}{16(\mathbf{\Sigma}_{11})^2}\right), \tag{85}$$

where the last inequality follows from (67).

Similarly,

$$\mathbb{P}\left(|\widehat{\mathbf{\Sigma}}_{12} - \mathbf{\Sigma}_{12}| > \epsilon\right) \tag{86}$$

$$= \mathbb{P}\left(\left|\frac{1}{n}\sum_{i=1}^{n} \widetilde{x}_{i1}\widetilde{x}_{i2} - \rho\right| > \frac{\epsilon}{(\mathbf{\Sigma}_{11}\mathbf{\Sigma}_{22})^{1/2}}\right) \tag{87}$$

$$= \mathbb{P}\left(\left|\frac{1}{n}\sum_{i=1}^{n} [(\widetilde{x}_{i1} + \widetilde{x}_{i2})^2 - 2(1+\rho) - (\widetilde{x}_{i1} - \widetilde{x}_{i2})^2 + 2(1-\rho)]\right| > \frac{4\epsilon}{(\mathbf{\Sigma}_{11}\mathbf{\Sigma}_{22})^{1/2}}\right) \tag{88}$$

$$\leq \mathbb{P}\left(\left|\frac{1}{n}\sum_{i=1}^{n} [(\widetilde{x}_{i1} + \widetilde{x}_{i2})^2 - 2(1+\rho)]\right| > \frac{2\epsilon}{(\mathbf{\Sigma}_{11}\mathbf{\Sigma}_{22})^{1/2}}\right) \tag{89}$$

$$+ \mathbb{P}\left(\left|\frac{1}{n}\sum_{i=1}^{n} [(\widetilde{x}_{i1} - \widetilde{x}_{i2})^2 - 2(1-\rho)]\right| > \frac{2\epsilon}{(\mathbf{\Sigma}_{11}\mathbf{\Sigma}_{22})^{1/2}}\right) \tag{90}$$

$$\leq \mathbb{P}\left(\left|\sum_{i=1}^{n} \left[\frac{(\widetilde{x}_{i1} + \widetilde{x}_{i2})^2}{2(1+\rho)}\right] - n\right| > \frac{n\epsilon}{(1+\rho)(\mathbf{\Sigma}_{11}\mathbf{\Sigma}_{22})^{1/2}}\right) \tag{91}$$

$$+ \mathbb{P}\left(\left|\sum_{i=1}^{n} \left[\frac{(\widetilde{x}_{i1} - \widetilde{x}_{i2})^2}{2(1-\rho)}\right] - n\right| > \frac{n\epsilon}{(1-\rho)(\mathbf{\Sigma}_{11}\mathbf{\Sigma}_{22})^{1/2}}\right). \tag{92}$$

Applying (67) on (91) and (92), we have

$$\mathbb{P}\left(|\widehat{\mathbf{\Sigma}}_{12} - \mathbf{\Sigma}_{12}| > \epsilon\right) \leq \exp\left(-\frac{3n\epsilon^2}{16(1+\rho)^2 \mathbf{\Sigma}_{11}\mathbf{\Sigma}_{22}}\right) + \exp\left(-\frac{3n\epsilon^2}{16(1-\rho)^2 \mathbf{\Sigma}_{11}\mathbf{\Sigma}_{22}}\right). \tag{93}$$

Combining (80) with (85) and (93), we get

$$\mathbb{P}\Big(\big|\widehat{\rho}-\rho\big|>t\Big) \leq 2\exp\left(-\frac{3n[t(1-r)-|\rho|r]^2}{16(1+|\rho|)^2}\right) + 2\exp\left(-\frac{3nr^2}{16}\right). \tag{94}$$

Since

$$r := \frac{t}{1+2|\rho|+t},$$

we have

$$\mathbb{P}\Big(\big|\widehat{\rho}-\rho\big|>t\Big) \leq 4\exp\left(-\frac{3nt^2}{64(1+|\rho|)^2}\right). \tag{95}$$

We thus complete the whole proof. $\square$

The following lemma bounds the difference between the sample correlation matrix and the true correlation matrix in elementwise sup-nrom.

**Lemma 7.** *Let $\widehat{\mathbf{R}}$ and $\mathbf{R}$ be the sample and population correlation matrices. We define the event*

$$\mathcal{A}_1 := \left\{\|\widehat{\mathbf{R}}-\mathbf{R}\|_{\max} \leq 18\sqrt{\frac{\log d}{n}}\right\}. \tag{96}$$

*Then, we have $\mathbb{P}(\mathcal{A}_1) \geq 1-1/d$.*

*Proof.* From Lemma 6, we have

$$\mathbb{P}\left(\|\widehat{\mathbf{R}}-\mathbf{R}\|_{\max} > t\right) \leq 4d^2\exp\left(-\frac{nt^2}{86}\right). \tag{97}$$

The result follows by choosing $t = 18\sqrt{\frac{\log d}{n}}$. $\square$

**Lemma 8.** *Let $\boldsymbol{\epsilon}_j := (\epsilon_{j1},\ldots,\epsilon_{jn})^T \in \mathbb{R}^n$ and $\boldsymbol{\epsilon}_j \sim N_n(\mathbf{0}, \sigma_j^2\mathbf{I}_n)$. We define the event*

$$\mathcal{A}_2 := \left\{\max_{1\leq j\leq d}\frac{\|\boldsymbol{\epsilon}_j\|_2^2}{n\sigma_j^2} \leq 1.4 \ \text{ and } \ \max_{1\leq j\leq d}\left|\frac{\|\boldsymbol{\epsilon}_j\|_2^2}{n\sigma_j^2}-1\right| \leq 3.5\sqrt{\frac{\log d}{n}}\right\}. \tag{98}$$

*We have $\mathbb{P}(\mathcal{A}_2) \geq 1 - d\exp\left(-\frac{n}{100}\right) - \frac{1}{d}$.*

*Proof.* Since

$$\frac{\|\boldsymbol{\epsilon}_j\|_2^2}{\sigma_j^2} = \sum_{i=1}^n \frac{\epsilon_{ij}^2}{\sigma_j^2} \sim \chi_n^2, \tag{99}$$

by Lemma 5, it is easy to see that for any constant $1 \leq w < 1.5$,

$$\mathbb{P}\Big(\frac{\|\boldsymbol{\epsilon}_j\|_2^2}{n\sigma_j^2} \leq w\Big) \geq 1-\exp\left(\frac{-3n(w-1)^2}{16}\right), \tag{100}$$

for all $j = 1, 2, \cdots, d$. By setting $w = 1.4$, we have

$$\mathbb{P}\left(\max_{1 \leq j \leq d} \frac{\|\boldsymbol{\epsilon}_j\|_2^2}{n\sigma_j^2} \leq 1.4\right) \geq 1 - d\exp\left(-\frac{n}{100}\right). \tag{101}$$

Similarly, for $t \in [0, 1/2)$, we have

$$\mathbb{P}\left(\left|\frac{\|\boldsymbol{\epsilon}_j\|_2^2}{n} - \sigma_j^2\right| > t\sigma_j^2\right) = \mathbb{P}\left(\left|\frac{\|\boldsymbol{\epsilon}_j\|_2^2}{\sigma_j^2} - n\right| > nt\right) \leq \exp\left(-\frac{3}{16}nt^2\right). \tag{102}$$

By setting $t = 3.5\sqrt{\frac{\log d}{n}}$, we have

$$\mathbb{P}\left(\max_{1 \leq j \leq d}\left|\frac{\|\boldsymbol{\epsilon}_j\|_2^2}{n\sigma_j^2} - 1\right| \leq 3.5\sqrt{\frac{\log d}{n}}\right) \geq 1 - \frac{1}{d}. \tag{103}$$

The desired result of this lemma follows from a union bound of (101) and (103). $\square$

The next lemma bounds the sample standard deviation of each marginal univariate Gaussian random variable.

**Lemma 9.** *Let $\widehat{\boldsymbol{\Sigma}}$ be the sample covariance matrix. Under the assumption that*

$$\limsup_{n \to \infty} \max_{1 \leq j \leq d} \boldsymbol{\Sigma}_{jj}\sqrt{\frac{\log d}{n}} < \frac{1}{2}, \tag{104}$$

*we define the event*

$$\mathcal{A}_3 := \left\{\frac{1}{2}\Lambda_{\min}(\boldsymbol{\Sigma}) \leq \min_{1 \leq j \leq d} \widehat{\boldsymbol{\Sigma}}_{jj} \leq \max_{1 \leq j \leq d} \widehat{\boldsymbol{\Sigma}}_{jj} \leq \frac{3}{2}\Lambda_{\max}(\boldsymbol{\Sigma})\right\}. \tag{105}$$

*We have $\mathbb{P}(\mathcal{A}_3) \geq 1 - 1/d$.*

*Proof.* By the definitions of $\Lambda_{\max}(\boldsymbol{\Sigma})$ and $\Lambda_{\min}(\boldsymbol{\Sigma})$, we have

$$\Lambda_{\max}(\boldsymbol{\Sigma}) \geq \max_{1 \leq j \leq d} \boldsymbol{\Sigma}_{jj} \geq \min_{1 \leq j \leq d} \boldsymbol{\Sigma}_{jj} \geq \Lambda_{\min}(\boldsymbol{\Sigma}). \tag{106}$$

By (85), we know that for any $j \in \{1, \ldots, d\}$ and $0 \leq \epsilon < 1/2$,

$$\mathbb{P}\left(|\widehat{\boldsymbol{\Sigma}}_{jj} - \boldsymbol{\Sigma}_{jj}| > \epsilon\right) \leq \exp\left(-\frac{3n\epsilon^2}{16(\boldsymbol{\Sigma}_{jj})^2}\right). \tag{107}$$

We choose $\epsilon = t\boldsymbol{\Sigma}_{jj}$, then

$$\mathbb{P}\left(\min_{1 \leq j \leq d} \boldsymbol{\Sigma}_{jj}(1-t) \leq \widehat{\boldsymbol{\Sigma}}_{jj} \leq (1+t)\max_{1 \leq j \leq d} \boldsymbol{\Sigma}_{jj}\right) \leq \exp\left(-\frac{3nt^2}{16}\right), \tag{108}$$

By the union bound, we have

$$\mathbb{P}\left((1-t)\Lambda_{\min}(\boldsymbol{\Sigma}) \leq \min_{1\leq j\leq d} \widehat{\boldsymbol{\Sigma}}_{jj} \leq \max_{1\leq j\leq d} \widehat{\boldsymbol{\Sigma}}_{jj} \leq (1+t)\Lambda_{\max}(\boldsymbol{\Sigma})\right) \leq d\exp\left(-\frac{3nt^2}{16}\right). \quad (109)$$

Under the assumption (104), we know that, for large enough $n$, there must be $\epsilon = t\Sigma_{jj} < 1/2$. The desired result of the lemma follows by setting $t = 3.5\sqrt{\frac{\log d}{n}}$. $\square$

Recall that we define

$$Q_j(\boldsymbol{\beta}_j) := \|\mathbf{Z}_{*j} - \mathbf{Z}_{*\backslash j}\boldsymbol{\beta}_j\|_2, \quad (110)$$

the next lemma provides theoretical justification to the choice of the tuning parameter $\lambda$.

**Lemma 10.** *Let* $\lambda = c\sqrt{\frac{2a\log d}{n}}$ *with* $c > 1$ *and* $a > 2$, *we define an event*

$$\mathcal{A}_4 := \left\{\max_{1\leq j\leq d} c\|\nabla Q_j(\boldsymbol{\beta}_j)\|_\infty \leq \lambda\sqrt{n}\right\}. \quad (111)$$

*Then*

$$\mathbb{P}(\mathcal{A}_4) \geq 1 - \sqrt{\frac{2}{\pi a\log d}} \cdot d^{2-a\left(1-2\sqrt{\frac{(a-1)\log d}{n}}\right)} - \frac{1}{d^{a-2}}. \quad (112)$$

*Proof.* Using the fact that $\mathbf{Z}_{*j} = \mathbf{X}_{*j}\widehat{\boldsymbol{\Gamma}}_{jj}^{-1/2}$ and $\mathbf{Z}_{*\backslash j} = \mathbf{X}_{*\backslash j}\widehat{\boldsymbol{\Gamma}}_{\backslash j,\backslash j}^{-1/2}$, we have

$$\mathbf{Z}_{*j} = \mathbf{Z}_{*\backslash j}\boldsymbol{\beta}_j + \widehat{\boldsymbol{\Gamma}}_{jj}^{-1/2}\boldsymbol{\epsilon}_j, \quad (113)$$

where $\boldsymbol{\epsilon}_j \sim N_n(\mathbf{0}, \sigma_j^2\mathbf{I}_n)$. We then have

$$\|\nabla Q_j(\boldsymbol{\beta}_j)\|_\infty = \frac{\|\mathbf{Z}_{*\backslash j}^T(\mathbf{Z}_{*j} - \mathbf{Z}_{*\backslash j}\boldsymbol{\beta}_j)\|_\infty}{\|\mathbf{Z}_{*j} - \mathbf{Z}_{*\backslash j}\boldsymbol{\beta}_j\|_2} \quad (114)$$

$$= \frac{\|\mathbf{Z}_{*\backslash j}^T\boldsymbol{\epsilon}_j\|_\infty}{\|\boldsymbol{\epsilon}_j\|_2} \quad (115)$$

From the properties of multivariate Gaussian, we know that $\mathbf{Z}_{*\backslash j}$ and $\boldsymbol{\epsilon}_j$ are independent. For any $\ell \neq j$, $\mathbf{Z}_{*\ell}^T\boldsymbol{\epsilon}_j$ follows $N(0, n\sigma_j^2)$ distribution when conditioning on $\mathbf{Z}_{*\backslash j}$. In the following argument, we suppose everything is conditioning on $\mathbf{Z}_{*\backslash j}$. Since $\|\boldsymbol{\epsilon}_j\|_2^2/\sigma_j^2 \sim \chi_n^2$, by Lemma 5, we know that, for any $0 \leq r_n < 1/2$,

$$\mathbb{P}\left(\max_{1\leq j\leq d} \frac{\|\boldsymbol{\epsilon}_j\|_2^2}{n\sigma_j^2} \leq 1 - r_n\right) \leq d\exp\left(-\frac{nr_n^2}{4}\right). \quad (116)$$

Let $\Phi(\cdot)$ and $\phi(\cdot)$ be the cumulative and density functions of a standard Gaussian random variable. For any $0 \leq r_n < 1/2$, we have

$$\mathbb{P}\left(\max_{1 \leq j \leq d} \frac{\|\mathbf{Z}_{*\backslash j}^T \boldsymbol{\epsilon}_j\|_\infty}{\|\boldsymbol{\epsilon}_j\|_2} > \sqrt{2a \log d}\right) \tag{117}$$

$$\leq d\mathbb{P}\left(\|\mathbf{Z}_{*\backslash j}^T \boldsymbol{\epsilon}_j\|_\infty > \sqrt{1-r_n}\sqrt{2an\sigma_j^2 \log d}\right) + \mathbb{P}\left(\max_{1 \leq j \leq d} \frac{\|\boldsymbol{\epsilon}_j\|_2^2}{n\sigma_j^2} \leq 1 - r_n\right) \tag{118}$$

$$\leq d\sum_{\ell \neq j} \mathbb{P}\left(|\mathbf{Z}_{*\ell}^T \boldsymbol{\epsilon}_j| > \sqrt{1-r_n}\sqrt{2an\sigma_j^2 \log d}\right) + d\exp\left(-\frac{nr_n^2}{4}\right) \tag{119}$$

$$\leq 2d^2\left(1 - \Phi(\sqrt{1-r_n}\sqrt{2a \log d})\right) + d\exp\left(-\frac{nr_n^2}{4}\right) \tag{120}$$

$$\leq 2d^2 \frac{d^{-a(1-r_n)}}{\sqrt{2\pi}\sqrt{1-r_n}\sqrt{2a \log d}} + d\exp\left(-\frac{nr_n^2}{4}\right) \tag{121}$$

$$= \frac{d^{2-a(1-r_n)}}{\sqrt{\pi(1-r_n)a \log d}} + d\exp\left(-\frac{nr_n^2}{4}\right). \tag{122}$$

where the second to last inequality follows from the fact that

$$1 - \Phi(t) \leq \frac{\phi(t)}{t} = \frac{1}{\sqrt{2\pi}t}\exp\left(-\frac{t^2}{2}\right), \tag{123}$$

whenever $t \geq 1$. Now let

$$r_n = 2\sqrt{\frac{(a-1)\log d}{n}}, \tag{124}$$

it can be seen that, when $n$ is large enough,

$$\mathbb{P}\left(\max_{1 \leq j \leq d} \frac{\|\mathbf{Z}_{*\backslash j}^T \boldsymbol{\epsilon}_j\|_\infty}{\|\boldsymbol{\epsilon}_j\|_2} \leq \sqrt{2a \log d}\right) \tag{125}$$

$$\geq 1 - \sqrt{\frac{2}{\pi a \log d}} \cdot d^{2-a\left(1-2\sqrt{\frac{(a-1)\log d}{n}}\right)} - \frac{1}{d^{a-2}}. \tag{126}$$

We finish the proof of this lemma. □

The next lemma shows that $\widehat{\boldsymbol{\beta}}_j - \boldsymbol{\beta}_j \in \boldsymbol{\Delta}_{\bar{c}}^{d-1}(k)$. Thus the error vector falls in a restricted set.

**Lemma 11.** *Let $\widehat{\boldsymbol{\beta}}_j$ be defined as in (27) and $\bar{c} = \frac{c+1}{c-1}$. Then, on the event $\mathcal{A}_4$, we have $\widehat{\boldsymbol{\beta}}_j - \boldsymbol{\beta}_j \in \boldsymbol{\Delta}_{\bar{c}}^{d-1}(k)$ for all $j = 1, \ldots, d$.*

*Proof.* Since $\widehat{\boldsymbol{\beta}}_j$ is the empirical minimizer of the objective function in (27), we have

$$\frac{1}{\sqrt{n}}\|\mathbf{Z}_{*j} - \mathbf{Z}_{*\backslash j}\widehat{\boldsymbol{\beta}}_j\|_2 + \lambda\|\widehat{\boldsymbol{\beta}}_j\|_1 \leq \frac{1}{\sqrt{n}}\|\mathbf{Z}_{*j} - \mathbf{Z}_{*\backslash j}\boldsymbol{\beta}_j\|_2 + \lambda\|\boldsymbol{\beta}_j\|_1. \tag{127}$$

Let $S := \{\ell : \beta_{j\ell} \neq 0\}$, it is obvious that $|S| \leq k$. From (127), we get

$$\frac{1}{\sqrt{n}}\|\mathbf{Z}_{*j} - \mathbf{Z}_{*\backslash j}\widehat{\boldsymbol{\beta}}_j\|_2 - \frac{1}{\sqrt{n}}\|\mathbf{Z}_{*j} - \mathbf{Z}_{*\backslash j}\boldsymbol{\beta}_j\|_2 \tag{128}$$

$$\leq \lambda\|\boldsymbol{\beta}_j\|_1 - \lambda\|\widehat{\boldsymbol{\beta}}_j\|_1 \tag{129}$$

$$\leq \lambda\|(\boldsymbol{\beta}_j)_S\|_1 - \lambda\|(\widehat{\boldsymbol{\beta}}_j)_S\|_1 - \lambda\|(\widehat{\boldsymbol{\beta}}_j)_{S^c}\|_1 \tag{130}$$

$$\leq \lambda\Big(\|(\boldsymbol{\beta}_j - \widehat{\boldsymbol{\beta}}_j)_S\|_1 - \|(\boldsymbol{\beta}_j - \widehat{\boldsymbol{\beta}}_j)_{S^c}\|_1\Big). \tag{131}$$

On the event $\mathcal{A}_4$, we have $c\|\nabla Q_j(\boldsymbol{\beta}_j)\|_\infty \leq \lambda\sqrt{n}$, where

$$Q_j(\boldsymbol{\beta}_j) := \|\mathbf{Z}_{*j} - \mathbf{Z}_{*\backslash j}\boldsymbol{\beta}_j\|_2. \tag{132}$$

Therefore, using the fact that $Q_j(\cdot)$ is a convex function

$$-\lambda\Big(\|(\boldsymbol{\beta}_j - \widehat{\boldsymbol{\beta}}_j)_S\|_1 + \|(\boldsymbol{\beta}_j - \widehat{\boldsymbol{\beta}}_j)_{S^c}\|_1\Big) \tag{133}$$

$$= -\lambda\|\boldsymbol{\beta}_j - \widehat{\boldsymbol{\beta}}_j\|_1 \tag{134}$$

$$\leq -\frac{c}{\sqrt{n}}\|\nabla Q_j(\boldsymbol{\beta}_j)\|_\infty \|\boldsymbol{\beta}_j - \widehat{\boldsymbol{\beta}}_j\|_1 \tag{135}$$

$$\leq -\frac{c}{\sqrt{n}}(\nabla Q_j(\boldsymbol{\beta}_j))^T(\boldsymbol{\beta}_j - \widehat{\boldsymbol{\beta}}_j) \tag{136}$$

$$\leq \frac{c}{\sqrt{n}}\big(Q_j(\widehat{\boldsymbol{\beta}}_j) - Q_j(\boldsymbol{\beta}_j)\big) \tag{137}$$

$$= c\Big(\frac{1}{\sqrt{n}}\|\mathbf{Z}_{*j} - \mathbf{Z}_{*\backslash j}\widehat{\boldsymbol{\beta}}_j\|_2 - \frac{1}{\sqrt{n}}\|\mathbf{Z}_{*j} - \mathbf{Z}_{*\backslash j}\boldsymbol{\beta}_j\|_2\Big). \tag{138}$$

By combining the above analysis, we have

$$-\Big(\|(\boldsymbol{\beta}_j - \widehat{\boldsymbol{\beta}}_j)_S\|_1 + \|(\boldsymbol{\beta}_j - \widehat{\boldsymbol{\beta}}_j)_{S^c}\|_1\Big) \leq c\Big(\|(\boldsymbol{\beta}_j - \widehat{\boldsymbol{\beta}}_j)_S\|_1 - \|(\boldsymbol{\beta}_j - \widehat{\boldsymbol{\beta}}_j)_{S^c}\|_1\Big). \tag{139}$$

Therefore,

$$\|(\boldsymbol{\beta}_j - \widehat{\boldsymbol{\beta}}_j)_{S^c}\|_1 \leq \frac{c+1}{c-1}\|(\boldsymbol{\beta}_j - \widehat{\boldsymbol{\beta}}_j)_S\|_1 = \bar{c}\|(\boldsymbol{\beta}_j - \widehat{\boldsymbol{\beta}}_j)_S\|_1. \tag{140}$$

We finish the proof of this lemma. $\square$

# B  Proof of Matrix Restricted Eigenvalue Conditions

The following lemma bounds the restricted eigenvalue of the sample correlation matrix $\widehat{\mathbf{R}}$.

**Lemma 12.** *Let the event $\mathcal{A}_3$ be defined as in Lemma 9. We assume $k\log d = o(n)$ and define an event*

$$\mathcal{B}_1 := \Bigg\{\inf_{\boldsymbol{\beta}\in\boldsymbol{\Delta}_{\bar{c}}^d(k)} \frac{\sqrt{k\boldsymbol{\beta}^T\widehat{\mathbf{R}}\boldsymbol{\beta}}}{\|\boldsymbol{\beta}\|_1} \geq \frac{1}{5(1+\bar{c})\xi_{\max}^{1/2}}\Bigg\}. \tag{141}$$

*Then, there exist constants $c_1$ and $c_2$, such that $\mathbb{P}(\mathcal{B}_1 \mid \mathcal{A}_3) \geq 1 - c_1\exp(-c_2 n)$.*

*Proof.* For any $S \subset \{1, 2, \cdots, d\}$ with $|S| \leq k$, we have, for any $\boldsymbol{\beta} \in \boldsymbol{\Delta}_{\bar{c}}^d(S)$,

$$\|\boldsymbol{\beta}\|_1 \leq (1+\bar{c})\|\boldsymbol{\beta}_S\|_1 \leq (1+\bar{c})\sqrt{k}\|\boldsymbol{\beta}_S\|_2 \leq (1+\bar{c})\sqrt{k}\|\boldsymbol{\beta}\|_2, \tag{142}$$

and

$$\boldsymbol{\beta}^T\boldsymbol{\Sigma}\boldsymbol{\beta} \geq \Lambda_{\min}(\boldsymbol{\Sigma})\|\boldsymbol{\beta}\|_2^2 \geq \Lambda_{\min}(\boldsymbol{\Sigma})\|\boldsymbol{\beta}_S\|_2^2 \geq \Lambda_{\min}(\boldsymbol{\Sigma})\frac{\|\boldsymbol{\beta}\|_1^2}{k(1+\bar{c})^2}, \tag{143}$$

where the last inequality uses the fact that $\|\boldsymbol{\beta}\|_1 \leq (1+\bar{c})\sqrt{k}\|\boldsymbol{\beta}_S\|_2$. This result is obtained from (142).

Recall that $\mathbf{X} \in \mathbb{R}^{n \times d}$ is a matrix and the rows of $\mathbf{X}$ are independent $N(0, \boldsymbol{\Sigma})$ Gaussian random vectors, where $\boldsymbol{\Sigma}$ is the $d \times d$ covariance matrix. Let $\widehat{\boldsymbol{\Sigma}}$ be the sample covariance matrix of $\mathbf{X}$. From Theorem 1 of Raskutti et al. (2010), we know that there exist two positive constants $c_1$ and $c_2$ such that

$$\mathbb{P}\left(\sqrt{\boldsymbol{\beta}^T\widehat{\boldsymbol{\Sigma}}\boldsymbol{\beta}} \geq \frac{1}{4}\sqrt{\boldsymbol{\beta}^T\boldsymbol{\Sigma}\boldsymbol{\beta}} - 9\max_{1\leq j \leq d}\sqrt{\boldsymbol{\Sigma}_{jj}}\sqrt{\frac{\log d}{n}}\|\boldsymbol{\beta}\|_1, \forall \boldsymbol{\beta} \in \mathbb{R}^d\right) \geq 1 - c_1\exp(-c_2 n). \tag{144}$$

Let $\widehat{\boldsymbol{\Gamma}} := \text{diag}(\widehat{\boldsymbol{\Sigma}})$, we have, for any $\boldsymbol{\beta} \in \mathbb{R}^d$,

$$\mathbb{P}\left(\sqrt{(\widehat{\boldsymbol{\Gamma}}^{-1/2}\boldsymbol{\beta})^T\widehat{\boldsymbol{\Sigma}}(\widehat{\boldsymbol{\Gamma}}^{-1/2}\boldsymbol{\beta})} \geq \frac{1}{4}\sqrt{(\widehat{\boldsymbol{\Gamma}}^{-1/2}\boldsymbol{\beta})^T\boldsymbol{\Sigma}(\widehat{\boldsymbol{\Gamma}}^{-1/2}\boldsymbol{\beta})} - 9\max_{1\leq j \leq d}\sqrt{\boldsymbol{\Sigma}_{jj}}\sqrt{\frac{\log d}{n}}\|\widehat{\boldsymbol{\Gamma}}^{-1/2}\boldsymbol{\beta}\|_1\right)$$
$$\geq 1 - c_1\exp(-c_2 n) \tag{145}$$

Since $\widehat{\boldsymbol{\Sigma}} = \widehat{\boldsymbol{\Gamma}}^{1/2}\widehat{\mathbf{R}}\widehat{\boldsymbol{\Gamma}}^{1/2}$, we have

$$(\widehat{\boldsymbol{\Gamma}}^{-1/2}\boldsymbol{\beta})^T\widehat{\boldsymbol{\Sigma}}(\widehat{\boldsymbol{\Gamma}}^{-1/2}\boldsymbol{\beta}) = \boldsymbol{\beta}^T\widehat{\mathbf{R}}\boldsymbol{\beta}. \tag{146}$$

On the event $\mathcal{A}_3$ as defined in Lemma 9, we have

$$\frac{1}{2}\Lambda_{\min}(\boldsymbol{\Sigma}) \leq \min_{1\leq j \leq d}\widehat{\boldsymbol{\Sigma}}_{jj} \leq \max_{1\leq j \leq d}\widehat{\boldsymbol{\Sigma}}_{jj} \leq \frac{3}{2}\Lambda_{\max}(\boldsymbol{\Sigma}). \tag{147}$$

It is easy to see that

$$\|\widehat{\boldsymbol{\Gamma}}^{-1/2}\boldsymbol{\beta}\|_1 \leq \|\boldsymbol{\beta}\|_1 \min_{1\leq j \leq d}(\widehat{\boldsymbol{\Sigma}}_{jj})^{-1/2} \leq \sqrt{\frac{2}{\Lambda_{\min}(\boldsymbol{\Sigma})}}\|\boldsymbol{\beta}\|_1, \tag{148}$$

$$\|\widehat{\boldsymbol{\Gamma}}^{-1/2}\boldsymbol{\beta}\|_2 \geq \|\boldsymbol{\beta}\|_2 \max_{1\leq j \leq d}(\widehat{\boldsymbol{\Sigma}}_{jj})^{-1/2} \geq \sqrt{\frac{2}{3\Lambda_{\max}(\boldsymbol{\Sigma})}}\|\boldsymbol{\beta}\|_2. \tag{149}$$

Therefore,

$$\sqrt{(\widehat{\boldsymbol{\Gamma}}^{-1/2}\boldsymbol{\beta})^T\boldsymbol{\Sigma}(\widehat{\boldsymbol{\Gamma}}^{-1/2}\boldsymbol{\beta})} \geq \sqrt{\Lambda_{\min}(\boldsymbol{\Sigma})}\|\widehat{\boldsymbol{\Gamma}}^{-1/2}\boldsymbol{\beta}\|_2 \geq \sqrt{\frac{2\Lambda_{\min}(\boldsymbol{\Sigma})}{3\Lambda_{\max}(\boldsymbol{\Sigma})}}\|\boldsymbol{\beta}\|_2. \tag{150}$$

Plugging (146), (148) and (150) into (145), we have

$$\mathbb{P}\left(\sqrt{\boldsymbol{\beta}^T\widehat{\mathbf{R}}\boldsymbol{\beta}} \geq \frac{1}{4}\sqrt{\frac{2\Lambda_{\min}(\boldsymbol{\Sigma})}{3\Lambda_{\max}(\boldsymbol{\Sigma})}}\|\boldsymbol{\beta}\|_2 - 9\sqrt{\frac{\max_{1\leq j\leq d}\boldsymbol{\Sigma}_{jj}}{\Lambda_{\min}(\boldsymbol{\Sigma})}}\sqrt{\frac{2\log d}{n}}\|\boldsymbol{\beta}\|_1,\ \forall \beta\in\mathbb{R}^d\right) \qquad (151)$$
$$\geq 1 - c_1\exp(-c_2 n).$$

By (142), for any $\boldsymbol{\beta} \in \boldsymbol{\Delta}_{\bar{c}}^d(k)$, we have

$$\|\boldsymbol{\beta}\|_2 \geq \frac{1}{(1+\bar{c})\sqrt{k}}\|\boldsymbol{\beta}\|_1. \qquad (152)$$

Therefore,

$$\mathbb{P}\left(\inf_{\boldsymbol{\beta}\in\boldsymbol{\Delta}_{\bar{c}}^d(k)}\frac{\sqrt{k\boldsymbol{\beta}^T\widehat{\mathbf{R}}\boldsymbol{\beta}}}{\|\boldsymbol{\beta}\|_1} \geq \frac{1}{4(1+\bar{c})}\sqrt{\frac{2\Lambda_{\min}(\boldsymbol{\Sigma})}{3\Lambda_{\max}(\boldsymbol{\Sigma})}} - 9\xi_{\max}^{1/2}\sqrt{\frac{2k\log d}{n}}\right) \geq 1 - c_1\exp(-c_2 n).$$

Since we assume $k\log d = o(n)$, for $n$ large enough, we have

$$\frac{1}{4(1+\bar{c})}\sqrt{\frac{2\Lambda_{\min}(\boldsymbol{\Sigma})}{3\Lambda_{\max}(\boldsymbol{\Sigma})}} - 9\xi_{\max}^{1/2}\sqrt{\frac{2k\log d}{n}} \geq \frac{1}{5(1+\bar{c})\xi_{\max}^{1/2}}. \qquad (153)$$

We finish the proof of this lemma. $\square$

We can see that when $k\log d = o(n)$ and $n$ large enough, with high probability,

$$\eta(\widehat{\mathbf{R}}) := \inf_{\boldsymbol{\beta}\in\boldsymbol{\Delta}_{\bar{c}}^d(k)}\frac{\sqrt{k\boldsymbol{\beta}^T\widehat{\mathbf{R}}\boldsymbol{\beta}}}{\|\boldsymbol{\beta}\|_1} \geq \frac{1}{5(1+\bar{c})\xi_{\max}^{1/2}} \qquad (154)$$

is bounded away from zero by a positive constant. This implies the restricted eigenvalue condition required by the SQRT-Lasso method, which is provided in the next lemma.

**Lemma 13.** *We define the event*

$$\mathcal{B}_2 := \left\{\max_{1\leq j\leq d}\frac{\|\mathbf{Z}_{*\setminus j}(\widehat{\boldsymbol{\beta}}_j - \boldsymbol{\beta}_j)\|_2}{\tau_j} \leq C\cdot\sqrt{k\log d}\right\}. \qquad (155)$$

*Then* $\mathbb{P}(\mathcal{B}_2\,|\,\mathcal{B}_1) \geq 1 - 1/d$.

*Proof.* We only need to verify that the $L_1$-restricted eigenvalue condition of the SQRT-Lasso is satisfied. More specifically, we need to verify the conditions in Theorem 1 of Belloni et al. (2012). Since $\widehat{\mathbf{R}} = \frac{1}{n}\mathbf{Z}\mathbf{Z}^T$, where $\mathbf{Z}$ is defined as in Section 3, we have proved that, on $\mathcal{B}_1$,

$$\inf_{\boldsymbol{\beta}\in\boldsymbol{\Delta}_{\bar{c}}^d(k)}\frac{\sqrt{k}\|\mathbf{Z}\boldsymbol{\beta}\|_2}{\sqrt{n}\|\boldsymbol{\beta}\|_1} = \inf_{\boldsymbol{\beta}\in\boldsymbol{\Delta}_{\bar{c}}^d(k)}\frac{\sqrt{k\boldsymbol{\beta}^T\widehat{\mathbf{R}}\boldsymbol{\beta}}}{\|\boldsymbol{\beta}\|_1} \geq \frac{1}{5(1+\bar{c})\xi_{\max}^{1/2}}. \qquad (156)$$

Recall that $\mathbf{Z}_{*\backslash j}$ is the $n \times (d-1)$ submatrix of $\mathbf{Z}$ with the $j^{\text{th}}$ column removed, then

$$\min_{1 \leq j \leq d} \inf_{\boldsymbol{\beta} \in \boldsymbol{\Delta}_{\bar{c}}^{d-1}(k)} \frac{\sqrt{k}\|\mathbf{Z}_{*\backslash j}\boldsymbol{\beta}\|_2}{\sqrt{n}\|\boldsymbol{\beta}\|_1} \geq \inf_{\boldsymbol{\beta} \in \boldsymbol{\Delta}_{\bar{c}}^d(k)} \frac{\sqrt{k}\|\mathbf{Z}\boldsymbol{\beta}\|_2}{\sqrt{n}\|\boldsymbol{\beta}\|_1} \geq \frac{1}{5(1+\bar{c})\xi_{\max}^{1/2}}. \tag{157}$$

Therefore, on the event $\mathcal{B}_1$, the $L_1$-restricted eigenvalue condition holds for all the SQRT-Lasso subproblem defined in (27). The desired result follows from Theorem 1 of Belloni et al. (2012). $\square$

## C Main Lemmas

We define the event $\mathcal{E} := \mathcal{A}_1 \bigcap \mathcal{A}_2 \bigcap \mathcal{A}_3 \bigcap \mathcal{A}_4 \bigcap \mathcal{B}_1 \bigcap \mathcal{B}_2$. It is easy to see that $\mathbb{P}(\mathcal{E}) \geq 1 - o(1)$. To prove the main results, we separately analyze the diagonal and off-diagonal elements of $\widehat{\boldsymbol{\Theta}} - \boldsymbol{\Theta}$. In the next subsection, we first control the diagonal elements.

### C.1 Analyzing the Diagonal Elements

**Lemma 14.** *On the event $\mathcal{E}$, we have*

$$\max_{1 \leq j \leq d} \left|\widehat{\boldsymbol{\Theta}}_{jj} - \boldsymbol{\Theta}_{jj}\right| \leq C \cdot \|\boldsymbol{\Theta}\|_2 \sqrt{\frac{\log d}{n}}, \tag{158}$$

*for large enough $n$.*

*Proof.* Recall that $\mathbf{Z}_{*j} = \mathbf{Z}_{*\backslash j}\boldsymbol{\beta}_j + \widehat{\boldsymbol{\Gamma}}_{jj}^{-1/2}\boldsymbol{\epsilon}_j$, we have

$$\left|(\widehat{\boldsymbol{\Theta}}_{jj})^{-1} - (\boldsymbol{\Theta}_{jj})^{-1}\right| \tag{159}$$

$$= \left|\widehat{\boldsymbol{\Gamma}}_{jj}\widehat{\tau}_j^2 - \widehat{\boldsymbol{\Gamma}}_{jj}\tau_j^2\right| \tag{160}$$

$$= \left|\widehat{\boldsymbol{\Gamma}}_{jj}\frac{\|\mathbf{Z}_{*j} - \mathbf{Z}_{*\backslash j}\widehat{\boldsymbol{\beta}}_j\|_2^2}{n} - \widehat{\boldsymbol{\Gamma}}_{jj}\tau_j^2\right| \tag{161}$$

$$= \left|\widehat{\boldsymbol{\Gamma}}_{jj}\frac{\|\mathbf{Z}_{*\backslash j}(\boldsymbol{\beta}_j - \widehat{\boldsymbol{\beta}}_j) + \widehat{\boldsymbol{\Gamma}}_{jj}^{-1/2}\boldsymbol{\epsilon}_j\|_2^2}{n} - \sigma_j^2\right| \tag{162}$$

$$\leq \left|\frac{\|\boldsymbol{\epsilon}_j\|_2^2}{n} - \sigma_j^2\right| + \widehat{\boldsymbol{\Gamma}}_{jj}\frac{\|\mathbf{Z}_{*\backslash j}(\widehat{\boldsymbol{\beta}}_j - \boldsymbol{\beta}_j)\|_2^2}{n} + 2\widehat{\boldsymbol{\Gamma}}_{jj}^{1/2}\frac{|(\widehat{\boldsymbol{\beta}}_j - \boldsymbol{\beta}_j)^T\mathbf{Z}_{*\backslash j}^T\boldsymbol{\epsilon}_j|}{n}. \tag{163}$$

On the event $\mathcal{E}$, we have

$$\frac{\|\boldsymbol{\epsilon}_j\|_2^2}{n} \leq 1.4\sigma_j^2 \text{ and } \left|\frac{\|\boldsymbol{\epsilon}_j\|_2^2}{n} - \sigma_j^2\right| \leq 3.5\sigma_j^2\sqrt{\frac{\log d}{n}}, \tag{164}$$

$$\|\mathbf{Z}_{*\backslash j}(\widehat{\boldsymbol{\beta}}_j - \boldsymbol{\beta}_j)\|_2 \leq C \cdot \tau_j\sqrt{k\log d}, \tag{165}$$

$$\widehat{\boldsymbol{\Gamma}}_{jj}^{1/2} \leq \max_{1 \leq j \leq d}(\widehat{\boldsymbol{\Sigma}}_{jj})^{1/2} \leq \sqrt{\frac{3\Lambda_{\max}(\boldsymbol{\Sigma})}{2}}, \tag{166}$$

$$\widehat{\boldsymbol{\Gamma}}_{jj}^{-1/2} = (\widehat{\boldsymbol{\Sigma}}_{jj})^{-1/2} \leq \sqrt{\frac{2}{\Lambda_{\min}(\boldsymbol{\Sigma})}}. \tag{167}$$

Futhermore, since $\tau_j^2 = \sigma_j^2 \widehat{\Gamma}_{jj}^{-1}$, we have

$$\widehat{\Gamma}_{jj} \| \mathbf{Z}_{*\backslash j}(\widehat{\boldsymbol{\beta}}_j - \boldsymbol{\beta}_j) \|_2^2 \leq C^2 \cdot \sigma_j^2 \cdot k \log d. \tag{168}$$

We also have

$$2\widehat{\Gamma}_{jj}^{1/2} |(\widehat{\boldsymbol{\beta}}_j - \boldsymbol{\beta}_j)^T \mathbf{Z}_{*\backslash j}^T \boldsymbol{\epsilon}_j| \leq 2\widehat{\Gamma}_{jj}^{1/2} \|\widehat{\boldsymbol{\beta}}_j - \boldsymbol{\beta}_j\|_1 \cdot \|\mathbf{Z}_{*\backslash j}^T \boldsymbol{\epsilon}_j\|_\infty. \tag{169}$$

Since $\widehat{\boldsymbol{\beta}}_j - \boldsymbol{\beta}_j \in \boldsymbol{\Delta}_{\bar{c}}^{d-1}(k)$, by the definition of the $L_1$-restricted eigenvalue, we know that

$$\widehat{\Gamma}_{jj}^{1/2} \|\widehat{\boldsymbol{\beta}}_j - \boldsymbol{\beta}_j\|_1 \tag{170}$$

$$\leq \widehat{\Gamma}_{jj}^{1/2} \cdot 5(1+\bar{c}) \xi_{\max}^{1/2} \cdot \sqrt{k(\widehat{\boldsymbol{\beta}}_j - \boldsymbol{\beta}_j)^T \widehat{\mathbf{R}}_{\backslash j, \backslash j}(\widehat{\boldsymbol{\beta}}_j - \boldsymbol{\beta}_j)} \tag{171}$$

$$\leq \widehat{\Gamma}_{jj}^{1/2} \cdot 5(1+\bar{c}) \xi_{\max}^{1/2} \cdot \sqrt{\frac{k}{n}} \cdot \|\mathbf{Z}_{*\backslash j}(\widehat{\boldsymbol{\beta}}_j - \boldsymbol{\beta}_j)\|_2 \tag{172}$$

$$\leq \widehat{\Gamma}_{jj}^{1/2} \cdot 5(1+\bar{c}) \xi_{\max}^{1/2} \cdot \sqrt{\frac{k}{n}} \cdot C \cdot \sigma_j \Gamma_{jj}^{-1/2} \sqrt{k \log d} \tag{173}$$

$$\leq 5(1+\bar{c}) \cdot C \cdot \xi_{\max}^{1/2} \cdot \sigma_j \cdot k \sqrt{\frac{\log d}{n}}. \tag{174}$$

Similarly, on the event $\mathcal{E}$, we have

$$\|\mathbf{Z}_{*\backslash j}^T \boldsymbol{\epsilon}_j\|_\infty \leq \frac{1}{c} \|\boldsymbol{\epsilon}_j\|_2 \lambda \sqrt{n} = \sigma_j \sqrt{2.8 n \log d}. \tag{175}$$

Therefore, on $\mathcal{E}$,

$$\left| (\widehat{\boldsymbol{\Theta}}_{jj})^{-1} - (\boldsymbol{\Theta}_{jj})^{-1} \right| \tag{176}$$

$$\leq 3.5\sigma_j^2 \sqrt{\frac{\log d}{n}} + C^2 \sigma_j^2 \cdot \frac{k \log d}{n} + 10\sqrt{2.8} \cdot (1+\bar{c}) \cdot C \xi_{\max}^{1/2} \sigma_j^2 \frac{k \log d}{n}. \tag{177}$$

Since $k\sqrt{\frac{\log d}{n}} = o(1)$, there exists a constant $C$ such that, for large enough $n$,

$$\left| (\widehat{\boldsymbol{\Theta}}_{jj})^{-1} - (\boldsymbol{\Theta}_{jj})^{-1} \right| \leq C\sigma_j^2 \cdot \sqrt{\frac{\log d}{n}}. \tag{178}$$

Multiplying $\boldsymbol{\Theta}_{jj}$ on both sides and using the fact that $\boldsymbol{\Theta}_{jj} = \sigma_j^{-2}$, we have

$$\left| \boldsymbol{\Theta}_{jj}(\widehat{\boldsymbol{\Theta}}_{jj})^{-1} - 1 \right| \leq C\sqrt{\frac{\log d}{n}}. \tag{179}$$

This implies that

$$\left(1 + C\sqrt{\frac{\log d}{n}}\right)^{-1} \leq \frac{\widehat{\boldsymbol{\Theta}}_{jj}}{\boldsymbol{\Theta}_{jj}} \leq \left(1 - C\sqrt{\frac{\log d}{n}}\right)^{-1} \tag{180}$$

Since, for large enough $n$,

$$1 - C\sqrt{\frac{\log d}{n}} \leq \left(1 + C\sqrt{\frac{\log d}{n}}\right)^{-1} \text{ and } \left(1 - C\sqrt{\frac{\log d}{n}}\right)^{-1} \leq 1 + 2C\sqrt{\frac{\log d}{n}}, \quad (181)$$

we have

$$\left(1 - C\sqrt{\frac{\log d}{n}}\right) \leq \frac{\widehat{\Theta}_{jj}}{\Theta_{jj}} \leq \left(1 + C\sqrt{\frac{\log d}{n}}\right). \quad (182)$$

This implies that

$$\max_{1 \leq j \leq d} |\widehat{\Theta}_{jj} - \Theta_{jj}| \leq C \cdot \|\Theta\|_2 \sqrt{\frac{\log d}{n}}. \quad (183)$$

The last inequality follows from the fact that $\max_{1 \leq j \leq d} \Theta_{jj} \leq \|\Theta\|_2$. $\square$

## C.2 Analyzing the Off-diagonal Elements in $L_1$-norm Error

We first bound the $L_1$-norm of each column of the off-diagonal elements of $\widehat{\Theta} - \Theta$.

**Lemma 15.** *On the event $\mathcal{E}$, we have*

$$\max_{1 \leq j \leq d} \|\widehat{\Theta}_{\setminus j, j} - \Theta_{\setminus j, j}\|_1 \leq C(\|\Theta\|_2 k + \|\Theta\|_1)\sqrt{\frac{\log d}{n}}, \quad (184)$$

$$\max_{1 \leq j \leq d} \|\widehat{\boldsymbol{\beta}}_j - \boldsymbol{\beta}_j\|_1 \leq 5\sqrt{2}C(1 + \bar{c}).\xi_{\max} k \sqrt{\frac{\log d}{n}}, \quad (185)$$

*for large enough $n$.*

*Proof.* We recall that

$$\Theta_{\setminus j, j} = -\Theta_{jj} \widehat{\Gamma}_{\setminus j, \setminus j}^{-1/2} \widehat{\Gamma}_{jj}^{1/2} \boldsymbol{\beta}_j. \quad (186)$$

Since $\tau_j^{-2} := \sigma_j^{-2}\widehat{\Gamma}_{jj} = \Theta_{jj}\widehat{\Gamma}_{jj}$, we have

$$\|\widehat{\Theta}_{\setminus j, j} - \Theta_{\setminus j, j}\|_1 \quad (187)$$
$$= \|\widehat{\tau}_j^{-2}\widehat{\Gamma}_{jj}^{-1/2}\widehat{\Gamma}_{\setminus j, \setminus j}^{-1/2}\widehat{\boldsymbol{\beta}}_j - \tau_j^{-2}\widehat{\Gamma}_{jj}^{-1/2}\widehat{\Gamma}_{\setminus j, \setminus j}^{-1/2}\boldsymbol{\beta}_j\|_1 \quad (188)$$
$$= \|\widehat{\Gamma}_{\setminus j, \setminus j}^{-1/2}\widehat{\Gamma}_{jj}^{1/2}(\widehat{\Theta}_{jj}\widehat{\boldsymbol{\beta}}_j - \Theta_{jj}\boldsymbol{\beta}_j)\|_1 \quad (189)$$
$$\leq \|\widehat{\Gamma}_{\setminus j, \setminus j}^{-1/2}\|_1 |\widehat{\Gamma}_{jj}^{1/2}\widehat{\Theta}_{jj}| \|\widehat{\boldsymbol{\beta}}_j - \boldsymbol{\beta}_j\|_1 + \widehat{\Gamma}_{jj}^{1/2}|\widehat{\Theta}_{jj} - \Theta_{jj}| \|\widehat{\Gamma}_{\setminus j, \setminus j}^{-1/2}\boldsymbol{\beta}_j\|_1 \quad (190)$$
$$= \|\widehat{\Gamma}_{\setminus j, \setminus j}^{-1/2}\|_1 |\widehat{\Gamma}_{jj}^{1/2}\widehat{\Theta}_{jj}| \|\widehat{\boldsymbol{\beta}}_j - \boldsymbol{\beta}_j\|_1 + |\widehat{\Theta}_{jj} - \Theta_{jj}| \|\Theta_{\setminus j, j}\Theta_{jj}^{-1}\|_1 \quad (191)$$
$$\leq \|\widehat{\Gamma}_{\setminus j, \setminus j}^{-1/2}\|_1 |\widehat{\Gamma}_{jj}^{1/2}\widehat{\Theta}_{jj}| \|\widehat{\boldsymbol{\beta}}_j - \boldsymbol{\beta}_j\|_1 + \left|\frac{\widehat{\Theta}_{jj}}{\Theta_{jj}} - 1\right| \|\Theta_{\setminus j, j}\|_1. \quad (192)$$

On the event $\mathcal{E}$, using (179) and the analysis of Lemma 14, we have, for large enough $n$,

$$\|\widehat{\mathbf{\Gamma}}_{\backslash j,\backslash j}^{-1/2}\|_1 \leq \max_{1\leq j\leq d} \widehat{\mathbf{\Gamma}}_{jj}^{-1/2} = \max_{1\leq j\leq d}(\widehat{\mathbf{\Sigma}}_{jj})^{-1/2} \leq \sqrt{\frac{2}{\Lambda_{\min}(\mathbf{\Sigma})}}, \tag{193}$$

$$\widehat{\mathbf{\Gamma}}_{jj}^{1/2} \leq \sqrt{\frac{3}{2}\Lambda_{\max}(\mathbf{\Sigma})}, \tag{194}$$

$$\widehat{\mathbf{\Theta}}_{jj} \leq \left(1 + C\sqrt{\frac{\log d}{n}}\right)\mathbf{\Theta}_{jj} \leq 2\|\mathbf{\Theta}\|_2, \tag{195}$$

$$\left|\frac{\widehat{\mathbf{\Theta}}_{jj}}{\mathbf{\Theta}_{jj}} - 1\right| \leq C\sqrt{\frac{\log d}{n}}. \tag{196}$$

By Lemma 11, $\widehat{\boldsymbol{\beta}}_j - \boldsymbol{\beta}_j \in \mathbf{\Delta}_{\bar{c}}^{d-1}(k)$. By the definition of the $L_1$-restricted eigenvalue, we know that, on the event $\mathcal{B}_2$,

$$\begin{aligned}
\|\widehat{\boldsymbol{\beta}}_j - \boldsymbol{\beta}_j\|_1 &\leq 5(1+\bar{c})\xi_{\max}^{1/2} \cdot \sqrt{k(\widehat{\boldsymbol{\beta}}_j - \boldsymbol{\beta}_j)^T \widehat{\mathbf{R}}_{\backslash j,\backslash j}(\widehat{\boldsymbol{\beta}}_j - \boldsymbol{\beta}_j)} & (197)\\
&\leq 5(1+\bar{c})\xi_{\max}^{1/2} \cdot \sqrt{\frac{k}{n}} \cdot \|\mathbf{Z}_{*\backslash j}(\widehat{\boldsymbol{\beta}}_j - \boldsymbol{\beta}_j)\|_2 & (198)\\
&\leq 5(1+\bar{c})\xi_{\max}^{1/2} \cdot \sqrt{\frac{k}{n}} \cdot C \cdot \sigma_j \widehat{\mathbf{\Gamma}}_{jj}^{-1/2}\sqrt{k\log d} & (199)\\
&= 5(1+\bar{c}) \cdot C \cdot \xi_{\max}^{1/2} \cdot \sigma_j \cdot \widehat{\mathbf{\Gamma}}_{jj}^{-1/2} \cdot k\sqrt{\frac{\log d}{n}} & (200)\\
&\leq 5\sqrt{2}C(1+\bar{c})\cdot\xi_{\max}k\sqrt{\frac{\log d}{n}}, & (201)
\end{aligned}$$

where the last inequality follows from the fact that

$$\sigma_j \widehat{\mathbf{\Gamma}}_{jj}^{-1/2} \leq \sqrt{\mathbf{\Sigma}_{jj}} \cdot \sqrt{\frac{2}{\Lambda_{\min}(\mathbf{\Sigma})}} \leq \sqrt{\frac{2\Lambda_{\max}(\mathbf{\Sigma})}{\Lambda_{\min}(\mathbf{\Sigma})}} \leq \sqrt{2}\xi_{\max}^{1/2}. \tag{202}$$

Therefore

$$\|\widehat{\mathbf{\Gamma}}_{\backslash j,\backslash j}^{-1/2}\|_1 |\widehat{\mathbf{\Gamma}}_{jj}^{1/2}\widehat{\mathbf{\Theta}}_{jj}| \|\widehat{\boldsymbol{\beta}}_j - \boldsymbol{\beta}_j\|_1 \leq 10(1+\bar{c})\sqrt{3}\xi_{\max} \cdot C \cdot \sigma_j \|\mathbf{\Theta}\|_2 \cdot \widehat{\mathbf{\Gamma}}_{jj}^{-1/2} \cdot k\sqrt{\frac{\log d}{n}} \tag{203}$$

and

$$\left|\frac{\widehat{\mathbf{\Theta}}_{jj}}{\mathbf{\Theta}_{jj}} - 1\right| \|\mathbf{\Theta}_{\backslash j,j}\|_1 \leq C \cdot \|\mathbf{\Theta}\|_1 \sqrt{\frac{\log d}{n}}. \tag{204}$$

Combining all the above analysis, we have

$$\|\widehat{\mathbf{\Theta}}_{\backslash j,j} - \mathbf{\Theta}_{\backslash j,j}\|_1 \leq C\|\mathbf{\Theta}\|_2 k\sqrt{\frac{\log d}{n}} + C \cdot \|\mathbf{\Theta}\|_1 \sqrt{\frac{\log d}{n}}. \tag{205}$$

We thus prove the desired result of the lemma. $\square$

## C.3 Analyzing the off-diagonal Elements in Sup-norm Error

To conduct sup-norm analysis of each of the column of the off-diagonal elements, we first present a technical lemma:

**Lemma 16.** *On the event $\mathcal{E}$, there exists a constant $C$ such that*

$$\left\|\frac{\widehat{\boldsymbol{\Gamma}}_{jj}^{1/2}}{n}\mathbf{Z}_{*\backslash j}^T(\mathbf{Z}_{*\backslash j}\widehat{\boldsymbol{\beta}}_j - \mathbf{Z}_{*\backslash j}\boldsymbol{\beta}_j)\right\|_\infty \leq C\sigma_j\sqrt{\frac{\log d}{n}} \quad \text{for all } j=1,\ldots,d. \tag{206}$$

*Proof.* Recall that we define

$$Q_j(\boldsymbol{\gamma}) = \left\|\mathbf{Z}_{*j} - \mathbf{Z}_{*\backslash j}\boldsymbol{\gamma}\right\|_2, \tag{207}$$

let $\lambda = c\sqrt{\dfrac{2a\log d}{n}}$ with $c > 1$ and $a > 2$, since $\widehat{\boldsymbol{\beta}}_j$ is defined as:

$$\widehat{\boldsymbol{\beta}}_j := \operatorname*{argmin}_{\boldsymbol{\gamma}\in\mathbb{R}^{d-1}}\left\{\frac{1}{\sqrt{n}}Q_j(\boldsymbol{\gamma}) + \lambda\|\boldsymbol{\gamma}\|_1\right\}, \tag{208}$$

we know that $c\|\nabla Q_j(\widehat{\boldsymbol{\beta}}_j)\|_\infty \leq \sqrt{n}\lambda$. This means

$$\left\|\mathbf{Z}_{*,\backslash j}^T(\mathbf{Z}_{*j} - \mathbf{Z}_{*\backslash j}\widehat{\boldsymbol{\beta}}_j)\right\|_\infty \leq \sqrt{2a\log d}\|\mathbf{Z}_{*j} - \mathbf{Z}_{*\backslash j}\widehat{\boldsymbol{\beta}}\|_2. \tag{209}$$

Using the fact that $\mathbf{Z}_{*j} = \mathbf{Z}_{*\backslash j}\boldsymbol{\beta}_j + \widehat{\boldsymbol{\Gamma}}_{jj}^{-1/2}\boldsymbol{\epsilon}_j$, we have

$$\left\|\mathbf{Z}_{*\backslash j}^T(\mathbf{Z}_{*\backslash j}(\boldsymbol{\beta}_j - \widehat{\boldsymbol{\beta}}_j) + \widehat{\boldsymbol{\Gamma}}_{jj}^{-1/2}\boldsymbol{\epsilon}_j)\right\|_\infty \leq \sqrt{2a\log d}\|\mathbf{Z}_{*\backslash j}(\boldsymbol{\beta}_j - \widehat{\boldsymbol{\beta}}_j) + \widehat{\boldsymbol{\Gamma}}_{jj}^{-1/2}\boldsymbol{\epsilon}_j\|_2. \tag{210}$$

Therefore

$$\left\|\mathbf{Z}_{*\backslash j}^T\mathbf{Z}_{*\backslash j}(\boldsymbol{\beta}_j - \widehat{\boldsymbol{\beta}}_j)\right\|_\infty \tag{211}$$
$$\leq \sqrt{2a\log d}\|\mathbf{Z}_{*\backslash j}(\boldsymbol{\beta}_j - \widehat{\boldsymbol{\beta}}_j)\|_2 + \sqrt{2a\log d}\widehat{\boldsymbol{\Gamma}}_{jj}^{-1/2}\|\boldsymbol{\epsilon}_j\|_2 + \widehat{\boldsymbol{\Gamma}}_{jj}^{-1/2}\|\mathbf{Z}_{*\backslash j}^T\boldsymbol{\epsilon}_j\|_\infty. \tag{212}$$

On the event $\mathcal{E}$, we have

$$\|\boldsymbol{\epsilon}_j\|_2 \leq \sigma_j \cdot \sqrt{1.4n}, \tag{213}$$
$$\|\mathbf{Z}_{*\backslash j}^T\boldsymbol{\epsilon}_j\|_\infty \leq \sqrt{2a\log d}\|\boldsymbol{\epsilon}_j\|_2 \leq \sigma_j \cdot \sqrt{2.8an\log d}, \tag{214}$$
$$\|\mathbf{Z}_{*\backslash j}(\boldsymbol{\beta}_j - \widehat{\boldsymbol{\beta}}_j)\|_2 \leq C\tau_j\sqrt{k\log d} \leq C \cdot \sigma_j \cdot \widehat{\boldsymbol{\Gamma}}_{jj}^{-1/2}\sqrt{k\log d}. \tag{215}$$

By piecing all these terms together, we have

$$\left\|\mathbf{Z}_{*\backslash j}^T\mathbf{Z}_{*\backslash j}(\boldsymbol{\beta}_j - \widehat{\boldsymbol{\beta}}_j)\right\|_\infty \tag{216}$$
$$\leq C \cdot \sigma_j \cdot \widehat{\boldsymbol{\Gamma}}_{jj}^{-1/2}\sqrt{2ak} \cdot \log d + 2\sigma_j \cdot \widehat{\boldsymbol{\Gamma}}_{jj}^{-1/2}\sqrt{2.8an\log d}. \tag{217}$$

Therefore
$$\left\|\frac{\widehat{\mathbf{\Gamma}}_{jj}^{1/2}}{n}\mathbf{Z}_{*\backslash j}^T(\mathbf{Z}_{*\backslash j}\widehat{\boldsymbol{\beta}}_j - \mathbf{Z}_{*\backslash j}\boldsymbol{\beta}_j)\right\|_\infty \leq C \cdot \sigma_j \cdot \sqrt{2ak} \cdot \frac{\log d}{n} + 2\sigma_j\sqrt{\frac{2.8a\log d}{n}}. \tag{218}$$

The result follows from the fact that
$$\frac{\sqrt{k}\log d}{n} \leq \sqrt{\frac{\log d}{n}}$$
for large enough $n$. We finish the proof of this lemma. $\square$

**Lemma 17.** *On the event $\mathcal{E}$, we have*
$$\max_{1\leq j\leq d}\|\widehat{\boldsymbol{\Theta}}_{\backslash j,j} - \boldsymbol{\Theta}_{\backslash j,j}\|_\infty \leq C\|\boldsymbol{\Theta}\|_1\sqrt{\frac{\log d}{n}}, \tag{219}$$
*for large enough $n$.*

*Proof.* We have the following decomposition
$$\|\widehat{\boldsymbol{\Theta}}_{\backslash j,j} - \boldsymbol{\Theta}_{\backslash j,j}\|_\infty \tag{220}$$
$$= \|\widehat{\tau}_j^{-2}\widehat{\mathbf{\Gamma}}_{jj}^{-1/2}\widehat{\mathbf{\Gamma}}_{\backslash j,\backslash j}^{-1/2}\widehat{\boldsymbol{\beta}}_j - \tau_j^{-2}\widehat{\mathbf{\Gamma}}_{jj}^{-1/2}\widehat{\mathbf{\Gamma}}_{\backslash j,\backslash j}^{-1/2}\boldsymbol{\beta}_j\|_\infty \tag{221}$$
$$= \|\widehat{\mathbf{\Gamma}}_{\backslash j,\backslash j}^{-1/2}\widehat{\mathbf{\Gamma}}_{jj}^{1/2}(\widehat{\boldsymbol{\Theta}}_{jj}\widehat{\boldsymbol{\beta}}_j - \boldsymbol{\Theta}_{jj}\boldsymbol{\beta}_j)\|_\infty \tag{222}$$
$$\leq |\widehat{\mathbf{\Gamma}}_{jj}^{1/2}\widehat{\boldsymbol{\Theta}}_{jj}|\|\widehat{\mathbf{\Gamma}}_{\backslash j,\backslash j}^{-1/2}(\widehat{\boldsymbol{\beta}}_j - \boldsymbol{\beta}_j)\|_\infty + \widehat{\mathbf{\Gamma}}_{jj}^{1/2}|\widehat{\boldsymbol{\Theta}}_{jj} - \boldsymbol{\Theta}_{jj}|\|\widehat{\mathbf{\Gamma}}_{\backslash j,\backslash j}^{-1/2}\boldsymbol{\beta}_j\|_\infty \tag{223}$$
$$= |\widehat{\mathbf{\Gamma}}_{jj}^{1/2}\widehat{\boldsymbol{\Theta}}_{jj}|\|\widehat{\mathbf{\Gamma}}_{\backslash j,\backslash j}^{-1/2}(\widehat{\boldsymbol{\beta}}_j - \boldsymbol{\beta}_j)\|_\infty + |\widehat{\boldsymbol{\Theta}}_{jj} - \boldsymbol{\Theta}_{jj}|\|\boldsymbol{\Theta}_{\backslash j,j}\boldsymbol{\Theta}_{jj}^{-1}\|_\infty \tag{224}$$
$$\leq |\widehat{\mathbf{\Gamma}}_{jj}^{1/2}\widehat{\boldsymbol{\Theta}}_{jj}|\|\widehat{\mathbf{\Gamma}}_{\backslash j,\backslash j}^{-1/2}(\widehat{\boldsymbol{\beta}}_j - \boldsymbol{\beta}_j)\|_\infty + \left|\frac{\widehat{\boldsymbol{\Theta}}_{jj}}{\boldsymbol{\Theta}_{jj}} - 1\right|\|\boldsymbol{\Theta}_{\backslash j,j}\|_\infty. \tag{225}$$

Following a similar argument as in Lemma 15, we have, on the event $\mathcal{E}$,
$$\left|\frac{\widehat{\boldsymbol{\Theta}}_{jj}}{\boldsymbol{\Theta}_{jj}} - 1\right|\|\boldsymbol{\Theta}_{\backslash j,j}\|_\infty \leq C \cdot \|\boldsymbol{\Theta}\|_1\sqrt{\frac{\log d}{n}}. \tag{226}$$

Now for the first term in (225), we have
$$|\widehat{\mathbf{\Gamma}}_{jj}^{1/2}|\|\widehat{\mathbf{\Gamma}}_{\backslash j,\backslash j}^{-1/2}(\widehat{\boldsymbol{\beta}}_j - \boldsymbol{\beta}_j)\|_\infty \tag{227}$$
$$= \|\widehat{\mathbf{\Gamma}}_{\backslash j,\backslash j}^{-1/2}\mathbf{R}_{\backslash j,\backslash j}^{-1}\mathbf{R}_{\backslash j,\backslash j}\widehat{\mathbf{\Gamma}}_{jj}^{1/2}(\widehat{\boldsymbol{\beta}}_j - \boldsymbol{\beta}_j)\|_\infty \tag{228}$$
$$\leq \|\widehat{\mathbf{\Gamma}}_{\backslash j,\backslash j}^{-1/2}\mathbf{R}_{\backslash j,\backslash j}^{-1}\|_\infty\|\mathbf{R}_{\backslash j,\backslash j}\widehat{\mathbf{\Gamma}}_{jj}^{1/2}(\widehat{\boldsymbol{\beta}}_j - \boldsymbol{\beta}_j)\|_\infty \tag{229}$$
$$\leq \|\widehat{\mathbf{\Gamma}}_{\backslash j,\backslash j}^{-1/2}\|_\infty\|\mathbf{R}_{\backslash j,\backslash j}^{-1}\|_\infty\|\mathbf{R}_{\backslash j,\backslash j}\widehat{\mathbf{\Gamma}}_{jj}^{1/2}(\widehat{\boldsymbol{\beta}}_j - \boldsymbol{\beta}_j)\|_\infty. \tag{230}$$

To analyze the term $\|\mathbf{R}_{\backslash j,\backslash j}\widehat{\mathbf{\Gamma}}_{jj}^{1/2}(\widehat{\boldsymbol{\beta}}_j - \boldsymbol{\beta}_j)\|_\infty$, we have
$$\|\mathbf{R}_{\backslash j,\backslash j}\widehat{\mathbf{\Gamma}}_{jj}^{1/2}(\widehat{\boldsymbol{\beta}}_j - \boldsymbol{\beta}_j)\|_\infty \tag{231}$$
$$\leq \|(\mathbf{R}_{\backslash j,\backslash j} - \frac{1}{n}\mathbf{Z}_{*\backslash j}^T\mathbf{Z}_{*\backslash j})\widehat{\mathbf{\Gamma}}_{jj}^{1/2}(\widehat{\boldsymbol{\beta}}_j - \boldsymbol{\beta}_j)\|_\infty + \|\frac{1}{n}\mathbf{Z}_{*\backslash j}^T\mathbf{Z}_{*\backslash j}\widehat{\mathbf{\Gamma}}_{jj}^{1/2}(\widehat{\boldsymbol{\beta}}_j - \boldsymbol{\beta}_j)\|_\infty \tag{232}$$
$$\leq \|\mathbf{R}_{\backslash j,\backslash j} - \frac{1}{n}\mathbf{Z}_{*\backslash j}^T\mathbf{Z}_{*\backslash j}\|_{\max}|\widehat{\mathbf{\Gamma}}_{jj}^{1/2}|\|\widehat{\boldsymbol{\beta}}_j - \boldsymbol{\beta}_j\|_1 + \|\frac{\widehat{\mathbf{\Gamma}}_{jj}^{1/2}}{n}\mathbf{Z}_{*\backslash j}^T\mathbf{Z}_{*\backslash j}(\widehat{\boldsymbol{\beta}}_j - \boldsymbol{\beta}_j)\|_\infty. \tag{233}$$

From (185), we have, on the event $\mathcal{E}$:

$$\left\|\widehat{\boldsymbol{\beta}}_j - \boldsymbol{\beta}_j\right\|_1 \leq 5\sqrt{2}C(1+\bar{c}).\xi_{\max}k\sqrt{\frac{\log d}{n}}. \tag{234}$$

By Lemma 7, we have, on the event $\mathcal{E}$:

$$\left\|\mathbf{R}_{\setminus j,\setminus j} - \frac{1}{n}\mathbf{Z}_{*\setminus j}^T\mathbf{Z}_{*\setminus j}\right\|_{\max} \leq \left\|\widehat{\mathbf{R}} - \mathbf{R}\right\|_{\max} \leq 18\sqrt{\frac{\log d}{n}}. \tag{235}$$

On the other hand, let $\mathbf{A}$ be an invertible matrix, we have

$$1 = \left\|\mathbf{A}\mathbf{A}^{-1}\right\|_\infty \leq \left\|\mathbf{A}\right\|_\infty\left\|\mathbf{A}^{-1}\right\|_\infty, \tag{236}$$

which implies that

$$\frac{1}{\left\|\mathbf{A}\right\|_\infty} \leq \left\|\mathbf{A}^{-1}\right\|_\infty. \tag{237}$$

Therefore,

$$\|\mathbf{R}_{\setminus j,\setminus j}^{-1}\|_\infty = \min_{\boldsymbol{\beta}\neq\mathbf{0},\boldsymbol{\beta}\in\mathbb{R}^{d-1}}\frac{\|\boldsymbol{\beta}\|_\infty}{\|\mathbf{R}_{\setminus j,\setminus j}\boldsymbol{\beta}\|_\infty} \tag{238}$$

$$\leq \min_{\boldsymbol{\beta}\neq\mathbf{0},\boldsymbol{\beta}\in\mathbb{R}^d}\frac{\|\boldsymbol{\beta}\|_\infty}{\|\mathbf{R}\boldsymbol{\beta}\|_\infty} \tag{239}$$

$$\leq \left\|\mathbf{R}^{-1}\right\|_\infty \tag{240}$$

$$\leq \max_{1\leq j\leq d}\boldsymbol{\Sigma}_{jj}\left\|\boldsymbol{\Theta}\right\|_1 \tag{241}$$

$$\leq \Lambda_{\max}(\boldsymbol{\Sigma})\left\|\boldsymbol{\Theta}\right\|_1. \tag{242}$$

From Lemma 16, we have

$$\left\|\frac{\widehat{\boldsymbol{\Gamma}}_{jj}^{1/2}}{n}\mathbf{Z}_{*\setminus j}^T\mathbf{Z}_{*\setminus j}(\widehat{\boldsymbol{\beta}}_j - \boldsymbol{\beta}_j)\right\|_\infty \leq C\sigma_j\sqrt{\frac{\log d}{n}}. \tag{243}$$

Therefore, by piecing all these terms together:

$$\left\|\mathbf{R}_{\setminus j,\setminus j}\widehat{\boldsymbol{\Gamma}}_{jj}^{1/2}(\widehat{\boldsymbol{\beta}}_j - \boldsymbol{\beta}_j)\right\|_\infty \tag{244}$$

$$\leq \widehat{\boldsymbol{\Gamma}}_{jj}^{1/2}\cdot 18\sqrt{\frac{\log d}{n}}\cdot 5\sqrt{2}C(1+\bar{c}).\xi_{\max}k\sqrt{\frac{\log d}{n}} + C\sigma_j\sqrt{\frac{\log d}{n}} \tag{245}$$

$$\leq Ck\frac{\log d}{n}\widehat{\boldsymbol{\Gamma}}_{jj}^{1/2} + C\sigma_j\sqrt{\frac{\log d}{n}}. \tag{246}$$

Again, on the event $\mathcal{E}$, we have

$$\widehat{\boldsymbol{\Gamma}}_{jj} \leq \frac{3}{2}\Lambda_{\max}(\boldsymbol{\Sigma}), \tag{247}$$

$$\|\widehat{\boldsymbol{\Gamma}}_{\backslash j,\backslash j}^{-1/2}\|_\infty \leq \frac{1}{\min_{1\leq j\leq d}(\widehat{\boldsymbol{\Sigma}}_{jj})^{1/2}} \leq \sqrt{\frac{2}{\Lambda_{\min}(\boldsymbol{\Sigma})}}, \tag{248}$$

$$\widehat{\boldsymbol{\Theta}}_{jj} \leq \left(1 + C\sqrt{\frac{\log d}{n}}\right)\boldsymbol{\Theta}_{jj} \leq 2\|\boldsymbol{\Theta}\|_2 = \frac{2}{\Lambda_{\min}(\boldsymbol{\Sigma})}, \tag{249}$$

$$\sigma_j = \frac{1}{\sqrt{\boldsymbol{\Theta}_{jj}}} \leq \frac{1}{\sqrt{\Lambda_{\min}(\boldsymbol{\Theta})}} = \sqrt{\Lambda_{\max}(\boldsymbol{\Sigma})}. \tag{250}$$

Putting all the terms in (242), (246), (248), (249) together, we get, for large enough $n$:

$$|\widehat{\boldsymbol{\Gamma}}_{jj}^{1/2}\widehat{\boldsymbol{\Theta}}_{jj}|\|\widehat{\boldsymbol{\Gamma}}_{\backslash j,\backslash j}^{-1/2}(\widehat{\boldsymbol{\beta}}_j - \boldsymbol{\beta}_j)\|_\infty \tag{251}$$

$$\leq \frac{2}{\Lambda_{\min}(\boldsymbol{\Sigma})}\sqrt{\frac{2}{\Lambda_{\min}(\boldsymbol{\Sigma})}} \cdot \Lambda_{\max}(\boldsymbol{\Sigma})\|\boldsymbol{\Theta}\|_1 \cdot \left(Ck\frac{\log d}{n}\widehat{\boldsymbol{\Gamma}}_{jj}^{1/2} + C\sigma_j\sqrt{\frac{\log d}{n}}\right) \tag{252}$$

$$\leq 2\xi_{\max}\sqrt{\frac{2}{\Lambda_{\min}(\boldsymbol{\Sigma})}}\|\boldsymbol{\Theta}\|_1 \cdot \left(Ck\frac{\log d}{n}\cdot\sqrt{\frac{3}{2}\Lambda_{\max}(\boldsymbol{\Sigma})} + C\sqrt{\Lambda_{\max}(\boldsymbol{\Sigma})}\sqrt{\frac{\log d}{n}}\right) \tag{253}$$

$$\leq 2\sqrt{3}\xi_{\max}^{3/2}\cdot C\cdot\|\boldsymbol{\Theta}\|_1 \cdot k\frac{\log d}{n} + 2\sqrt{2}\xi_{\max}^{3/2}\cdot C\cdot\|\boldsymbol{\Theta}\|_1 \cdot \sqrt{\frac{\log d}{n}} \tag{254}$$

$$\leq 2(\sqrt{3}+\sqrt{2})\xi_{\max}^{3/2}\cdot C\cdot\|\boldsymbol{\Theta}\|_1 \cdot \sqrt{\frac{\log d}{n}}, \tag{255}$$

where the last inequality follows from the fact that $k\sqrt{\log d} \leq \sqrt{n}$ for large enough $n$. Therefore, the lemma is proved. $\square$

## D  Proof of the Main Theorems

In this section, we prove the main results based on the previous technical lemmas.

## D.1 Proof of Theorem 2

*Proof.* By piecing together the results of Lemma 14 and Lemma 15, we have

$$\begin{align}
\left\|\widehat{\boldsymbol{\Theta}} - \boldsymbol{\Theta}\right\|_1 &= \max_{1 \leq j \leq d} \|\widehat{\boldsymbol{\Theta}}_{*j} - \boldsymbol{\Theta}_{*j}\|_1 \tag{256}\\
&\leq \max_{1 \leq j \leq d} |\widehat{\boldsymbol{\Theta}}_{jj} - \boldsymbol{\Theta}_{jj}| + \max_{1 \leq j \leq d} \|\widehat{\boldsymbol{\Theta}}_{\setminus j,j} - \boldsymbol{\Theta}_{\setminus j,j}\|_1 \tag{257}\\
&\leq C \cdot \|\boldsymbol{\Theta}\|_2 \sqrt{\frac{\log d}{n}} + C(\|\boldsymbol{\Theta}\|_2 k + \|\boldsymbol{\Theta}\|_1) \sqrt{\frac{\log d}{n}} \tag{258}\\
&\leq C(\|\boldsymbol{\Theta}\|_2 k + \|\boldsymbol{\Theta}\|_1) \sqrt{\frac{\log d}{n}} \tag{259}\\
&\leq C\left(k \|\boldsymbol{\Theta}\|_2 \sqrt{\frac{\log d}{n}}\right). \tag{260}
\end{align}$$

The last inequality follows from the fact that

$$\|\boldsymbol{\Theta}\|_1 \leq k \|\boldsymbol{\Theta}\|_{\max} \leq k \|\boldsymbol{\Theta}\|_2. \tag{261}$$

We complete the proof of this theorem. □

## D.2 Proof of Theorem 3

*Proof.* By piecing together the results of Lemma 14 and Lemma 17, we have

$$\left\|\widehat{\boldsymbol{\Theta}} - \boldsymbol{\Theta}\right\|_{\max} = \max_{1 \leq j \leq d} |\widehat{\boldsymbol{\Theta}}_{jj} - \boldsymbol{\Theta}_{jj}| + \max_{1 \leq j \leq d} \|\widehat{\boldsymbol{\Theta}}_{\setminus j,j} - \boldsymbol{\Theta}_{\setminus j,j}\|_\infty \leq C \|\boldsymbol{\Theta}\|_1 \sqrt{\frac{\log d}{n}}. \tag{262}$$

We complete the proof of this theorem. □

# References

AKAIKE, H. (1973). Information theory and an extension of the maximum likelihood principle. *Second International Symposium on Information Theory* 267–281.

BACH, F. R. (2008). Bolasso: model consistent lasso estimation through the bootstrap. In *In Proceedings of the Twenty-fifth International Conference on Machine Learning (ICML)*.

BANERJEE, O., GHAOUI, L. E. and D'ASPREMONT, A. (2008). Model selection through sparse maximum likelihood estimation. *Journal of Machine Learning Research* **9** 485–516.

BARABASI, A. L. and ALBERT, R. (1999). Emergence of scaling in random networks. *Science (New York, N.Y.)* **286** 509–512.

Belloni, A., Chernozhukov, V. and Wang, L. (2012). Square-root lasso: Pivotal recovery of sparse signals via conic programming. *Biometrika* **98** 791–806.

Ben-david, S., Luxburg, U. V. and Pal, D. (2006). A sober look at clustering stability. In *Proceedings of the Annual Conference of Learning Theory*. Springer.

Bishop, C., Spiegelhalter, D. and Winn, J. (2003). VIBES: A variational inference engine for Bayesian networks. In *Advances in Neural Information Processing Systems 15* (S. Becker, S. Thrun and K. Obermayer, eds.). MIT Press, Cambridge, MA, 777–784.

Boos, D. D., Stefanski, L. A. and Wu, Y. (2009). Fast fsr variable selection with applications to clinical trials. *Biometrics* **65** 692–700.

Cai, T., Liu, W. and Luo, X. (2011a). A constrained $\ell_1$ minimization approach to sparse precision matrix estimation. *Journal of the American Statistical Association* **106** 594–607.

Cai, T., Liu, W. and Zhou, H. (2011b). Minimax rates of convergence for sparse inverse covariance matrix estimation. *Technical Report, University of Pennsylvania* .

Chen, J. and Chen, Z. (2008). Extended bayesian information criteria for model selection with large model spaces. *Biometrika* **95** 759–771.

Chen, J. and Chen, Z. (2012). Extended bic for small-n-large-p sparse glm. *Statistica Sinica* **22** 555–574.

Chen, S., Donoho, D. and Saunders, M. (1998). Atomic decomposition by basis pursuit. *SIAM Journal on Scientific Computing* **20** 33–61.

Dempster, A. (1972). Covariance selection. *Biometrics* **28** 157–175.

Drton, M. and Perlman, M. (2007). Multiple testing and error control in Gaussian graphical model selection. *Statistical Science* **22** 430–449.

Drton, M. and Perlman, M. (2008). A SINful approach to Gaussian graphical model selection. *Journal of Statistical Planning and Inference* **138** 1179–1200.

Efron, B. (1982). *The jackknife, the bootstrap and other resampling plans*. SIAM [Society for Industrial and Applied Mathematics].

Efron, B. (2004). The estimation of prediction error: Covariance penalties and cross-validation. *Journal of the American Statistical Association* **99** 619–632.

Foster, D. P. and George, E. I. (1994). The risk inflation criterion for multiple regression. *The Annals of Statistics* **22** 1947–1975.

Foygel, R. and Drton, M. (2010). Extended bayesian information criteria for gaussian graphical models. In *Advances in Neural Information Processing Systems 23*.

Friedman, J., Hastie, T. and Tibshirani, R. (2008). Sparse inverse covariance estimation with the graphical lasso. *Biostatistics* **9** 432–441.

Friedman, J., T. Hastie, H. H. and Tibshirani, R. (2007). Pathwise coordinate optimization. *Annals of Applied Statistics* **1** 302–332.

Jalali, A., Johnson, C. and Ravikumar, P. (2012). High-dimensional sparse inverse covariance estimation using greedy methods. *International Conference on Artificial Intelligence and Statistics* .

Johnstone., I. (2000). *Chi-square oracle inequalities*. Lecture Notes-Monograph Series.

Lafferty, J., Liu, H. and Wasserman, L. (2012). Sparse Nonparametric Graphical Models. *Statistical Science* To appear.

Lam, C. and Fan, J. (2009). Sparsistency and rates of convergence in large covariance matrix estimation. *Annals of Statistics* **37** 42–54.

Lange, T., Roth, V., Braun, M. L. and Buhmann, J. M. (2004). Stability-based validation of clustering solutions. *Neural Computation* **16** 1299–1323.

Laurent, B. and Massart, P. (1998). Adaptive estimation of a quadratic functional by model selection. *Annals of Statistics* **28** 1303–1338.

Liu, H., Chen, X., Lafferty, J. and Wasserman, L. (2010a). Graph-valued regression. In *Proceedings of the Twenty-Third Annual Conference on Neural Information Processing Systems (NIPS)*.

Liu, H., Han, F., Yuan, M., Lafferty, J. and Wasserman, L. (2012). High dimensional semiparametric gaussian copula graphical models. *The Annals of Statistics (to appear)* .

Liu, H., Roeder, K. and Wasserman, L. (2010b). Stability approach to regularization selection (stars) for high dimensional graphical models. In *Proceedings of the Twenty-Third Annual Conference on Neural Information Processing Systems (NIPS)*.

Liu, H., Xu, M., Gu, H., Gupta, A., Lafferty, J. and Wasserman, L. (2011). Forest density estimation. *Journal of Machine Learning Research* **12** 907–951. A short version has appeared in the 23rd Annual Conference on Learning Theory (COLT).

Liu, W. and Luo, X. (2012). High-dimensional sparse precision matrix estimation via sparse column inverse operator. *arXiv/1203.3896* .

Lysen, S. (2009). Permuted inclusion criterion: A variable selection technique. *Publicly accessible Penn Dissertations. Paper 28* .

Mallows, C. L. (1973). Some comments on $C_p$. *Technometrics* **15** 661–675.

Meinshausen, N. and Bühlmann, P. (2006). High dimensional graphs and variable selection with the lasso. *Annals of Statistics* **34** 1436–1462.

Meinshausen, N. and Bühlmann, P. (2010). Stability selection. *Journal of the Royal Statistical Society, Series B, Methodological* **72** 417–473.

Raskutti, G., Wainwright, M. J. and Yu, B. (2010). Restricted eigenvalue properties for correlated gaussian designs. *Journal of Machine Learning Research* **99** 2241–2259.

Ravikumar, P., Wainwright, M., Raskutti, G. and Yu, B. (2011). High-dimensional covariance estimation by minimizing $\ell_1$-penalized log-determinant divergence. *Electronic Journal of Statistics* **5** 935–980.

Rothman, A., Bickel, P., Levina, E. and Zhu, J. (2008). Sparse permutation invariant covariance estimation. *Electronic Journal of Statistics* **2** 494–515.

Schäfer, J. and Strimmer, K. (2005). A Shrinkage Approach to Large-Scale Covariance Matrix Estimation and Implications for Functional Genomics. *Statistical Applications in Genetics and Molecular Biology* **4**.

Schwarz, G. (1978). Estimating the dimension of a model. *The Annals of Statistics* **6** 461–464.

Shen, X., Pan, W. and Zhu, Y. (2012). ). likelihood-based selection and sharp parameter estimation. *Journal of the American Statistical Association* To appear.

Sun, T. and Zhang, C.-H. (2012). Sparse matrix inversion with scaled lasso. Tech. rep., Department of Statistics, Rutgers University.

Tibshirani, R. (1996). Regression shrinkage and selection via the lasso. *Journal of the Royal Statistical Society, Series B* **58** 267–288.

Wille, A., Zimmermann, P., Vranova, E., Frholz, A., Laule, O., Bleuler, S., Hennig, L., Prelic, A., von Rohr, P., Thiele, L., Zitzler, E., Gruissem, W. and Bühlmann, P. (2004). Sparse graphical gaussian modeling of the isoprenoid gene network in arabidopsis thaliana. *Genome Biology* **5** R92.

Wu, Y., Boos, D. D. and Stefanski, L. A. (2007). Controlling variable selection by the addition of pseudovariables. *Journal of the American Statistical Association* **102** 235–243.

Yuan, M. (2010). High dimensional inverse covariance matrix estimation via linear programming. *Journal of Machine Learning Research* **11** 2261–2286.

Yuan, M. and Lin, Y. (2007). Model selection and estimation in the gaussian graphical model. *Biometrika* **94** 19–35.

Zhao, T., Liu, H., Roeder, K., Lafferty, J. and Wasserman, L. (2012). The huge package for high-dimensional undirected graph estimation in r. *Journal of Machine Learning Research* To appear.